\documentclass[12pt,a4paper]{article}
\usepackage{graphicx}
\usepackage{subeqnarray}
\begin{document}
%
%
%
%
%
\def\astrobj#1{#1}
\newenvironment{lefteqnarray}{\arraycolsep=0pt\begin{eqnarray}}
{\end{eqnarray}\protect\aftergroup\ignorespaces}
\newenvironment{lefteqnarray*}{\arraycolsep=0pt\begin{eqnarray*}}
{\end{eqnarray*}\protect\aftergroup\ignorespaces}
\newenvironment{leftsubeqnarray}{\arraycolsep=0pt\begin{subeqnarray}}
{\end{subeqnarray}\protect\aftergroup\ignorespaces}
\newcommand{\diff}{{\rm\,d}}
\newcommand{\pprime}{{\prime\prime}}
\newcommand{\ppprime}{{\prime\prime\prime}}
\newcommand{\szeta}{\mskip 3mu /\mskip-10mu \zeta}
\newcommand{\FC}{\mskip 0mu {\rm F}\mskip-10mu{\rm C}}
\newcommand{\appleq}{\stackrel{<}{\sim}}
\newcommand{\appgeq}{\stackrel{>}{\sim}}
\newcommand{\Int}{\mathop{\rm Int}\nolimits}
\newcommand{\Nint}{\mathop{\rm Nint}\nolimits}
\newcommand{\sgn}{\mathop{\rm sgn}\nolimits}
\newcommand{\range}{{\rm -}}
\newcommand{\displayfrac}[2]{\frac{\displaystyle #1}{\displaystyle #2}}
\newcommand{\mmatrix}[1]{\left|\left|\matrix{#1}\right|\right|}
\def\astrobj#1{#1}
%
\title{Bivariate least squares linear regression: \\
towards a unified analytic formalism. \\
II. Extreme structural models}
\author{{R.~Caimmi}\footnote{
{\it Physics and Astronomy Department, Padua Univ., Vicolo Osservatorio 3/2,
I-35122 Padova, Italy}
email: roberto.caimmi@unipd.it~~~
fax: 39-049-8278212}
\phantom{agga}}
%
%
\maketitle
\begin{quotation}
\section*{}
\begin{Large}
\begin{center}

Abstract

\end{center}
\end{Large}
\begin{small}

\noindent\noindent
Concerning bivariate least squares linear regression, the
classical results obtained for extreme structural
models in earlier attempts (Isobe et al., 1990;
Feigelson and Babu, 1992) are
reviewed using a new formalism in terms
of deviation (matrix) traces
which, for homoscedastic data, reduce to usual
quantities leaving aside an
unessential (but dimensional) multiplicative factor.
Within the framework of classical error models,
the dependent variable relates to the independent
variable according to a variant of the usual additive model.
The classes of linear models considered are
regression lines in the limit of
uncorrelated errors in $X$ and in $Y$.
The following models are 
considered in detail: (Y) errors
in $X$ negligible (ideally null) with
respect to errors in $Y$; (X) errors
in $Y$ negligible (ideally null) with
respect to errors in $X$; (C) oblique
regression; (O) orthogonal regression;
(R) reduced major-axis regression;
(B) bisector regression.
For homoscedastic data, the results are
taken from earlier attempts and rewritten
using a more compact notation.   For
heteroscedastic data, the results are
inferred from a procedure related to
functional models (York, 1966; Caimmi, 2011).
An example of astronomical
application is considered, concerning
the [O/H]-[Fe/H] empirical relations
deduced from five samples related to
different stars and/or different methods
of oxygen abundance determination.
For low-dispersion samples and assigned
methods, different regression models
yield results which are in agreement
within the errors $(\mp\sigma)$ for both
heteroscedastic and homoscedastic data,
while the contrary holds for large-dispersion
samples.    In any case,
samples related to different
methods produce discrepant results,
due to the presence of (still undetected)
systematic errors, which implies no
definitive statement can be made at
present.   Asymptotic expressions
approximate regression line slope
and intercept variance estimators,
for normal residuals, to a better
extent with respect to earlier
attempts.   Related fractional
discrepancies are not exceeding
a few percent for low-dispersion
data, which grows up to about 10\%
for large-dispersion data.
An extension of the formalism to
generic structural models
is left to a forthcoming paper.

\noindent
{\it keywords - 
galaxies: evolution - stars: formation; evolution - methods: data analysis -
methods: statistical.}

pacs codes: 98.62.-g; 97.10.Cv; 02.50.-r
\end{small}
\end{quotation}

\section{Introduction} \label{intro}

\noindent\noindent

Linear regression is a fundamental and frequently
used statistical tool in almost all branches of
science, among which astronomy.
The related problem is twofold:
regression line slope and intercept estimators
are expressed involving minimizing or maximizing
some function of the data;
on the other hand, regression line slope and
intercept variance estimators are expressed
requiring knowledge of the error distributions
of the data.
The complexity mainly arises from the
occurrence of intrinsic dispersion in
addition to the dispersion related to
the measurement processes (hereafter
quoted as instrumental dispersion),
where the distribution corresponding
to the former can be different from
the distribution corresponding to the
latter i.e. non Gaussian (non normal).

In statistics, problems where the true
points have fixed but unknown coordinates
are called functional regression
models, while problems where the true points
have random (i.e. obeying their own intrinsic
distribution) and unknown coordinates are called
structural regression models.
Accordingly, functional regression
models may be conceived as structural
regression models where the intrinsic
dispersion is negligible (ideally null)
with respect to the instrumental
dispersion.   Conversely, structural
regression models where the instrumental
dispersion is negligible (ideally null)
with respect to the intrinsic dispersion,
can be defined as extreme structural
models (Caimmi, 2011, hereafter quoted
as C11).   A distinction between
functional and structural  modelling
is currently preferred, where the former
can be affected by intrinsic scatter
but with no or only minimal assumptions
on related distributions, while the
latter implies (usually parametric)
models are placed on the above mentioned
distributions.   For further details
refer to specific textbooks (e.g.,
Carroll et al., 2006, Chap.\,2, \S 2.1).
In addition, models
where the instrumental dispersion
is the same from point to point for
each variable, are called homoscedastic
models, while models where the
instrumental dispersion is (in general)
different from point to point, are called
heteroscedastic models.   Similarly,
related data are denoted as homoscedastic
and heteroscedastic, respectively.

In general,
problems where the true points lie precisely
on an expected line relate to functional
regression models, while problems where the
true points are (intrinsically) scattered
about an expected line relate to structural
regression models
(e.g., Feigelson and Babu, 1992, erratum,
2011, hereafter quoted together as FB92 if
not otherwise specified).

Bivariate least squares linear regression
related to heteroscedastic
functional models with uncorrelated and
correlated errors, following
Gaussian distributions, were
analysed and formulated in two classical
papers (York, 1966; 1969; hereafter quoted
as Y66 and Y69, respectively), where
regression line slope and intercept variance
estimators are determined using the method
of partial differentiation (Y69).   On the
contrary, the method of moments estimator
is used to this aim in later attempts [e.g.,
Fuller, 1987 (hereafter quoted as F87),
Chap.\,1, \S 1.3.2, Eq.\,(1.3.7) therein; FB92].

Bivariate
least squares linear regression
related to extreme structural
models, where the instrumental dispersion
is negligible (ideally null) with respect
to intrinsic dispersion, was exhaustively
treated in two classical papers
(Isobe et al., 1990, hereafter quoted as
Ia90; FB92) and extended to generic
structural models in a later attempt
(Akritas and Bershady, 1996, hereafter
quoted as AB96).

The above mentioned
papers provide the simplest description
of linear regression.   In reality,
biases and additional effects must be
taken into consideration, which implies
much more complicated description and
formulation, as it can be seen in specific
monographies (e.g., F87; Carroll
et al., 2006; Buonaccorsi, 2010).
Restricting to the astronomical literature,
a recent investigation (Kelly, 2007) is
particularly relevant in that 
it is the first example (in the
field under discussion) where
linear regression is considered
following the modern (since about
half a century ago) approach based
on likelihoods rather than the old
(up to about a century ago) least-squares
approach.   More specifically, a
hierarchical measurement error
model is set up therein, the
complicated likelihood is written
down, and a variety of minimum
least-squares and Bayesan solutions
are shown, which can treat functional,
structural, multivariate, truncated
and censored mesaurement error
regression problems.

Even in dealing with the simplest
homoscedastic (or heteroscedastic)
functional and structural models,
still no unified analytic formalism has been
developed (to the knowledge of the
author) where (i) structural
heteroscedastic models with
instrumental and intrinsic
dispersion of comparable order
in both variables, are considered;
(ii) previous results are recovered
in the limit of dominant instrumental
dispersion; and (iii) previous results
are recovered in the limit of dominant
intrinsic dispersion.   A related
formulation may be useful also for
computational methods, in the sense
that both the general case and
limiting situations can be described
by a single numerical code.

A first step towards a unified analytic
formalism
of bivariate least squares linear regression
involving functional models has been
performed in an earlier attempt (C11),
where the least-squares approach developed
in two classical papers (Y66; Y69) has been
reviewed and reformulated
by definition and use of deviation
(matrix) traces.   The current investigation
aims at making a second step along the same
direction, in dealing with extreme structural
models.

More specifically, the results
found in two classical papers (Ia90; FB92)
shall be reformulated in terms of deviation
traces for homoscedastic models, and extended
to the general case of heteroscedastic models
by analogy with their counterparts related to
functional models, within the framework
of classical error models where
the dependent variable relates to the independent
variable according to a variant of the classical
additive error model.

In this view, homoscedastic structural
models are conceived as models where
both the instrumental and the intrinsic dispersion
are the same from point to point.   Conversely,
models where the instrumental and/or the intrinsic
dispersion are (in general) different from point to
point, are conceived as heteroscedastic structural models.

Regression line slope and intercept estimators,
and related variance estimators, are expressed
in terms of deviation traces for different
homoscedastic models (Ia90; FB92) in section
\ref{lesmo}, where an extension to corresponding
heteroscedastic models is also performed, and
both normal and non normal residuals are considered.
An example of astronomical application
is outlined in section \ref{apfm}.
The discussion is presented in section
\ref{disc}.   Finally, the conclusion
is shown in section \ref{conc}.
Some points are developed with more
detail in the Appendix.
An extension of the formalism to generic
structural models is
left to a forthcoming paper.

\section{Least-squares fitting of a straight line} \label{lesmo}

\subsection{General considerations} \label{geco}

\noindent\noindent

Attention shall be restricted to the classical
problem of least-squares fitting of a straight
line, where both variables are measured with
errors.   Without loss of generality, structural
models can be conceived as related to an ideal
situation where the variables obey a linear relation, as:
\begin{equation}
\label{eq:rlf}
y_i^\ast=ax_i^\ast+b~~;\qquad1\le i\le n~~;
\end{equation}
in connection with true points, ${\sf P}_i^\ast\equiv(x_i^\ast, y_i^\ast)$,
$1\le i\le n$.   The occurrence of random (measure independent) processes
makes true points shift
outside or along the ideal straight line, inferred from Eq.\,(\ref{eq:rlf}),
towards actual points, ${\sf P}_{{\rm S}i}^\ast\equiv(x_{{\rm S}i},
y_{{\rm S}i})$.   The occurrence of mesaurement processes makes the actual
points shift towards the observed points, ${\sf P}_i\equiv(X_i,Y_i)$.

In this view, the least squares fitting of a straight line is conceptually
similar for functional (in absence of intrinsic scatter) and structural
(in presence of intrinsic scatter) models: ``What is the best line fitting to a
sample of observed points, ${\sf P}_i$, $1\le i\le n$?''  It is worth noticing
the correspondence between true points, ${\sf P}_i^\ast$, and observed points,
${\sf P}_i$, is not one-to-one unless it is assumed all points are shifted
along the same direction.   More specifically, two observed points,
${\sf P}_i$, ${\sf P}_j$, with equal coordinates, $(X_i,Y_i)=(X_j,Y_j)$,
relate to true points, ${\sf P}_i^\ast$, ${\sf P}_j^\ast$, with (in general)
different coordinates, $(x_i^\ast, y_i^\ast)\ne(x_j^\ast, y_j^\ast)$, both in
presence and in absence of intrinsic scatter.
The least-square estimator and the loss function have the same formal
expression for functional and structural models, but in the latter case the
``statistical distances'' (e.g., F87, Chap.\,1, \S1.3.3) depend on the total
(instrumental + intrinsic) scatter.

The observed points and the actual points are related as:
\begin{lefteqnarray}
\label{eq:erm}
&& Z_i=z_{{\rm S}i}+(\xi_{{\rm F}_z})_i~~;\quad Z=X,Y~~;\quad z=x,y~~;\quad
1\le i\le n~~;
\end{lefteqnarray}
where $(\xi_{{\rm F}_x})_i$, $(\xi_{{\rm F}_y})_i$,
are the instrumental
(i.e. due to the intrumental scatter) errors on
$x_{{\rm S}i}$, $y_{{\rm S}i}$, respectively, assumed
to obey Gaussian distributions with null
expectation values and known variances,
$[(\sigma_{xx})_{\rm F}]_i$, $[(\sigma_{yy})_{\rm F}]_i$,
and covariance, $[(\sigma_{xy})_{\rm F}]_i$.

The actual points and the true points on the ideal straight line are related
as:
\begin{lefteqnarray}
\label{eq:csiS}
&& z_{{\rm S}i}=z_i^\ast+(\xi_{{\rm S}_z})_i~~;\quad z=x,y;\quad1\le i\le n~~;
\end{lefteqnarray}
where $(\xi_{{\rm S}_x})_i$, $(\xi_{{\rm S}_y})_i$,
are the intrinsic
(i.e. due to the intrinsic scatter) errors on
$x_i^\ast$, $y_i^\ast$, respectively, assumed
to obey specified distributions with null
expectation values and finite variances,
$[(\sigma_{xx})_{\rm S}]_i$, $[(\sigma_{yy})_{\rm S}]_i$,
and covariance, $[(\sigma_{xy})_{\rm S}]_i$.

The observed points and the true points on the ideal straight line are related
as:
\begin{lefteqnarray}
\label{eq:er}
&& Z_i=z_i^\ast+\xi_{z_i}~~;\quad Z=X,Y;\quad z=x,y~~;\quad1\le i\le n~~;
\end{lefteqnarray}
where the (instrumental + intrinsic)
errors, $\xi_{x_i}$, $\xi_{y_i}$, are
defined as:
\begin{lefteqnarray}
\label{eq:csi}
&& \xi_{z_i}=(\xi_{{\rm F}_z})_i+(\xi_{{\rm S}_z})_i~~;\quad z=x,y~~;\quad
1\le i\le n~~;
\end{lefteqnarray}
which obey specified distributions
with null expectation values and
finite variances, $(\sigma_{xx})_i$,
$(\sigma_{yy})_i$, and covariance,
$(\sigma_{xy})_i$.   The further
restriction that $(\xi_{{\rm F}_z})_i$,
$(\xi_{{\rm S}_z})_i$, $z=x,y$,
$1\le i\le n$, are independent,
implies the relation (AB96):
\begin{equation}
\label{eq:szz}
(\sigma_{zz})_i=[(\sigma_{zz})_
{\rm F}]_i+[(\sigma_{zz})_{\rm S}]_i~~;\qquad
(\sigma_{xy})_i=[(\sigma_{xy})_{\rm F}]_i+
[(\sigma_{xy})_{\rm S}]_i~~;
\end{equation}
where the intrinsic covariance matrixes
are unknown and must be assigned or
estimated, which will be supposed in the following.

Then the error model is defined by Eqs.\,(\ref{eq:rlf})-(\ref{eq:szz}), where 
%
%
both instrumental errors, $(\xi_{{\rm F}_z})_i$, and intrinsic errors,
$(\xi_{{\rm S}_z})_i$, are assumed to be independent of true values,
$z_i^\ast$, for given instrumental covariance matrixes,
$(\Sigma_{\rm F})_i=\mmatrix{[(\sigma_{xy})_{\rm F}]_i}$, intrinsic
covariance matrixes,
$(\Sigma_{\rm S})_i=\mmatrix{[(\sigma_{xy})_{\rm S}]_i}$,
respectively, and (total) covariance matrixes,
$\Sigma_i=\mmatrix{(\sigma_{xy})_i}$, hence
$\Sigma_i=(\Sigma_{\rm F})_i+(\Sigma_{\rm S})_i$,
$1\le i\le n$.   It may be considered as a variant of the classical additive
error model (e.g., AB96; Carrol et al., 2006, Chap.\,1, \S1.2, Chap.\,3,
\S3.2.1; Kelly, 2007, 2011; Buonaccorsi, 2010, Chap.\,4, \S4.3).

In the case under discussion, the regression
estimator minimizes the loss function, defined as the sum (over the $n$
observations) of squared residuals (e.g.,
Y69), or statistical distances of the
observed points, ${\sf P}_i\equiv(X_i,Y_i)$,
from the estimated line in the unknown
parameters, $a, b, x_1, ..., x_n$ (e.g.,
F87, Chap.\,1, \S 1.3.3).   Under
restrictive assumptions, the regression
estimator is the functional maximum
likelihood estimator (e.g., Carroll et al.,
2006, Chap.\,3, \S 3.4.2).

The coordinates, $(x_i, y_i)$, may be
conceived as the adjusted values of
related observations, $(X_i, Y_i)$,
on the estimated regression line
(Y66; Y69)
and, in addition, as estimators of the
coordinates, $(x_i^\ast, y_i^\ast)$,
on the true regression line i.e. the ideal straight line.
The line of adjustment, $\overline
{{\sf P}_i\hat{{\sf P}}_i}$ (e.g.,
Y69), may be conceived as an estimator
of the statistical distance, $\overline
{{\sf P}_i{\sf P}_i^\ast}$ (e.g.,
F87, Chap.\,1, \S 1.3.3),
where $\hat{\sf P}_i(x_i, y_i)$ is the
adjusted point on the estimated
regression line:
\begin{equation}
\label{eq:esl}
y_i=\hat{a}x_i+\hat{b}~~;\qquad1\le i\le n~~;
\end{equation}
where, in general, estimators are
denoted by hats, and
${\sf P}_i^\ast(x_i^\ast,
y_i^\ast)$ is the true point on the
ideal straight line, Eq.\,(\ref{eq:rlf}).

To the knowledge of the author, only
classical error models are considered for
astronomical applications, and for this
reason different error models such as
Berkson models and mixture error models
(e.g., Carroll et al., 2006, Chap.\,3, Sect.\,3.2)
shall not be dealt with in the current
attempt.   From this point on, investigation
shall be limited to extreme structural models and
least-squares regression estimators for the
following reasons.   First, they are important
models in their own right, furnishing an
approximation to real world situations.
Second, a careful examination of these
simple models helps for understanding
the theoretical underpinnings of methods
for other models of greater complexity
such as hierarchical models (e.g., Kelly, 2007, 2011).

\subsection{Extreme structural models} \label{esmo}

\noindent\noindent

With regard to extreme structural models, bivariate
least squares linear regression were analysed in two
classical papers in the special case of oblique
regression i.e. constant variance ratio,
$(\sigma_{yy})_i/(\sigma_{xx})_i=c^2$, $1\le i\le n$,
and constant correlation coefficients, $r_i=r$, $1\le
i\le n$.   More specifically, orthogonal $(c^2=1)$
and oblique regression were analysed in the earlier
(Ia90) and in the latter (FB92) paper, respectively.
In absence of additional information, homoscedastic
models are used (Ia90) unless the intrinsic dispersion
is estimated (AB96), from which related weights may be
determined and the least squares estimator together
with the loss function may be expressed for both homoscedastic
and heteroscedastic models (AB96; Kelly, 2011).

The (dimensionless) squared weighted residuals can be defined
as in the case of functional models (Y69):
\begin{leftsubeqnarray}
\slabel{eq:R2ga}
&& (\widetilde{R}_i)^2=\frac{w_{x_i}(X_i-x_i)^2+w_{y_i}(Y_i-y_i)^2-2r_i
\sqrt{w_{x_i}w_{y_i}}(X_i-x_i)(Y_i-y_i)}{1-r_i^2}~~;\qquad \\
\slabel{eq:R2gb}
&& r_i=\frac{(\sigma_{xy})_i}{[(\sigma_{xx})_i(\sigma_{yy})_i]^{1/2}}~~;
\qquad\vert r_i\vert\le1~~;\qquad1\le i\le n~~;
\label{seq:R2g}
\end{leftsubeqnarray}
where $w_{x_i}$, $w_{y_i}$, are the
weights of the various measurements
(or observations) and $r_i$ the correlation
coefficients.   The terms, $w_{x_i}(X_i-x_i)^2$,
$w_{y_i}(Y_i-y_i)^2$, $r_i$, $1\le i\le n$, are
dimensionless by definition.   An equivalent formulation
in matrix formalism can be found in specific
textbooks, where weighted true residuals are
conceived as (dimensionless) ``statistical
distances'' from data points to related
points on the regression line [e.g., F87,
Chap.\,1, \S 1.3.3, Eq.\,(1.3.16)].

Accordingly, the least-squares regression
estimator and the loss function can be expressed as in the case
of functional models (C11) but the weights,
$w_{x_i}$, $w_{y_i}$, and the correlation
coefficients, $r_i$, are related to intrinsic
scatter instead of instrumental scatter.
Then the regression line slope and
intercept estimators take the same formal
expression with respect to their counterparts
related to functional models, while (in general)
the contrary holds for regression line slope
and intercept variance estimators.

Classical results on extreme structural models
(Ia90; FB92) are restricted to oblique
regression for homoscedastic data
with constant correlation coefficients
($w_{x_i}=w_x$, $w_{y_i}=w_y$, $r_i=r$, $1\le i\le n$).
In the following subsections, the above mentioned
results extended to heteroscedastic data
shall be expressed in terms of weighted
deviation (matrix) traces (C11):
\begin{lefteqnarray}
\label{eq:wgc}
&& \widetilde{Q}_{pq}=\sum_{i=1}^nQ_i(w_{x_i},w_{y_i},r_i)
(X_i-\widetilde{X})^p(Y_i-\widetilde{Y})^q~~; \\
\label{eq:w00}
&& \widetilde{Q}_{00}=\sum_{i=1}^nQ_i(w_{x_i},w_{y_i},r_i)=n\overline{Q}~~;
\end{lefteqnarray}
where $\widetilde{Q}_{pq}$ are the (weighted)
pure ($p=0$ and/or $q=0$)
and mixed ($p>0$ and $q>0$) deviation
traces, and $\widetilde{X}$, $\widetilde{Y}$,
are weighted means:
\begin{lefteqnarray}
\label{eq:wZa}
&& \widetilde{Z}=\displayfrac{\sum_{i=1}^nW_iZ_i}{\sum_{i=1}^nW_i}~~;\qquad
Z=X,Y~~; \\
\label{eq:Wi}
&& W_i=\frac{w_{x_i}\Omega_i^2}{1+a^2\Omega_i^2-2ar_i\Omega_i}~~;\qquad
1\le i\le n~~; \\
\label{eq:wOg}
&& \Omega_i=\sqrt{\frac{w_{y_i}}{w_{x_i}}}~~;\qquad1\le i\le n~~;
\end{lefteqnarray}
in the limit of homoscedastic data with equal
correlation coefficients,
$w_{x_i}=w_x$, $w_{y_i}=w_y$, $r_i=r$, $1\le i\le n$, which
implies $Q_i(w_{x_i},w_{y_i},r_i)=Q(w_x,w_y,r)=Q$,
Eqs.\,(\ref{eq:wgc}), (\ref{eq:w00}), (\ref{eq:wZa}),
(\ref{eq:Wi}), and (\ref{eq:wOg}) reduce to:
\begin{lefteqnarray}
\label{eq:Qun}
&& \widetilde{Q}_{pq}=QS_{pq}~~; \\
\label{eq:Spq}
&& S_{pq}=\sum_{i=1}^n(X_i-\overline{X})^p(Y_i-\overline{Y})^q~~; \\
\label{eq:Q00}
&& \widetilde{Q}_{00}=QS_{00}~~; \\
\label{eq:S00}
&& S_{00}=n~~; \\
\label{eq:Zpa}
&& \widetilde{Z}=\overline{Z}~~;\qquad Z=X,Y~~; \\
\label{eq:Wia}
&& W_i=W=\frac{w_x\Omega^2}{1+a^2\Omega^2-2ar\Omega}~~;\qquad
1\le i\le n~~; \\
\label{eq:wOa}
&& \Omega_i=\Omega=\sqrt{\frac{w_y}{w_x}}~~;\qquad1\le i\le n~~;
\end{lefteqnarray}
where $S_{pq}$ are the (unweighted)
pure ($p=0$ and/or $q=0$)
and mixed ($p>0$ and $q>0$) deviation
traces.

Turning to the general case and using
the weighted squared error loss function,
$T_{\widetilde{R}}=\sum_{i=1}^n(\widetilde{R}_i)^2$,
yields for regression line slope and
intercept estimators the same expression
with respect to functional models (C11).
Accordingly, regression line slope
and intercept estimators may be conceived
similarly to state functions in thermodynamics:
for an assigned true point,
${\sf P}_i^\ast\equiv(x_i^\ast, y_i^\ast)$,
what is relevant is the related observed point,
${\sf P}_i\equiv(X_i,Y_i)$, regardless of
the path followed via instrumental and/or
intrinsic scatter.   More specifically,
the regression line intercept estimator
obeys the equation (e.g., Y69; C11):
\begin{equation}
\label{eq:bar}
\hat{b}=\widetilde{Y}-\hat{a}\widetilde{X}~~;
\end{equation}
which implies the ``barycentre'' of the data, 
$\widetilde{\sf P}\equiv(\widetilde{X}, \widetilde{Y})$,
lies on the estimated regression line, inferred from
Eq.\,(\ref{eq:esl}), and the regression
line slope estimator is one among three
real solutions of a pseudo cubic equation
or two real solutions of a pseudo quadratic
equation, where the coefficients are weakly
dependent on the unknown slope.   For
further details refer to earlier attempts
(Y66; Y69; C11).
The above mentioned equations have the same
formal expression for functional and structural
models, which also holds for the regression
line slope and intercept estimators.

The regression line slope and intercept
variance estimators for functional models,
calculated using the method of partial
differentiation (e.g., Y69) and the method
of moments estimators [e.g., F87,
Chap.\,1, \S 1.3.2, Eq.\,(1.3.7) therein]
yield, in general, different results (C11).
The same is expected to hold, a fortiori,
for structural models, for which the method
of moments estimators and the $\delta$-method
have been exploited in classical investigations
(e.g., Ia90; FB92).   Accordingly, related
results shall be considered and expressed
in terms of unweighted deviation traces
for homoscedastic data with equal
correlation coefficients and extended
in terms of weighted deviation traces for
heteroscedastic data, with regard to a
number of special cases considered in
earlier attempts in the limit of
uncorrelated errors in $X$ and in $Y$
(Ia90; FB92).   With this restriction,
the pseudo cubic equation reduces to:
\begin{equation}
\label{eq:pccu}
\widetilde{V}_{20}a^3-2\widetilde{V}_{11}a^2-(\widetilde{W}_{20}-
\widetilde{V}_{02})a+\widetilde{W}_{11}=0~~;
\end{equation}
where the deviation traces are defined
by Eq.\,(\ref{eq:wgc}), via Eq.\,(\ref{eq:Wi})
and $V_i=W_i^2/w_{x_i}$.   For further
details refer to the parent paper (Y66)
and to a recent attempt (C11).   A
formulation of Euclidean and statistical
squared residual sum for homoscedastic
and heteroscedastic data is expressed
in Appendix \ref{a:esr2}.

\subsection{Errors in $X$ negligible with respect to errors in $Y$}
\label{XnrY}

\noindent\noindent

In the limit of errors in $X$ negligible with
respect to errors in $Y$, $a^2(\sigma_{xx})_i
\ll(\sigma_{yy})_i$, $a(\sigma_{xy})_i
\ll(\sigma_{yy})_i$, $1\le i\le n$.
Ideally, $(\sigma_{xx})_i\to0$, $(\sigma_{xy})_i\to0$,
$1\le i\le n$,
which implies $r_i\to0$, $w_{x_i}\to+\infty$,
$\Omega_i\to0$, $W_i\to w_{y_i}$, $1\le i\le n$.
Accordingly, the errors in $X$ and in $Y$ are
uncorrelated.

For homoscedastic data, $w_{x_i}=w_x$, $w_{y_i}=w_y$,
$1\le i\le n$, the regression line slope and intercept
estimators are (Ia90; C11):
\begin{lefteqnarray}
\label{eq:aYu}
&& \hat{a}_{\rm Y}=\frac{S_{11}}{S_{20}}~~; \\
\label{eq:bYu}
&& \hat{b}_{\rm Y}=\overline{Y}-\hat{a}_{\rm Y}\overline{X}~~;
\end{lefteqnarray}
where the index, Y, stands for
OLS(Y$|$X) i.e. ordinary least square regression
or, in general, WLS(Y$|$X) i.e. weighted least
square regression of the dependent
variable, $Y$, against the independent variable,
$X$ (Ia90).   Accordingly, related models shall
be quoted as Y models.

The regression line slope and intercept variance
estimators, in the special case of normal residuals
may be calculated using different methods and/or
models [e.g., F87, Chap.\,1, \S 1.3.2, Eq.\,(1.3.7)
therein; FB92; C11].  The result is:
\begin{lefteqnarray}
\label{eq:vaYu}
&& [(\hat{\sigma}_{\hat{a}_{\rm Y}})_{\rm N}]^2=\frac{(\hat{a}_{\rm Y})^2}
{n-2}\left[\frac{(n-2)R_{\rm Y}}{\hat{a}_{\rm Y}S_{11}}+
\Theta(\hat{a}_{\rm Y},\hat{a}_{\rm Y},\hat{a}_{\rm X})\right] \nonumber \\
&& \phantom{[(\hat{\sigma}_{\hat{a}_{\rm Y}})_{\rm N}]^2}=
\frac{(\hat{a}_{\rm Y})^2}{n-2}\left[
\frac{\hat{a}_{\rm X}-\hat{a}_{\rm Y}}{\hat{a}_{\rm Y}}+
\Theta(\hat{a}_{\rm Y},\hat{a}_{\rm Y},\hat{a}_{\rm X})\right]~~; \\
\label{eq:vbYu}
&& [(\hat{\sigma}_{\hat{b}_{\rm Y}})_{\rm N}]^2=\left[\frac1{\hat{a}_{\rm Y}}
\frac{S_{11}}{S_{00}}+(\overline{X})^2\right][(\hat{\sigma}_{\hat{a}_
{\rm Y}})_{\rm N}]^2-\frac{\hat{a}_{\rm Y}}{n-2}\frac{S_{11}}{S_{00}}
\Theta(\hat{a}_{\rm Y},\hat{a}_{\rm Y},\hat{a}_{\rm X})~~;
\end{lefteqnarray}
where the index, N, denotes normal residuals, $R$
is defined in Appendix \ref{a:esr2}, and
$\hat{a}_{\rm X}=S_{02}/S_{11}$.
The funcion, $\Theta(\hat{a}_
{\rm Y},\hat{a}_{\rm Y},\hat{a}_{\rm X})$, is a special
case of a more general function, $\Theta(\hat{a}_
{\rm C},\hat{a}_{\rm Y},\hat{a}_{\rm X})$ which, in
turn, depends on the method and/or model used.  For
further details refer to Appendix  \ref{a:Theta}.

The regression line slope and intercept variance
estimators, in the general case of non normal residuals
may be calculated using the $\delta$-method (Ia90).
The result is:
\begin{lefteqnarray}
\label{eq:vaYv}
&& (\hat{\sigma}_{\hat{a}_{\rm Y}})^2=\frac{S_{22}+
(\hat{a}_{\rm Y})^2S_{40}-2\hat{a}_{\rm Y}S_{31}}{(S_{20})^2}~~; \\
\label{eq:vbYv}
&& (\hat{\sigma}_{\hat{b}_{\rm Y}})^2=\frac{\hat{a}_{\rm Y}}n\frac
{\hat{a}_{\rm X}-\hat{a}_{\rm Y}}{\hat{a}_{\rm Y}}\frac{S_{11}}{S_{00}}
+(\overline{X})^2(\hat{\sigma}_{\hat{a}_{\rm Y}})^2-\frac2n\overline{X}
\hat{\sigma}_{\hat{b}_{\rm Y}\hat{a}_{\rm Y}}~~; \\
\label{eq:SbaY}
&& \hat{\sigma}_{\hat{b}_{\rm Y}\hat{a}_{\rm Y}}=\frac{S_{12}+(\hat{a}_
{\rm Y})^2S_{30}-2\hat{a}_{\rm Y}S_{21}}{S_{20}}~~;
\end{lefteqnarray}
where Eqs.\,(\ref{eq:vaYv})-(\ref{eq:SbaY}) are
equivalent to their counterparts expressed in the
parent paper (Ia90).

The application of the $\delta$-method provides
asymptotic formulae which understimate the true
regression coefficient uncertainty
in samples with low $(n\appleq50)$
or weakly correlated population (FB92).
In the special case of normal and data-independent
residuals, $\Theta(\hat{a}_
{\rm Y},\hat{a}_{\rm Y},\hat{a}_{\rm X})\to0$,
Eqs.\,(\ref{eq:vaYv}), (\ref{eq:vbYv}),
must necessarily reduce to (\ref{eq:vaYu}),
(\ref{eq:vbYu}), respectively, which implies
an additional factor, $n/(n-2)$, in
the first term on the right-hand side of
Eqs.\,(\ref{eq:vaYv})-(\ref{eq:SbaY}).
For further details
refer to Appendix \ref{a:daire}.

The expression of the regression line slope and
intercept estimators and related variance
estimators for normal residuals, Eqs.\,(\ref{eq:aYu}),
(\ref{eq:bYu}), (\ref{eq:vaYu}), (\ref{eq:vbYu}),
coincide with their counterparts determined for Y
models in classical and recent attempts [e.g., FB92,
Eq.\,(4) therein in the limit $c^2=\sigma_{yy}/\sigma_{xx}
\to+\infty$; Lavagnini and Magno, 2007, Eqs.\,(3)-(7)
therein].

For heteroscedastic data, the regression line slope and
intercept estimators are (C11):
\begin{lefteqnarray}
\label{eq:aYw}
&& \hat{a}_{\rm Y}=\frac{(\widetilde{w_y})_{11}}{(\widetilde{w_y})_{20}}~~; \\
\label{eq:bYw}
&& \hat{b}_{\rm Y}=\widetilde{Y}-\hat{a}_{\rm Y}\widetilde{X}~~;
\end{lefteqnarray}
where the weighted means, $\widetilde{X}$ and
$\widetilde{Y}$, are defined by
Eqs.\,(\ref{eq:wZa})-(\ref{eq:wOg}).

For functional models, regression line slope and
intercept variance estimators in the general case
of heteroscedastic data reduce to their counterparts
in the special case of homoscedastic data, as
$\{\hat{\sigma}_{\hat{a}_{\rm Y}}[(\widetilde{w_y})_{pq}]\}^2
\to[\hat{\sigma}_{\hat{a}_{\rm Y}}(w_yS_{pq})]^2$,
$\{\hat{\sigma}_{\hat{b}_{\rm Y}}[(\widetilde{w_y})_{pq}]\}^2
\to[\hat{\sigma}_{\hat{b}_{\rm Y}}(w_yS_{pq})]^2$,
via Eq.\,(\ref{eq:wgc}) where $Q_i=(w_y)_i=w_y$, $1\le i\le n$.
For further details refer to an earlier attempt (C11).

Under the assumption that the same holds for
extreme structural models, Eqs.\,(\ref{eq:vaYu})-(\ref{eq:SbaY})
take the general expression:
\begin{lefteqnarray}
\label{eq:vaYw}
&& [(\hat{\sigma}_{\hat{a}_{\rm Y}})_{\rm N}]^2=\frac{(\hat{a}_{\rm Y})^2}
{n-2}\left[\frac{n-2}n\frac{R_{\rm Y}}{\hat{a}_{\rm Y}}\frac
{(\widetilde{w_y})_{00}}{(\widetilde{w_y})_{11}}+
\Theta(\hat{a}_{\rm Y},\hat{a}_{\rm Y},\hat{a}_{\rm X}^\prime)\right]
\nonumber \\
&& \phantom{[(\hat{\sigma}_{\hat{a}_{\rm Y}})_{\rm N}]^2}=
\frac{(\hat{a}_{\rm Y})^2}{n-2}\left[\frac{\hat{a}_{\rm X}^\prime-\hat{a}_
{\rm Y}}{\hat{a}_{\rm Y}}+\Theta(\hat{a}_{\rm Y},\hat{a}_{\rm Y},\hat{a}_
{\rm X}^\prime)\right]~~; \\
\label{eq:vbYw}
&& [(\hat{\sigma}_{\hat{b}_{\rm Y}})_{\rm N}]^2=\left[\frac1{\hat{a}_{\rm Y}}
\frac{(\widetilde{w_y})_{11}}{(\widetilde{w_y})_{00}}+
(\widetilde{X})^2\right][(\hat{\sigma}_{\hat{a}_{\rm Y}})_{\rm N}]^2-\frac
{\hat{a}_{\rm Y}}{n-2}\frac{(\widetilde{w_y})_{11}}{(\widetilde{w_y})_{00}}
\Theta(\hat{a}_{\rm Y},\hat{a}_{\rm Y},\hat{a}_{\rm X}^\prime)~~;\qquad  \\
\label{eq:vaYx}
&& (\hat{\sigma}_{\hat{a}_{\rm Y}})^2=\frac{(\widetilde{w_y})_{00}}n
\frac{(\widetilde{w_y})_{22}+(\hat{a}_{\rm Y})^2
(\widetilde{w_y})_{40}-2\hat{a}_{\rm Y}(\widetilde{w_y})_{31}}
{[(\widetilde{w_y})_{20}]^2}~~; \\
\label{eq:vbYx}
&& (\hat{\sigma}_{\hat{b}_{\rm Y}})^2=\frac{\hat{a}_{\rm Y}}n\frac
{\hat{a}_{\rm X}^\prime-\hat{a}_{\rm Y}}{\hat{a}_{\rm Y}}\frac
{(\widetilde{w_y})_{11}}{(\widetilde{w_y})_{00}}
+(\widetilde{X})^2(\hat{\sigma}_{\hat{a}_{\rm Y}})^2-\frac2n\widetilde{X}
\hat{\sigma}_{\hat{b}_{\rm Y}\hat{a}_{\rm Y}}~~; \\
\label{eq:WbaY}
&& \hat{\sigma}_{\hat{b}_{\rm Y}\hat{a}_{\rm Y}}=\frac
{(\widetilde{w_y})_{12}+(\hat{a}_{\rm Y})^2(\widetilde{w_y})_{30}-2\hat{a}_
{\rm Y}(\widetilde{w_y})_{21}}{(\widetilde{w_y})_{20}}~~; 
\end{lefteqnarray}
where $\hat{a}_{\rm X}^\prime=(\widetilde{w_y})_{02}/(\widetilde{w_y})_{11}$,
$R$ is defined in Appendix \ref{a:esr2} and
$\Theta$ is expressed in terms of
$n(\widetilde{w_y})_{pq}/(\widetilde{w_y})_{00}$
instead of $S_{pq}$.

In the special case of normal and data-independent
residuals, $\Theta(\hat{a}_
{\rm Y},\hat{a}_{\rm Y},\hat{a}_{\rm X}^\prime)\to0$,
Eqs.\,(\ref{eq:vaYx}), (\ref{eq:vbYx}),
must necessarily reduce to (\ref{eq:vaYw}),
(\ref{eq:vbYw}), respectively, which implies
an additional factor, $n/(n-2)$, in
the first term on the right-hand side of
Eqs.\,(\ref{eq:vaYx})-(\ref{eq:WbaY}).

In absence of a rigorous proof,
Eqs.\,(\ref{eq:vaYw})-(\ref{eq:WbaY}) must be
considered as approximate results.

\subsection{Errors in $Y$ negligible with respect to errors in $X$}
\label{YnrX}

\noindent\noindent

In the limit of errors in $Y$ negligible with
respect to errors in $X$, $(\sigma_{yy})_i
\ll a^2(\sigma_{xx})_i$, $(\sigma_{xy})_i
\ll a(\sigma_{xx})_i$, $1\le i\le n$.
Ideally, $(\sigma_{yy})_i\to0$, $(\sigma_{xy})_i\to0$,
$1\le i\le n$,
which implies $r_i\to0$, $w_{y_i}\to+\infty$,
$\Omega_i\to+\infty$, $W_i\to w_{x_i}$, $1\le i\le n$.
Accordingly, the errors in $X$ and in $Y$ are
uncorrelated.   As outlined in an earlier paper
(C11), the model under discussion can be related to
the inverse regression, which has a large
associate literature (e.g., Miller, 1966; Garden
et al., 1980; Osborne, 1991; Brown, 1993; Lavagnini
and Magno, 2007).

For homoscedastic data, $w_{x_i}=w_x$, $w_{y_i}=w_y$,
$1\le i\le n$, the regression line slope and intercept
estimators are (Ia90; C11):
\begin{lefteqnarray}
\label{eq:aXu}
&& \hat{a}_{\rm X}=\frac{S_{02}}{S_{11}}~~; \\
\label{eq:bXu}
&& \hat{b}_{\rm X}=\overline{Y}-\hat{a}_{\rm X}\overline{X}~~;
\end{lefteqnarray}
where the index, X, stands for
OLS(X$|$Y) i.e. ordinary least square regression
or, in general, WLS(X$|$Y) i.e. weighted least
square regression of the dependent
variable, $X$, against the independent variable,
$Y$ (Ia90).   Accordingly, related models shall
be quoted as X models.

The regression line slope and intercept variance
estimators, in the special case of normal residuals
may be calculated using different methods and/or models
[e.g., F87, Chap.\,1, \S 1.3.2, Eq.\,(1.3.7) therein;
FB92; C11].  The result is:
\begin{lefteqnarray}
\label{eq:vaXu}
&& [(\hat{\sigma}_{\hat{a}_{\rm X}})_{\rm N}]^2=\frac{(\hat{a}_{\rm X})^2}
{n-2}\left[\frac{(n-2)R_{\rm X}}{\hat{a}_{\rm X}S_{11}}+
\Theta(\hat{a}_{\rm X},\hat{a}_{\rm Y},\hat{a}_{\rm X})\right] \nonumber \\
&& \phantom{[(\hat{\sigma}_{\hat{a}_{\rm X}})_{\rm N}]^2}=
\frac{(\hat{a}_{\rm X})^2}
{n-2}\left[\frac{\hat{a}_{\rm X}-\hat{a}_{\rm Y}}{\hat{a}_{\rm Y}}+
\Theta(\hat{a}_{\rm X},\hat{a}_{\rm Y},\hat{a}_{\rm X})\right]~~; \\
\label{eq:vbXu}
&& [(\hat{\sigma}_{\hat{b}_{\rm X}})_{\rm N}]^2=\left[\frac1
{\hat{a}_{\rm X}}\frac{S_{11}}{S_{00}}+(\overline{X})^2\right]
[(\hat{\sigma}_{\hat{a}_{\rm X}})_{\rm N}]^2-\frac{\hat{a}_{\rm X}}{n-2}
\frac{S_{11}}{S_{00}}\Theta(\hat{a}_{\rm X},\hat{a}_{\rm Y},\hat{a}_{\rm X})
~~;
\end{lefteqnarray}
where the index, N, denotes normal residuals, $R$
is defined in Appendix \ref{a:esr2}, and
$\hat{a}_{\rm Y}=S_{11}/S_{20}$.
The funcion, $\Theta(\hat{a}_
{\rm X},\hat{a}_{\rm Y},\hat{a}_{\rm X})$, is a special
case of a more general function, $\Theta(\hat{a}_
{\rm C},\hat{a}_{\rm Y},\hat{a}_{\rm X})$ which, in
turn, depends on the method and/or model used.  For
further details refer to Appendix  \ref{a:Theta}.

The regression line slope and intercept variance
estimators, in the general case of non normal residuals
may be calculated using the $\delta$-method (Ia90).
The result is:
\begin{lefteqnarray}
\label{eq:vaXv}
&& (\hat{\sigma}_{\hat{a}_{\rm X}})^2=\frac{S_{04}+(\hat{a}_{\rm X})^2
S_{22}-2\hat{a}_{\rm X}S_{13}}{(S_{11})^2}~~; \\
\label{eq:vbXv}
&& (\hat{\sigma}_{\hat{b}_{\rm X}})^2=\frac{\hat{a}_{\rm X}}n\frac{\hat{a}_
{\rm X}-\hat{a}_{\rm Y}}{\hat{a}_{\rm Y}}\frac{S_{11}}{S_{00}}+
(\overline{X})^2(\hat{\sigma}_{\hat{a}_{\rm X}})^2-\frac2n\overline{X}
\hat{\sigma}_{\hat{b}_{\rm X}\hat{a}_{\rm X}}~~; \\
\label{eq:SbaX}
&& \hat{\sigma}_{\hat{b}_{\rm X}\hat{a}_{\rm X}}=\frac{S_{03}+(\hat{a}_
{\rm X})^2S_{21}-2\hat{a}_{\rm X}S_{12}}{S_{11}}~~; 
\end{lefteqnarray}
where Eqs.\,(\ref{eq:vaXv})-(\ref{eq:SbaX}) are
equivalent to their counterparts expressed in the
parent paper (Ia90).

The application of the $\delta$-method provides
asymptotic formulae which understimate the true
regression coefficient uncertainty
in samples with low $(n\appleq50)$
or weakly correlated population (FB92).
In the special case of normal and data-independent
residuals, $\Theta(\hat{a}_
{\rm X},\hat{a}_{\rm Y},\hat{a}_{\rm X})\to0$,
Eqs.\,(\ref{eq:vaXv}), (\ref{eq:vbXv}),
must necessarily reduce to (\ref{eq:vaXu}),
(\ref{eq:vbXu}), respectively, which implies
an additional factor, $n/(n-2)$, in
the first term on the right-hand side of
Eqs.\,(\ref{eq:vaXv})-(\ref{eq:SbaX}).
For further details
refer to Appendix \ref{a:daire}.

For heteroscedastic data, the regression line slope and
intercept estimators are (C11):
\begin{lefteqnarray}
\label{eq:aXw}
&& \hat{a}_{\rm X}=\frac{(\widetilde{w_x})_{02}}{(\widetilde{w_x})_{11}}~~; \\
\label{eq:bXw}
&& \hat{b}_{\rm X}=\widetilde{Y}-\hat{a}_{\rm X}\widetilde{X}~~;
\end{lefteqnarray}
where the weighted means, $\widetilde{X}$ and
$\widetilde{Y}$, are defined by
Eqs.\,(\ref{eq:wZa})-(\ref{eq:wOg}).

For functional models, regression line slope and
intercept variance estimators in the general case
of heteroscedastic data reduce to their counterparts
in the special case of homoscedastic data, as
$\{\hat{\sigma}_{\hat{a}_{\rm X}}[(\widetilde{w_x})_{pq}]\}^2
\to[\hat{\sigma}_{\hat{a}_{\rm X}}(w_xS_{pq})]^2$,
$\{\hat{\sigma}_{\hat{b}_{\rm X}}[(\widetilde{w_x})_{pq}]\}^2
\to[\hat{\sigma}_{\hat{b}_{\rm X}}(w_xS_{pq})]^2$,
via Eq.\,(\ref{eq:wgc}) where $Q_i=(w_x)_i=w_x$, $1\le i\le n$.
For further details refer to an earlier attempt (C11).

Under the assumption that the same holds for
extreme structural models,
Eqs.\,(\ref{eq:vaXu})-(\ref{eq:SbaX})
take the general expression:
\begin{lefteqnarray}
\label{eq:vaXw}
&& [(\hat{\sigma}_{\hat{a}_{\rm X}})_{\rm N}]^2=\frac{(\hat{a}_{\rm X})^2}
{n-2}\left[\frac{n-2}n\frac{R_{\rm X}}{\hat{a}_{\rm X}}\frac
{(\widetilde{w_x})_{00}}{(\widetilde{w_x})_{11}}+
\Theta(\hat{a}_{\rm X},\hat{a}_{\rm Y}^\prime,\hat{a}_{\rm X})\right]
\nonumber \\
&& \phantom{[(\hat{\sigma}_{\hat{a}_{\rm X}})_{\rm N}]^2}=
\frac{(\hat{a}_{\rm X})^2}{n-2}\left[
\frac{\hat{a}_{\rm X}-\hat{a}_{\rm Y}^\prime}{\hat{a}_{\rm Y}^\prime}+
\Theta(\hat{a}_{\rm X},\hat{a}_{\rm Y}^\prime,\hat{a}_{\rm X})\right]~~; \\
\label{eq:vbXw}
&& [(\hat{\sigma}_{\hat{b}_{\rm X}})_{\rm N}]^2=\left[\frac1
{\hat{a}_{\rm X}}\frac{(\widetilde{w_x})_{11}}{(\widetilde{w_x})_{00}}+
(\widetilde{X})^2\right][(\hat{\sigma}_{\hat{a}_{\rm X}})_{\rm N}]^2-
\frac{\hat{a}_{\rm X}}{n-2}\frac{(\widetilde{w_x})_{11}}
{(\widetilde{w_x})_{00}}\Theta(\hat{a}_{\rm X},\hat{a}_{\rm Y}^\prime,
\hat{a}_{\rm X})~~;\qquad \\
\label{eq:vaXx}
&& (\hat{\sigma}_{\hat{a}_{\rm X}})^2=\frac{(\widetilde{w_x})_{00}}n
\frac{(\widetilde{w_x})_{04}+(\hat{a}_{\rm X})^2(\widetilde{w_x})_{22}-
2\hat{a}_{\rm X}(\widetilde{w_x})_{13}}{[(\widetilde{w_x})_{11}]^2}~~; \\
\label{eq:vbXx}
&& (\hat{\sigma}_{\hat{b}_{\rm X}})^2=\frac{\hat{a}_{\rm X}}n\frac
{\hat{a}_{\rm X}-\hat{a}_{\rm Y}^\prime}{\hat{a}_{\rm Y}^\prime}\frac
{(\widetilde{w_x})_{11}}{(\widetilde{w_x})_{00}}
+(\widetilde{X})^2(\hat{\sigma}_{\hat{a}_{\rm X}})^2-\frac2n\widetilde{X}
\hat{\sigma}_{\hat{b}_{\rm X}\hat{a}_{\rm X}}~~; \\
\label{eq:WbaX}
&& \hat{\sigma}_{\hat{b}_{\rm X}\hat{a}_{\rm X}}=\frac
{(\widetilde{w_x})_{03}+(\hat{a}_{\rm X})^2(\widetilde{w_x})_{21}-2\hat{a}_
{\rm X}(\widetilde{w_x})_{12}}{(\widetilde{w_x})_{11}}~~; 
\end{lefteqnarray}
where $\hat{a}_{\rm Y}^\prime=(\widetilde{w_x})_{11}/
(\widetilde{w_x})_{20}$, $R$
is defined in Appendix \ref{a:esr2}, and
$\Theta$ is formulated in terms of
$n(\widetilde{w_x})_{pq}/(\widetilde{w_x})_{00}$
instead of $S_{pq}$.

In the special case of normal and data-independent
residuals, $\Theta(\hat{a}_
{\rm X},\hat{a}_{\rm Y}^\prime,\hat{a}_{\rm X})\to0$,
Eqs.\,(\ref{eq:vaXx}), (\ref{eq:vbXx}),
must necessarily reduce to (\ref{eq:vaXw}),
(\ref{eq:vbXw}), respectively, which implies
an additional factor, $n/(n-2)$, in
the first term on the right-hand side of
Eqs.\,(\ref{eq:vaXx})-(\ref{eq:WbaX}).

In absence of a rigorous proof,
Eqs.\,(\ref{eq:vaXw})-(\ref{eq:WbaX}) must be
considered as approximate results.

\subsection{Oblique regression}
\label{YXc2}

\noindent\noindent

In the limit of constant $y$ to $x$ variance
ratios and constant correlation coefficients,
%
%
the following relations hold:
\begin{lefteqnarray}
\label{eq:xyc2}
&& \frac{(\sigma_{yy})_i}{(\sigma_{xx})_i}=c^2~;~~\frac{w_{x_i}}{w_{y_i}}=
\Omega_i^{-2}=c^2~;~~\frac{(\sigma_{xy})_i}{(\sigma_{xx})_i}=r_ic=
rc~~;~~1\le i\le n~~;\qquad \\
%
%
\label{eq:Wic2}
&& W_i=\frac{w_{x_i}}{a^2+c^2-2rac}~~;\qquad1\le i\le n~~;\qquad
\frac{(\widetilde{w_y})_{pq}}{(\widetilde{w_y})_{rs}}=
\frac{(\widetilde{w_x})_{pq}}{(\widetilde{w_x})_{rs}}~~;\qquad
\end{lefteqnarray}
where the weights are assumed to be inversely
proportional to related variances,
$w_{z_i}\propto1/(\sigma_{zz})_i$, $z=x,y$,
as usually done (e.g., FB92).   By definition,
$c$ has the dimensions of a slope, which
highly simplifies dimension checks throughout
equations, and for this reason it has been
favoured with respect to different choices
exploited in earlier attempts (e.g., Y66;
F87, Chap.\,1, \S 1.3; FB92).

It is worth
noticing that Eq.\,(\ref{eq:xyc2}) holds for
both homoscedastic and heteroscedastic data.
It can be seen that the lines of adjustment
are oriented along the same direction (York,
1967) but are perpendicular
to the regression line only in the special case
of orthogonal regression, $c^2=1$
(e.g., Carroll et al., 2006, Chap.\,3, \S 3.4.2).
Accordingly, the term ``oblique regression''
has been preferred with respect to ``generalized
orthogonal regression'' used in an earlier attempt
(C11).

The variance ratio, $c^2$, may be expressed in
terms of instrumental and intrinsic variance
ratios, $c_{\rm F}^2$, and $c_{\rm S}^2$,
respectively, as:
\begin{leftsubeqnarray}
\slabel{eq:CFSa}
&& c^2=\frac{[(\sigma_{xx})_i]_{\rm F}}{(\sigma_{xx})_i}c_{\rm F}^2+
       \frac{[(\sigma_{xx})_i]_{\rm S}}{(\sigma_{xx})_i}c_{\rm S}^2~~;
\qquad1\le i\le n~~; \\
\slabel{eq:CFSb}
&& c_{\rm F}^2=\frac{[(\sigma_{yy})_i]_{\rm F}}{[(\sigma_{xx})_i]_{\rm F}}~~;
\qquad
c_{\rm S}^2=\frac{[(\sigma_{yy})_i]_{\rm S}}{[(\sigma_{xx})_i]_{\rm S}}~~;
\qquad1\le i\le n~~;
\label{seq:CFS}
\end{leftsubeqnarray}
where $c_{\rm F}^2=c_{\rm S}^2$ implies
$c_{\rm F}^2=c_{\rm S}^2=c^2$;
$c^2\to c_{\rm F}^2$ for functional models,
$[(\sigma_{zz})_i]_{\rm S}\ll[(\sigma_{zz})_i]_{\rm F}$,
$z=x,y$, $1\le i\le n$;
$c^2\to c_{\rm S}^2$ for extreme structural models,
$[(\sigma_{zz})_i]_{\rm F}\ll[(\sigma_{zz})_i]_{\rm S}$,
$z=x,y$, $1\le i\le n$.

For homoscedastic data, $w_{x_i}=w_x$,
$w_{y_i}=w_y$, $1\le i\le n$, the regression
line slope and intercept estimators are
(FB92; C11):
\begin{lefteqnarray}
\label{eq:acS}
&& \hat{a}_{\rm C}=\frac{S_{02}-c^2S_{20}}{2S_{11}}\left\{1\mp\left[1+c^2
\left(\frac{S_{02}-c^2S_{20}}{2S_{11}}\right)^{-2}\right]^{1/2}\right\}
\nonumber \\
&& \phantom{\hat{a}_{\rm C}}=
\frac{\hat{a}_{\rm X}\hat{a}_{\rm Y}-c^2}{2\hat{a}_{\rm Y}}\left\{1\mp\left[
1+c^2\left(\frac{\hat{a}_{\rm X}\hat{a}_{\rm Y}-c^2}{2\hat{a}_{\rm Y}}
\right)^{-2}\right]^{1/2}\right\}~~; \\
\label{eq:bcS}
&& \hat{b}_{\rm C}=\overline{Y}-\hat{a}_{\rm C}\overline{X}~~;
\end{lefteqnarray}
where the index, C, denotes oblique
regression, $\hat{a}_{\rm Y}=
S_{11}/S_{20}$; $\hat{a}_{\rm X}=S_{02}/
S_{11}$; and the double
sign corresponds to the solutions of a
second-degree equation, where the
parasite solution must be disregarded.
Accordingly, related models shall be
quoted as C models or O models in the
special case of   orthogonal
regression $(c^2=1)$.
For further details refer to an earlier
attempt (C11).

The regression line slope and intercept variance
estimators, in the special case of normal residuals
may be calculated using different methods and/or
models [e.g., F87, Chap.\,1, \S
1.3.2, Eq.\,(1.3.7) therein; FB92; C11].  The result is:
\begin{lefteqnarray}
\label{eq:vacc2}
&& [(\hat{\sigma}_{\hat{a}_{\rm C}})_{\rm N}]^2=
\frac{(\hat{a}_{\rm C})^2}
{n-2}\left[\frac{(n-2)R_{\rm C}}{\hat{a}_{\rm C}S_{11}}+\Theta
(\hat{a}_{\rm C},\hat{a}_{\rm Y},\hat{a}_{\rm X})\right]\nonumber \\
&& \phantom{[(\hat{\sigma}_{\hat{a}_{\rm C}})_{\rm N}]^2}=
\frac{(\hat{a}_{\rm C})^2}
{n-2}\left[\frac{\hat{a}_{\rm X}-\hat{a}_{\rm C}}{\hat{a}_{\rm C}}+
\frac{\hat{a}_{\rm C}-\hat{a}_{\rm Y}}{\hat{a}_{\rm Y}}+\Theta
(\hat{a}_{\rm C},\hat{a}_{\rm Y},\hat{a}_{\rm X})\right]~~; \\
\label{eq:vbcc2}
&& [(\hat{\sigma}_{\hat{b}_{\rm C}})_{\rm N}]^2=\left[\frac1{\hat{a}_{\rm C}}
\frac{S_{11}}{S_{00}}+(\overline{X})^2\right][(\hat{\sigma}_{\hat{a}_
{\rm C}})_{\rm N}]^2-\frac{\hat{a}_{\rm C}}{n-2}\frac{S_{11}}{S_{00}}
\Theta(\hat{a}_{\rm C},\hat{a}_{\rm Y},\hat{a}_{\rm X})~~;
\end{lefteqnarray}
where $R$ is defined in Appendix \ref{a:esr2} and
$\Theta$ depends on the method and/or model
used, as shown in Table \ref{t:Theta}.
\begin{table}
\caption{Explicit expression of the function,
$\Theta(\hat{a}_{\rm C},\hat{a}_{\rm Y},
\hat{a}_{\rm X})$, appearing in the slope
and intercept variance
estimator formula for oblique
regression, Eq.\,(\ref{eq:vacc2}) and
(\ref{eq:vbcc2}), respectively, according
to different methods and/or models.   Symbol
captions: $A_{\rm UV}=\hat{a}_{\rm U}/
\hat{a}_{\rm V}-1$; (U,V) =
(X,C), (C,Y).   Method captions: AFD -
asymptotic formula determination; MME - method
of moments estimators; LSE - least squares
estimation;
MPD - method of partial differentiation.
Model captions: F - functional; S - structural;
E - extreme structural.   Case captions:
HM - homoscedastic; HT - heteroscedastic.
Reference
captions: F987 (F87, Chap.\,1, \S 1.3.2);
FB92 (Feigelson and Babu, 1992);
FB11 (FB92, erratum 2011); C011 (C11).}
\label{t:Theta}
\begin{center}
\begin{tabular}{lllll}
\hline
\multicolumn{1}{c}{$\Theta(\hat{a}_{\rm C},\hat{a}_{\rm Y},\hat{a}_{\rm X})$}
 &
\multicolumn{1}{c}{method} &
\multicolumn{1}{c}{model}  &
\multicolumn{1}{c}{case}   &
\multicolumn{1}{c}{source} \\

\hline

0                                                                        & AFD & E & HM & FB11 \\
$A_{\rm XC}A_{\rm CY}+\frac{(A_{\rm CY})^2}{n-1}$                        & MME & E & HM & FB92 \\
$A_{\rm XC}A_{\rm CY}+\frac{(A_{\rm CY})^2}{n-1}$                        & MME & S & HM & F987 \\
$\frac{A_{\rm XC}A_{\rm CY}}{n-1}+\left(\frac{A_{\rm CY}}{n-1}\right)^2$ & LSE & E & HM & FB11 \\
$2A_{\rm XC}A_{\rm CY}$                                                  & MPD & F & HT & C011 \\
\hline                            
\end{tabular}                     
\end{center}                      
\end{table}                       
For a formal demonstration, see
Appendix \ref{a:Theta}.    The
extreme situations, $a_{\rm C}\to a_{\rm Y}$,
$a_{\rm C}\to a_{\rm X}$, are directly
inferred from Table \ref{t:Theta} as
$\Theta(a_{\rm Y},a_{\rm Y},a_{\rm X})=0$,
$\Theta(a_{\rm X},a_{\rm Y},a_{\rm X})=0$,
respectively, in all cases.   More
specifically, the former relation
rigorously holds while the latter has to be
restricted to large ($n\gg1$, ideally
$n\to+\infty$) samples when appropriate.
In general, $\Theta$ can be neglected with
respect to the remaining terms in the
asymptotic expressions of
Eqs.\,(\ref{eq:vacc2}) and (\ref{eq:vbcc2}).
If the residuals are
independent of the data, $\Theta$ also
vanishes regardless of the sample
population.   For further details refer
to Appendix \ref{a:Theta} and \ref{a:daire}.

The regression line slope and intercept
variance estimators, in the general case
of non normal residuals, may be calculated
using the $\delta$-method (Ia90; FB92).
The result is:
\begin{lefteqnarray}
\label{eq:vaCv}
&& (\hat{\sigma}_{\hat{a}_{\rm C}})^2=(\hat{a}_{\rm C})^2\frac
{c^4(\hat{\sigma}_{\hat{a}_{\rm Y}})^2+(\hat{a}_{\rm Y})^4
(\hat{\sigma}_{\hat{a}_{\rm X}})^2+2(\hat{a}_{\rm Y})^2c^2
\hat{\sigma}_{\hat{a}_{\rm Y}\hat{a}_{\rm X}}}{(\hat{a}_{\rm Y})^2
[4(\hat{a}_{\rm Y})^2c^2+(\hat{a}_{\rm Y}\hat{a}_{\rm X}-c^2)^2]}~~; \\
\label{eq:vbCv}
&& (\hat{\sigma}_{\hat{b}_{\rm C}})^2=\frac{\hat{a}_{\rm C}}n\left[
\frac{\hat{a}_{\rm X}-\hat{a}_{\rm C}}{\hat{a}_{\rm C}}+
\frac{\hat{a}_{\rm C}-\hat{a}_{\rm Y}}{\hat{a}_{\rm Y}}\right]\frac{S_{11}}
{S_{00}}+(\overline{X})^2(\hat{\sigma}_{\hat{a}_{\rm C}})^2 \nonumber \\
&& \phantom{(\hat{\sigma}_{\hat{b}_{\rm C}})^2=}
-\frac2n\overline{X}
(\hat{\sigma}_{\hat{b}_{\rm Y}\hat{a}_{\rm C}}+\hat{\sigma}_
{\hat{b}_{\rm X}\hat{a}_{\rm C}})~~; \\
\label{eq:caYaX}
&& \hat{\sigma}_{\hat{a}_{\rm Y}\hat{a}_{\rm X}}=
\frac{S_{13}+\hat{a}_{\rm Y}\hat{a}_{\rm X}S_{31}
-(\hat{a}_{\rm Y}+\hat{a}_{\rm X})S_{22}}{S_{20}S_{11}}~~;\qquad \\
\label{eq:cbYaC}
&& \hat{\sigma}_{\hat{b}_{\rm Y}\hat{a}_{\rm C}}=\frac{\hat{a}_{\rm C}c^2}
{\hat{a}_{\rm Y}[4(\hat{a}_{\rm Y})^2c^2+(\hat{a}_{\rm Y}\hat{a}_{\rm X}-
c^2)^2]^{1/2}}\frac{S_{12}+\hat{a}_{\rm Y}\hat{a}_{\rm C}S_{30}
-(\hat{a}_{\rm Y}+\hat{a}_{\rm C})S_{21}}{S_{20}}~~;\qquad \\
\label{eq:cbXaC}
&& \hat{\sigma}_{\hat{b}_{\rm X}\hat{a}_{\rm C}}=\frac{\hat{a}_{\rm C}
\hat{a}_{\rm Y}}{[4(\hat{a}_{\rm Y})^2c^2+(\hat{a}_{\rm Y}\hat{a}_{\rm X}-
c^2)^2]^{1/2}}\frac{S_{03}+\hat{a}_{\rm X}\hat{a}_{\rm C}S_{21}-(\hat{a}_
{\rm X}+\hat{a}_{\rm C})S_{12}}{S_{11}}~~;
\end{lefteqnarray}
where Eqs.\,(\ref{eq:vaCv}) and (\ref{eq:vbCv}) in
the special case, $c^2=1$, are
equivalent to their counterparts expressed in the
parent paper (Ia90) provided absolute values
appearing therein are removed.   For a formal
discussion refer to Appendix \ref{a:limca}.
In addition, Eq.\,(\ref{eq:vaCv}) is equivalent
to its counterpart expressed in the parent paper
(FB92, erratum 2011).

The dependence on the variance ratio, $c^2$, in 
Eqs.\,(\ref{eq:vaCv}), (\ref{eq:cbYaC}),
(\ref{eq:cbXaC}), may be eliminated via
Eq.\,(\ref{eq:cS2}), Appendix \ref{a:Theta}.   The result is:
\begin{lefteqnarray}
\label{eq:vaCv2}
&& (\hat{\sigma}_{\hat{a}_{\rm C}})^2=(\hat{a}_{\rm C})^2 \nonumber \\
&& \times\frac{(\hat{a}_{\rm C})^4(A_{\rm XC})^2(\hat{\sigma}_
{\hat{a}_{\rm Y}})^2+
(\hat{a}_{\rm Y})^4(A_{\rm CY})^2(\hat{\sigma}_{\hat{a}_{\rm X}})^2+
2(\hat{a}_{\rm Y})^2(\hat{a}_{\rm C})^2A_{\rm XC}A_{\rm CY}
\hat{\sigma}_{\hat{a}_{\rm Y}\hat{a}_{\rm X}}}{(\hat{a}_{\rm Y})^2
\{4(\hat{a}_{\rm Y})^2(\hat{a}_{\rm C})^2A_{\rm XC}A_{\rm CY}+[\hat{a}_{\rm Y}
\hat{a}_{\rm X}A_{\rm CY}-(\hat{a}_{\rm C})^2A_{\rm XC}]^2\}};\qquad \\
\label{eq:cbYaC2}
&& \hat{\sigma}_{\hat{b}_{\rm Y}\hat{a}_{\rm C}}=\frac{(\hat{a}_{\rm C})^3
A_{\rm XC}}{\hat{a}_{\rm Y}\{4(\hat{a}_{\rm Y})^2
(\hat{a}_{\rm C})^2A_{\rm XC}A_{\rm CY}+[(\hat{a}_{\rm Y}\hat{a}_{\rm X}
A_{\rm CY}-(\hat{a}_{\rm C})^2A_{\rm XC}]^2\}^{1/2}} \nonumber \\
&& \phantom{\hat{\sigma}_{\hat{b}_{\rm Y}\hat{a}_{\rm C}}=}\times
\frac{S_{12}+\hat{a}_{\rm Y}\hat{a}_{\rm C}S_{30}-
(\hat{a}_{\rm Y}+\hat{a}_{\rm C})S_{21}}{S_{20}}~~; \\
\label{eq:cbXaC2}
&& \hat{\sigma}_{\hat{b}_{\rm X}\hat{a}_{\rm C}}=
\frac{\hat{a}_{\rm C}\hat{a}_{\rm Y}A_{\rm CY}}{\{4(\hat{a}_{\rm Y})^2
(\hat{a}_{\rm C})^2A_{\rm XC}A_{\rm CY}+[(\hat{a}_{\rm Y}\hat{a}_{\rm X}
A_{\rm CY}-(\hat{a}_{\rm C})^2A_{\rm XC}]^2\}^{1/2}} \nonumber \\
&& \phantom{\hat{\sigma}_{\hat{b}_{\rm X}\hat{a}_{\rm C}}=}\times
\frac{S_{03}+\hat{a}_{\rm X}\hat{a}_{\rm C}S_{21}-(\hat{a}_
{\rm X}+\hat{a}_{\rm C})S_{12}}{S_{11}} ~~; \\
\label{eq:AUV}
&& A_{\rm UV}=\frac{\hat{a}_{\rm U}}{\hat{a}_{\rm V}}-1~~;
\qquad\rm{(U,V)=(X,C),(C,Y),(X,Y)}~~;
\end{lefteqnarray}
in terms of slope estimators, variance
slope estimators, and deviation traces.

The application of the $\delta$-method provides
asymptotic formulae which understimate the true
regression coefficient uncertainty
in samples with low $(n\appleq50)$
or weakly correlated population (FB92).
In the special case of normal and data-independent
residuals, $\Theta(\hat{a}_
{\rm C},\hat{a}_{\rm Y},\hat{a}_{\rm X})\to0$,
Eqs.\,(\ref{eq:vaCv}), (\ref{eq:vbCv}),
must necessarily reduce to (\ref{eq:vacc2}),
(\ref{eq:vbcc2}), respectively, which implies
an additional factor, $n/(n-2)$, in
the first term on the right-hand side of
Eqs.\,(\ref{eq:vaYv}), (\ref{eq:vaXv}),
and (\ref{eq:vbCv})-(\ref{eq:cbXaC}).
For further details
refer to Appendix \ref{a:daire}.

For heteroscedastic data, the regression line
slope and intercept estimators are (C11):
\begin{lefteqnarray}
\label{eq:acw}
&& \hat{a}_{\rm C}=\frac{(\widetilde{w_x})_{02}-c^2(\widetilde{w_x})_{20}}
{2(\widetilde{w_x})_{11}}\left\{1\mp\left[1+c^2\left(\frac{(\widetilde{w_x})_
{02}-c^2(\widetilde{w_x})_{20}}{2(\widetilde{w_x})_{11}}\right)^{-2}\right]^
{1/2}\right\}\nonumber \\
&& \phantom{\hat{a}_{\rm C}}=
\frac{\hat{a}_{\rm X}\hat{a}_{\rm Y}^\prime-c^2}{2\hat{a}_{\rm Y}^\prime}
\left\{1\mp\left[1+c^2\left(\frac{\hat{a}_{\rm X}\hat{a}_{\rm Y}^\prime-c^2}
{2\hat{a}_{\rm Y}^\prime}\right)^{-2}\right]^{1/2}\right\}~~; \\
\label{eq:bcw}
&& \hat{b}_{\rm C}=\widetilde{Y}-\hat{a}_{\rm C}\widetilde{X}~~;
\end{lefteqnarray}
where $\hat{a}_{\rm Y}^\prime=(\widetilde{w_x})_{11}/
(\widetilde{w_x})_{20}$; $\hat{a}_{\rm X}=
(\widetilde{w_x})_{02}/(\widetilde{w_x})_{11}$;
and the weighted means, $\widetilde{X}$,
$\widetilde{Y}$, are defined by
Eqs.\,(\ref{eq:wZa})-(\ref{eq:wOg}).

For functional models, regression line slope and
intercept variance estimators in the general case
of heteroscedastic data reduce to their counterparts
in the special case of homoscedastic data, as
$\{\hat{\sigma}_{\hat{a}_{\rm C}}[(\widetilde{w_x})_{pq}]\}^2
\to[\hat{\sigma}_{\hat{a}_{\rm C}}(w_xS_{pq})]^2$,
$\{\hat{\sigma}_{\hat{b}_{\rm C}}[(\widetilde{w_x})_{pq}]\}^2
\to[\hat{\sigma}_{\hat{b}_{\rm C}}(w_xS_{pq})]^2$,
via Eq.\,(\ref{eq:wgc}) where $Q_i=(w_x)_i=w_x$, $1\le i\le n$.
For further details refer to an earlier attempt (C11).

Under the assumption that the same holds for
extreme structural models,
Eqs.\,(\ref{eq:vacc2})-(\ref{eq:cbXaC2})
take the general expression:
\begin{lefteqnarray}
\label{eq:vaCw}
&& [(\hat{\sigma}_{\hat{a}_{\rm C}})_{\rm N}]^2=\frac{(\hat{a}_{\rm C})^2}
{n-2}\left[\frac{n-2}n\frac{R_{\rm C}}{\hat{a}_{\rm C}}\frac
{(\widetilde{w_x})_{00}}{(\widetilde{w_x})_{11}}+
\Theta(\hat{a}_{\rm C},\hat{a}_{\rm Y}^\prime,\hat{a}_{\rm X})\right]
\nonumber \\
&& \phantom{[(\hat{\sigma}_{\hat{a}_{\rm C}})_{\rm N}]^2}=
\frac{(\hat{a}_{\rm C})^2}{n-2}\left[
\frac{\hat{a}_{\rm X}-\hat{a}_{\rm C}}{\hat{a}_{\rm C}}+
\frac{\hat{a}_{\rm C}-\hat{a}_{\rm Y}^\prime}{\hat{a}_{\rm Y}^\prime}+
\Theta(\hat{a}_{\rm C},\hat{a}_{\rm Y}^\prime,\hat{a}_{\rm X})\right]~~; \\
\label{eq:vbCw}
&& [(\hat{\sigma}_{\hat{b}_{\rm C}})_{\rm N}]^2=\left[\frac1
{\hat{a}_{\rm C}}\frac{(\widetilde{w_x})_{11}}{(\widetilde{w_x})_{00}}+
(\widetilde{X})^2\right][(\hat{\sigma}_{\hat{a}_{\rm C}})_{\rm N}]^2-
\frac{\hat{a}_{\rm C}}{n-2}\frac{(\widetilde{w_x})_{11}}
{(\widetilde{w_x})_{00}}\Theta(\hat{a}_{\rm C},\hat{a}_{\rm Y}^\prime,
\hat{a}_{\rm X})~~;\qquad \\
\label{eq:vaCx}
&& (\hat{\sigma}_{\hat{a}_{\rm C}})^2=(\hat{a}_{\rm C})^2\frac
{c^4(\hat{\sigma}_{\hat{a}_{\rm Y}})^2+(\hat{a}_{\rm Y})^4
(\hat{\sigma}_{\hat{a}_{\rm X}})^2+2(\hat{a}_{\rm Y})^2c^2
\hat{\sigma}_{\hat{a}_{\rm Y}\hat{a}_{\rm X}}}{(\hat{a}_{\rm Y})^2
[4(\hat{a}_{\rm Y})^2c^2+(\hat{a}_{\rm Y}\hat{a}_{\rm X}-c^2)^2]}~~; \\
\label{eq:vbCx}
&& (\hat{\sigma}_{\hat{b}_{\rm C}})^2=\frac{\hat{a}_{\rm C}}n\left[
\frac{\hat{a}_{\rm X}-\hat{a}_{\rm C}}{\hat{a}_{\rm C}}+
\frac{\hat{a}_{\rm C}-\hat{a}_{\rm Y}^\prime}{\hat{a}_{\rm Y}^\prime}
\right]\frac{(\widetilde{w_x})_{11}}{(\widetilde{w_x})_{00}}
+(\widetilde{X})^2(\hat{\sigma}_{\hat{a}_{\rm C}^\prime})^2 \nonumber \\
&& \phantom{(\hat{\sigma}_{\hat{b}_{\rm C}})^2=}
-\frac2n\widetilde{X}(\hat{\sigma}_{\hat{b}_{\rm Y}\hat{a}_{\rm C}}+
\hat{\sigma}_{\hat{b}_{\rm X}\hat{a}_{\rm C}})~~; \\
\label{eq:WaYaX}
&& \hat{\sigma}_{\hat{a}_{\rm Y}\hat{a}_{\rm X}}=\frac{(\widetilde{w_x})_{00}}
n\frac{(\widetilde{w_x})_{13}+\hat{a}_{\rm Y}\hat{a}_{\rm X}(\widetilde{w_x})_
{31}-(\hat{a}_{\rm Y}+\hat{a}_{\rm X})(\widetilde{w_x})_{22}}
{(\widetilde{w_x})_{20}(\widetilde{w_x})_{11}}~~;\qquad \\
\label{eq:WbaCY}
&& \hat{\sigma}_{\hat{b}_{\rm Y}\hat{a}_{\rm C}}=\frac{\hat{a}_{\rm C}c^2}
{\hat{a}_{\rm Y}[4(\hat{a}_{\rm Y})^2c^2+(\hat{a}_{\rm Y}\hat{a}_{\rm X}-c^2)^
2]^{1/2}} \nonumber \\
&& \phantom{\hat{\sigma}_{\hat{b}_{\rm Y}\hat{a}_{\rm C}}=}\times
\frac{(\widetilde{w_x})_{12}+\hat{a}_{\rm Y}\hat{a}_{\rm C}(\widetilde{w_x})_
{30}-(\hat{a}_{\rm Y}+\hat{a}_{\rm C})(\widetilde{w_x})_{21}}
{(\widetilde{w_x})_{20}}~~; \\
\label{eq:WbaCX}
&& \hat{\sigma}_{\hat{b}_{\rm X}\hat{a}_{\rm C}}=\frac{\hat{a}_{\rm C}
\hat{a}_{\rm Y}}{[4(\hat{a}_{\rm Y})^2c^2+(\hat{a}_{\rm Y}\hat{a}_{\rm X}-
c^2)^2]^{1/2}} \nonumber \\
&& \phantom{\hat{\sigma}_{\hat{b}_{\rm X}\hat{a}_{\rm C}}=}\times
\frac{(\widetilde{w_x})_{03}+\hat{a}_{\rm X}\hat{a}_{\rm C}
(\widetilde{w_x})_{21}-(\hat{a}_{\rm X}+\hat{a}_{\rm C})(\widetilde{w_x})_
{12}}{(\widetilde{w_x})_{11}}~~; \\
\label{eq:WaCv2}
&& (\hat{\sigma}_{\hat{a}_{\rm C}})^2=(\hat{a}_{\rm C})^2 \nonumber \\
&& \times\frac{(\hat{a}_{\rm C})^4(A_{\rm XC})^2(\hat{\sigma}_
{\hat{a}_{\rm Y}})^2+
(\hat{a}_{\rm Y})^4(A_{\rm CY^\prime})^2(\hat{\sigma}_{\hat{a}_{\rm X}})^2+
2(\hat{a}_{\rm Y})^2(\hat{a}_{\rm C})^2A_{\rm XC}A_{\rm CY^\prime}
\hat{\sigma}_{\hat{a}_{\rm Y}\hat{a}_{\rm X}}}{(\hat{a}_{\rm Y})^2
\{4(\hat{a}_{\rm Y})^2(\hat{a}_{\rm C})^2A_{\rm XC}A_{\rm CY^\prime}+
[\hat{a}_{\rm Y}\hat{a}_{\rm X}A_{\rm CY^\prime}-(\hat{a}_{\rm C})^2
A_{\rm XC}]^2\}};\qquad \\
\label{eq:WbaCY2}
&& \hat{\sigma}_{\hat{b}_{\rm Y}\hat{a}_{\rm C}}=\frac{(\hat{a}_{\rm C})^3
A_{\rm XC}}{\hat{a}_{\rm Y}\{4(\hat{a}_{\rm Y})^2
(\hat{a}_{\rm C})^2A_{\rm XC}A_{\rm CY^\prime}+[(\hat{a}_{\rm Y}
\hat{a}_{\rm X}A_{\rm CY^\prime}-(\hat{a}_{\rm C})^2A_{\rm XC}]^2\}^{1/2}}
\nonumber \\
&& \phantom{\hat{\sigma}_{\hat{b}_{\rm Y}\hat{a}_{\rm C}}=}\times
\frac{(\widetilde{w_x})_{12}+\hat{a}_{\rm Y}\hat{a}_{\rm C}
(\widetilde{w_x})_{30}-(\hat{a}_{\rm Y}+\hat{a}_{\rm C})
(\widetilde{w_x})_{21}}{(\widetilde{w_x})_{20}}~~; \\
\label{eq:WbaCX2}
&& \hat{\sigma}_{\hat{b}_{\rm X}\hat{a}_{\rm C}}=
\frac{\hat{a}_{\rm C}\hat{a}_{\rm Y}A_{\rm CY^\prime}}{\{4(\hat{a}_{\rm Y})^2
(\hat{a}_{\rm C})^2A_{\rm XC}A_{\rm CY^\prime}+[(\hat{a}_{\rm Y}
\hat{a}_{\rm X}A_{\rm CY^\prime}-(\hat{a}_{\rm C})^2A_{\rm XC}]^2\}^{1/2}}
\nonumber \\
&& \phantom{\hat{\sigma}_{\hat{b}_{\rm X}\hat{a}_{\rm C}}=}\times
\frac{(\widetilde{w_x})_{03}+\hat{a}_{\rm X}\hat{a}_{\rm C}
(\widetilde{w_x})_{21}-(\hat{a}_{\rm X}+\hat{a}_{\rm C})
(\widetilde{w_x})_{12}}{(\widetilde{w_x})_{11}} ~~;
\end{lefteqnarray}
and, in addition:
\begin{lefteqnarray}
\label{eq:vaCxp}
&& (\hat{\sigma}_{\hat{a}_{\rm C}^\prime})^2=(\hat{a}_{\rm C})^2\frac
{c^4(\hat{\sigma}_{\hat{a}_{\rm Y}^\prime})^2+(\hat{a}_{\rm Y})^4
(\hat{\sigma}_{\hat{a}_{\rm X}})^2+2(\hat{a}_{\rm Y})^2c^2
\hat{\sigma}_{\hat{a}_{\rm Y}\hat{a}_{\rm X}}}{(\hat{a}_{\rm Y})^2
[4(\hat{a}_{\rm Y})^2c^2+(\hat{a}_{\rm Y}\hat{a}_{\rm X}-c^2)^2]}~~; \\
\label{eq:vaYxp}
&& (\hat{\sigma}_{\hat{a}_{\rm Y}^\prime})^2=\frac{(\widetilde{w_x})_{00}}n
\frac{(\widetilde{w_x})_{22}+(\hat{a}_{\rm Y})^2
(\widetilde{w_x})_{40}-2\hat{a}_{\rm Y}(\widetilde{w_x})_{31}}
{[(\widetilde{w_x})_{20}]^2}~~;
\end{lefteqnarray}
where $\hat{a}_{\rm Y}=(\widetilde{w_y})_{11}/
(\widetilde{w_y})_{20}$; $\hat{a}_{\rm Y}^\prime=
(\widetilde{w_x})_{11}/(\widetilde{w_x})_{20}$;
$\hat{a}_{\rm X}=(\widetilde{w_x})_{02}/
(\widetilde{w_x})_{11}$; $A_{\rm CY^\prime}=
\hat{a}_{\rm C}/\hat{a}_{\rm Y}^\prime-1$; $R$
is defined in Appendix \ref{a:esr2}, and $\Theta$
is formulated in terms of $n(\widetilde{w_x})_{pq}/
(\widetilde{w_x})_{00}$ instead of $S_{pq}$.

In the special case of normal and data-independent
residuals, $\Theta(\hat{a}_
{\rm C},\hat{a}_{\rm Y}^\prime,\hat{a}_{\rm X})\to0$,
Eqs.\,(\ref{eq:vaCx}), (\ref{eq:vbCx}),
must necessarily reduce to (\ref{eq:vaCw}),
(\ref{eq:vbCw}), respectively, which implies
an additional factor, $n/(n-2)$, in
the first term on the right-hand side of
Eqs.\,(\ref{eq:vaYx}), (\ref{eq:vaXx}),
and (\ref{eq:vbCx})-(\ref{eq:WbaCX}).

In absence of a rigorous proof,
Eqs.\,(\ref{eq:vaCw})-(\ref{eq:WbaCX}) must be
considered as approximate results.

\subsection{Reduced major-axis regression}
\label{rema}

\noindent\noindent

The reduced major-axis regression may be
considered as a special case of oblique
regression, where $c^2=a_{\rm X}a_{\rm Y}$.
Accordingly, Eqs.\,(\ref{eq:xyc2}) and (\ref{eq:Wic2})
also hold.

For homoscedastic data, $w_{x_i}=w_x$,
$w_{y_i}=w_y$, $1\le i\le n$, the regression
line slope and intercept estimators, via
Eqs.\,(\ref{eq:acS}) and (\ref{eq:bcS}) are:
\begin{lefteqnarray}
\label{eq:arS}
&& \hat{a}_{\rm R}=\mp\sqrt{\frac{S_{02}}{S_{20}}}=
\mp\sqrt{\hat{a}_{\rm X}\hat{a}_{\rm Y}}~~; \\
\label{eq:brS}
&& \hat{b}_{\rm R}=\overline{Y}-\hat{a}_{\rm R}\overline{X}~~;
\end{lefteqnarray}
where the index, R, denotes reduced
major-axis regression, $\hat{a}_{\rm Y}
=S_{11}/S_{20}$; $\hat{a}_{\rm X}=S_{02}/
S_{11}$; and the double
sign corresponds to the solutions of
the square root, where the
parasite solution must be disregarded.
Accordingly, related models shall be
quoted as R models.   For further
details refer to an earlier attempt (C11).

The regression line slope and intercept variance
estimators may be directly inferred from
Eqs.\,(\ref{eq:vacc2}), (\ref{eq:vbcc2}), for
normal residuals,
in the limit, $\hat{a}_{\rm C}\to\hat{a}_{\rm R}=
\sqrt{\hat{a}_{\rm X}\hat{a}_{\rm Y}}$.
The result is:
\begin{lefteqnarray}
\label{eq:vacr2}
&& [(\hat{\sigma}_{\hat{a}_{\rm R}})_{\rm N}]^2=
\frac{(\hat{a}_{\rm R})^2}
{n-2}\left[\frac{(n-2)R_{\rm R}}{\hat{a}_{\rm R}S_{11}}+\Theta
(\hat{a}_{\rm R},\hat{a}_{\rm Y},\hat{a}_{\rm X})\right]\nonumber \\
&& \phantom{[(\hat{\sigma}_{\hat{a}_{\rm R}})_{\rm N}]^2}=
\frac{(\hat{a}_{\rm R})^2}
{n-2}\left[\frac{\hat{a}_{\rm X}-\hat{a}_{\rm R}}{\hat{a}_{\rm R}}+
\frac{\hat{a}_{\rm R}-\hat{a}_{\rm Y}}{\hat{a}_{\rm Y}}+\Theta
(\hat{a}_{\rm R},\hat{a}_{\rm Y},\hat{a}_{\rm X})\right]~~; \\
\label{eq:vbcr2}
&& [(\hat{\sigma}_{\hat{b}_{\rm R}})_{\rm N}]^2=\left[\frac1{\hat{a}_{\rm R}}
\frac{S_{11}}{S_{00}}+(\overline{X})^2\right][(\hat{\sigma}_{\hat{a}_
{\rm R}})_{\rm N}]^2-\frac{\hat{a}_{\rm R}}{n-2}\frac{S_{11}}{S_{00}}
\Theta(\hat{a}_{\rm R},\hat{a}_{\rm Y},\hat{a}_{\rm X})~~;
\end{lefteqnarray}
and for non normal residuals the application
of the $\delta$-method yields (Ia90):
\begin{lefteqnarray}
\label{eq:vaRv}
&& (\hat{\sigma}_{\hat{a}_{\rm R}})^2=(\hat{a}_{\rm R})^2\left[\frac14\frac
{(\hat{\sigma}_{\hat{a}_{\rm Y}})^2}{(\hat{a}_{\rm Y})^2}+\frac14\frac
{(\hat{\sigma}_{\hat{a}_{\rm X}})^2}{(\hat{a}_{\rm X})^2}+\frac12\frac
{\hat{\sigma}_{\hat{a}_{\rm Y}\hat{a}_{\rm X}}}{\hat{a}_{\rm Y}
\hat{a}_{\rm X}}\right]~~; \\
\label{eq:vbRv}
&& (\hat{\sigma}_{\hat{b}_{\rm R}})^2=\frac{\hat{a}_{\rm R}}n\left[
\frac{\hat{a}_{\rm X}-\hat{a}_{\rm R}}{\hat{a}_{\rm R}}+
\frac{\hat{a}_{\rm R}-\hat{a}_{\rm Y}}{\hat{a}_{\rm Y}}\right]\frac{S_{11}}
{S_{00}}+(\overline{X})^2(\hat{\sigma}_{\hat{a}_{\rm R}})^2 \nonumber \\
&& \phantom{(\hat{\sigma}_{\hat{b}_{\rm R}})^2=}
-\frac2n\overline{X}
(\hat{\sigma}_{\hat{b}_{\rm Y}\hat{a}_{\rm R}}+\hat{\sigma}_
{\hat{b}_{\rm X}\hat{a}_{\rm R}})~~; \\
%
%
\label{eq:cbYaR}
&& \hat{\sigma}_{\hat{b}_{\rm Y}\hat{a}_{\rm R}}=\frac12\left(\frac
{\hat{a}_{\rm X}}{\hat{a}_{\rm Y}}\right)^{1/2}\frac{S_{12}+\hat{a}_{\rm Y}
\hat{a}_{\rm R}S_{30}
-(\hat{a}_{\rm Y}+\hat{a}_{\rm R})S_{21}}{S_{20}}~~;\qquad \\
\label{eq:cbXaR}
&& \hat{\sigma}_{\hat{b}_{\rm X}\hat{a}_{\rm R}}=\frac12\left(\frac
{\hat{a}_{\rm Y}}{\hat{a}_{\rm X}}\right)^{1/2}\frac{S_{03}+\hat{a}_{\rm X}
\hat{a}_{\rm R}S_{21}-(\hat{a}_{\rm X}+\hat{a}_{\rm R})S_{12}}{S_{11}}~~;
\end{lefteqnarray}
where $\hat{\sigma}_{\hat{a}_{\rm Y}\hat{a}_{\rm X}}$
is defined by Eq.\,(\ref{eq:caYaX}) and
Eqs.\,(\ref{eq:vaRv}), (\ref{eq:vbRv}), are
equivalent to their counterparts expressed in the
parent paper (Ia90).
For further details refer to Appendix \ref{a:limca}.

The extension of the above results to heteroscedastic
data via Eqs.\,(\ref{eq:arS})-(\ref{eq:cbXaR})
reads:
\begin{lefteqnarray}
\label{eq:arw}
&& \hat{a}_{\rm R}=\mp\sqrt{\frac{(\widetilde{w_x})_{02}}
{(\widetilde{w_x})_{20}}}=\mp\sqrt{\hat{a}_{\rm X}\hat{a}_{\rm Y}^\prime}~~;
\\
\label{eq:brw}
&& \hat{b}_{\rm R}=\widetilde{Y}-\hat{a}_{\rm R}\widetilde{X}~~; \\
\label{eq:vaRw}
&& [(\hat{\sigma}_{\hat{a}_{\rm R}})_{\rm N}]^2=\frac{(\hat{a}_{\rm R})^2}
{n-2}\left[\frac{n-2}n\frac{R_{\rm R}}{\hat{a}_{\rm R}}\frac
{(\widetilde{w_x})_{00}}{(\widetilde{w_x})_{11}}+
\Theta(\hat{a}_{\rm R},\hat{a}_{\rm Y}^\prime,\hat{a}_{\rm X})\right]
\nonumber \\
&& \phantom{[(\hat{\sigma}_{\hat{a}_{\rm R}})_{\rm N}]^2}=
\frac{(\hat{a}_{\rm R})^2}{n-2}\left[
\frac{\hat{a}_{\rm X}-\hat{a}_{\rm R}}{\hat{a}_{\rm R}}+
\frac{\hat{a}_{\rm R}-\hat{a}_{\rm Y}^\prime}{\hat{a}_{\rm Y}^\prime}+
\Theta(\hat{a}_{\rm R},\hat{a}_{\rm Y}^\prime,\hat{a}_{\rm X})\right]~~; \\
\label{eq:vbRw}
&& [(\hat{\sigma}_{\hat{b}_{\rm R}})_{\rm N}]^2=\left[\frac1
{\hat{a}_{\rm R}}\frac{(\widetilde{w_x})_{11}}{(\widetilde{w_x})_{00}}+
(\widetilde{X})^2\right][(\hat{\sigma}_{\hat{a}_{\rm R}})_{\rm N}]^2-
\frac{\hat{a}_{\rm R}}{n-2}\frac{(\widetilde{w_x})_{11}}
{(\widetilde{w_x})_{00}}\Theta(\hat{a}_{\rm R},\hat{a}_{\rm Y}^\prime,
\hat{a}_{\rm X})~~;\qquad \\
\label{eq:vaRx}
&& (\hat{\sigma}_{\hat{a}_{\rm R}})^2=(\hat{a}_{\rm R})^2
\frac{\hat{a}_{\rm Y}}{\hat{a}_{\rm Y}^\prime}\left[\frac14\frac
{(\hat{\sigma}_{\hat{a}_{\rm Y}})^2}{(\hat{a}_{\rm Y})^2}+\frac14\frac
{(\hat{\sigma}_{\hat{a}_{\rm X}})^2}{(\hat{a}_{\rm X})^2}+\frac12\frac
{\hat{\sigma}_{\hat{a}_{\rm Y}\hat{a}_{\rm X}}}{\hat{a}_{\rm Y}
\hat{a}_{\rm X}}\right]~~; \\
\label{eq:vbRx}
&& (\hat{\sigma}_{\hat{b}_{\rm R}})^2=\frac{\hat{a}_{\rm R}}n\left[
\frac{\hat{a}_{\rm X}-\hat{a}_{\rm R}}{\hat{a}_{\rm R}}+
\frac{\hat{a}_{\rm R}-\hat{a}_{\rm Y}^\prime}{\hat{a}_{\rm Y}^\prime}
\right]\frac{(\widetilde{w_x})_{11}}{(\widetilde{w_x})_{00}}
+(\widetilde{X})^2(\hat{\sigma}_{\hat{a}_{\rm R}^\prime})^2 \nonumber \\
&& \phantom{(\hat{\sigma}_{\hat{b}_{\rm R}})^2=}
-\frac2n\widetilde{X}
(\hat{\sigma}_{\hat{b}_{\rm Y}\hat{a}_{\rm R}}+
\hat{\sigma}_{\hat{b}_{\rm X}\hat{a}_{\rm R}})~~; \\
\label{eq:WraYaX}
%
%
\label{eq:WbaRY}
&& \hat{\sigma}_{\hat{b}_{\rm Y}\hat{a}_{\rm R}}=\frac12\left(\frac
{\hat{a}_{\rm X}}{\hat{a}_{\rm Y}}\right)^{1/2}
\frac{(\widetilde{w_x})_{12}+\hat{a}_{\rm Y}\hat{a}_{\rm R}(\widetilde{w_x})_
{30}-(\hat{a}_{\rm Y}+\hat{a}_{\rm R})(\widetilde{w_x})_{21}}
{(\widetilde{w_x})_{20}}~~; \\
\label{eq:WbaRX}
&& \hat{\sigma}_{\hat{b}_{\rm X}\hat{a}_{\rm R}}=\frac12\left(\frac
{\hat{a}_{\rm Y}}{\hat{a}_{\rm X}}\right)^{1/2}
\frac{(\widetilde{w_x})_{03}+\hat{a}_{\rm X}\hat{a}_{\rm R}
(\widetilde{w_x})_{21}-(\hat{a}_{\rm X}+\hat{a}_{\rm R})(\widetilde{w_x})_
{12}}{(\widetilde{w_x})_{11}}~~;
\end{lefteqnarray}
and, in addition:
\begin{lefteqnarray}
\label{eq:vaRxp}
&& (\hat{\sigma}_{\hat{a}_{\rm R}^\prime})^2=\frac14\left[\frac
{\hat{a}_{\rm X}}{\hat{a}_{\rm Y}}(\hat{\sigma}_{\hat{a}_{\rm Y}^\prime})^2+
\frac{\hat{a}_{\rm Y}}{\hat{a}_{\rm X}}(\hat{\sigma}_{\hat{a}_{\rm X}})^2+2
\hat{\sigma}_{\hat{a}_{\rm Y}\hat{a}_{\rm X}}\right]~~;
\end{lefteqnarray}
where $\hat{a}_{\rm Y}=(\widetilde{w_y})_{11}/
(\widetilde{w_y})_{20}$; $\hat{a}_{\rm Y}^\prime=
(\widetilde{w_x})_{11}/(\widetilde{w_x})_{20}$;
$\hat{a}_{\rm X}=(\widetilde{w_x})_{02}/
(\widetilde{w_x})_{11}$;
$R$ is defined in Appendix \ref{a:esr2},
$\hat{\sigma}_{\hat{a}_{\rm Y}\hat{a}_{\rm X}}$,
$(\hat{\sigma}_{\hat{a}_{\rm Y}^\prime})^2$,
are expressed by Eqs.\,(\ref{eq:WaYaX}),
(\ref{eq:vaYxp}), respectively, and
$\Theta$ is formulated in terms of
$n(\widetilde{w_x})_{pq}/(\widetilde{w_x})_{00}$
instead of $S_{pq}$.

In absence of a rigorous proof,
Eqs.\,(\ref{eq:vaRw})-(\ref{eq:WbaRX}) must be
considered as approximate results.

\subsection{Bisector regression}
\label{bise}

\noindent\noindent

The bisector regression implies use of both
Y and X models for determining the angle
formed by related regression lines.   The
bisecting line is assumed to be the estimated
regression line of the model.

\begin{figure*}[t]
\begin{center}      
\includegraphics[scale=0.8]{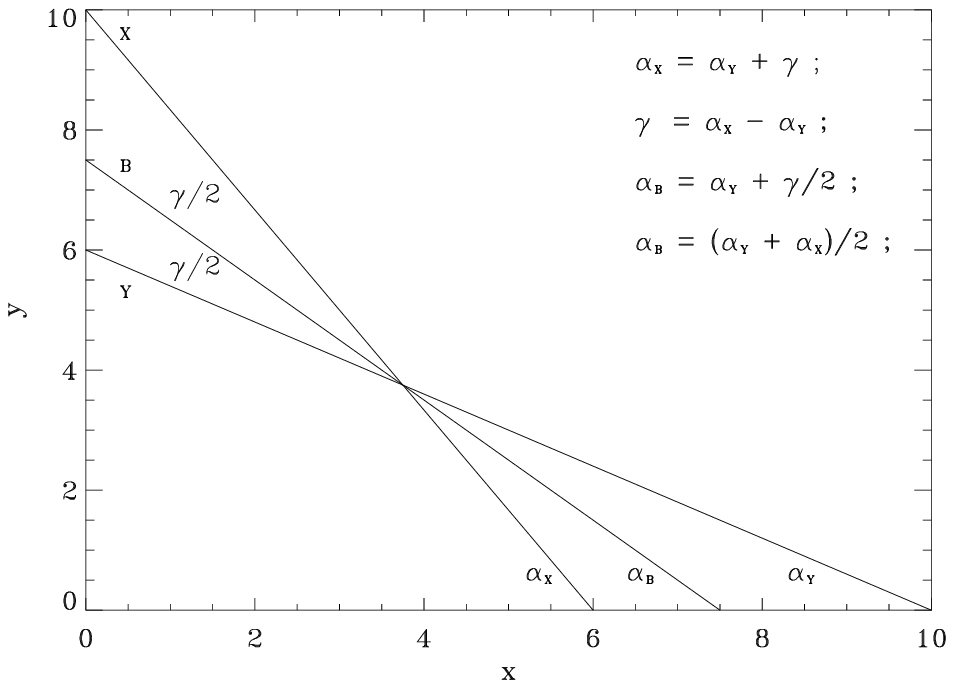}                      
\caption{Regression lines related to
Y, X, and B models, for an assigned
sample.   By definition, the B line
bisects the angle, $\gamma$, formed between
Y and X lines.   The angle, $\alpha_
{\rm B}$, formed between B line and $x$ axis,
is the arithmetic mean of the angles,
$\alpha_{\rm Y}$ and $\alpha_{\rm X}$,
formed between Y line and $x$ axis and
between $X$ line and $x$ axis, respectively.}
\label{f:bis}
\end{center}       
\end{figure*}

Let $\alpha_{\rm Y}$, $\alpha_{\rm X}$,
$\alpha_{\rm B}$, be the angles formed between
Y, X, B, regression line, respectively, and
$x$ axis, and $\gamma$ the angle formed
between Y and X regression lines, as
outlined in Fig.\,\ref{f:bis}.
Accordingly, $\gamma/2$ is the angle
formed between Y or X and B regression
lines.

The following relations can easily be
deduced from Fig.\,\ref{f:bis}:
$\alpha_{\rm X}=\alpha_{\rm Y}+\gamma;$
$\alpha_{\rm B}=\alpha_{\rm Y}+\gamma/2=
(\alpha_{\rm Y}+\alpha_{\rm X})/2$, and
the dimensionless slope of the regression
line is $\tan\alpha_{\rm B}$.   Using the
trigonometric formulae:
\begin{lefteqnarray*}
&& \tan(u+v)=\frac{\tan u+\tan v}{1-\tan u\tan v}~~;\qquad\tan\frac u2=\frac
{\sin u}{1+\cos u}~~;
\end{lefteqnarray*}
and the identity:
\begin{lefteqnarray*}
&& \frac{X(1+S_{\rm Y})+Y(1+S_{\rm X})}{(1+S_{\rm X})(1+S_{\rm Y})-XY}=
\frac{XY-1+S_{\rm X}S_{\rm Y}}{X+Y}~~; \\
&& X=\frac{a_{\rm X}}{a_u}~~;\quad Y=\frac{a_{\rm Y}}{a_u}~~;\quad
S_{\rm X}=\sqrt{1+X^2}~~;\quad S_{\rm Y}=\sqrt{1+Y^2}~~;
\end{lefteqnarray*}
the regression line slope estimator,
after some algebra, is expressed as
(Ia90):
\begin{lefteqnarray}
\label{eq:abS}
&& \hat{a}_{\rm B}=\frac{\hat{a}_{\rm Y}\hat{a}_{\rm X}-a_u^2+
\sqrt{a_u^2+(\hat{a}_{\rm Y})^2}\sqrt{a_u^2+(\hat{a}_{\rm X})^2}}
{\hat{a}_{\rm Y}+\hat{a}_{\rm X}}~~;
\end{lefteqnarray}
where $a_u$ is the unit slope,
$\hat{a}_{\rm Y}=S_{11}/S_{20}$;
$\hat{a}_{\rm X}=S_{02}/S_{11}$;
and the regression line intercept
estimator reads (Ia90):
\begin{lefteqnarray}
\label{eq:bbS}
&& \hat{b}_{\rm B}=\overline{Y}-\hat{a}_{\rm B}\overline{X}~~;
\end{lefteqnarray}
where the index, B, denotes bisector
regression.

The bisector regression may be
considered as a special case of
oblique regression
where the variance ratio, $c^2$,
is deduced from the combination
of Eqs.\,(\ref{eq:acS}) and (\ref{eq:abS}),
requiring $a_{\rm C}=
a_{\rm B}$.   After a lot
of algebra involving the roots
of a second-degree equation,
the result is:
\begin{lefteqnarray}
\label{eq:c2B}
&& c^2=(a_{\rm B})^2\frac{a_{\rm X}\mp a_{\rm B}}{a_{\rm B}}
\left(\frac{a_{\rm B}\mp a_{\rm Y}}{a_{\rm Y}}\right)^{-1}~~;
\end{lefteqnarray}
where the parasite solution
must be disregarded.   Accordingly,
Eqs.\,(\ref{eq:xyc2}) and
(\ref{eq:Wic2}) also hold.

For normal residuals and
homoscedastic data,
the regression line slope and
intercept variance estimators
may be directly inferred from
Eqs.\,(\ref{eq:vacc2}) and
(\ref{eq:vbcc2}) in the limit,
$\hat{a}_{\rm C}\to\hat{a}_{\rm B}$.
The result is:
\begin{lefteqnarray}
\label{eq:vacb2}
&& [(\hat{\sigma}_{\hat{a}_{\rm B}})_{\rm N}]^2=
\frac{(\hat{a}_{\rm B})^2}
{n-2}\left[\frac{(n-2)R_{\rm B}}{\hat{a}_{\rm B}S_{11}}+\Theta
(\hat{a}_{\rm B},\hat{a}_{\rm Y},\hat{a}_{\rm X})\right]\nonumber \\
&& \phantom{[(\hat{\sigma}_{\hat{a}_{\rm B}})_{\rm N}]^2}=
\frac{(\hat{a}_{\rm B})^2}
{n-2}\left[\frac{\hat{a}_{\rm X}-\hat{a}_{\rm B}}{\hat{a}_{\rm B}}+
\frac{\hat{a}_{\rm B}-\hat{a}_{\rm Y}}{\hat{a}_{\rm Y}}+\Theta
(\hat{a}_{\rm B},\hat{a}_{\rm Y},\hat{a}_{\rm X})\right]~~; \\
\label{eq:vbcb2}
&& [(\hat{\sigma}_{\hat{b}_{\rm B}})_{\rm N}]^2=\left[\frac1{\hat{a}_{\rm B}}
\frac{S_{11}}{S_{00}}+(\overline{X})^2\right][(\hat{\sigma}_{\hat{a}_
{\rm B}})_{\rm N}]^2-\frac{\hat{a}_{\rm B}}{n-2}\frac{S_{11}}{S_{00}}
\Theta(\hat{a}_{\rm B},\hat{a}_{\rm Y},\hat{a}_{\rm X})~~;
\end{lefteqnarray}
and for non normal residuals the application
of the $\delta$-method yields (Ia90):
\begin{lefteqnarray}
\label{eq:vaBv}
&& (\hat{\sigma}_{\hat{a}_{\rm B}})^2=\frac{(\hat{a}_{\rm B})^2}
{(\hat{a}_{\rm Y}+\hat{a}_{\rm X})^2}\left[\frac{a_u^2+(\hat{a}_{\rm X})^2}
{a_u^2+(\hat{a}_{\rm Y})^2}(\hat{\sigma}_{\hat{a}_{\rm Y}})^2+
\frac{a_u^2+(\hat{a}_{\rm Y})^2}
{a_u^2+(\hat{a}_{\rm X})^2}(\hat{\sigma}_{\hat{a}_{\rm X}})^2+2
\hat{\sigma}_{\hat{a}_{\rm Y}\hat{a}_{\rm X}}\right];\qquad \\
\label{eq:vbBv}
&& (\hat{\sigma}_{\hat{b}_{\rm B}})^2=\frac{\hat{a}_{\rm B}}n\left[
\frac{\hat{a}_{\rm X}-\hat{a}_{\rm B}}{\hat{a}_{\rm B}}+
\frac{\hat{a}_{\rm B}-\hat{a}_{\rm Y}}{\hat{a}_{\rm Y}}\right]\frac{S_{11}}
{S_{00}}+(\overline{X})^2(\hat{\sigma}_{\hat{a}_{\rm B}})^2 \nonumber \\
&& \phantom{(\hat{\sigma}_{\hat{b}_{\rm B}})^2=}
-\frac2n\overline{X}
(\hat{\sigma}_{\hat{b}_{\rm Y}\hat{a}_{\rm B}}+\hat{\sigma}_
{\hat{b}_{\rm X}\hat{a}_{\rm B}})~~; \\
\label{eq:baYaX}
%
%
\label{eq:cbYaB}
&& \hat{\sigma}_{\hat{b}_{\rm Y}\hat{a}_{\rm B}}=\frac{\hat{a}_{\rm B}\sqrt
{a_u^2+(\hat{a}_{\rm X})^2}}{(\hat{a}_{\rm Y}+\hat{a}_{\rm X})
\sqrt{a_u^2+(\hat{a}_{\rm Y})^2}}\frac{S_{12}+\hat{a}_{\rm Y}\hat{a}_{\rm B}
S_{30}-(\hat{a}_{\rm Y}+\hat{a}_{\rm B})S_{21}}{S_{20}}~~;\qquad \\
\label{eq:cbXaB}
&& \hat{\sigma}_{\hat{b}_{\rm X}\hat{a}_{\rm B}}=\frac{\hat{a}_{\rm B}\sqrt
{a_u^2+(\hat{a}_{\rm Y})^2}}{(\hat{a}_{\rm Y}+\hat{a}_{\rm X})
\sqrt{a_u^2+(\hat{a}_{\rm X})^2}}\frac{S_{03}+\hat{a}_{\rm X}\hat{a}_{\rm B}
S_{21}-(\hat{a}_{\rm X}+\hat{a}_{\rm B})S_{12}}{S_{11}}~~;
\end{lefteqnarray}
where $\hat{\sigma}_{\hat{a}_{\rm Y}\hat{a}_{\rm X}}$
is defined by Eq.\,(\ref{eq:caYaX}) and
Eqs.\,(\ref{eq:vaBv}), (\ref{eq:vbBv}), are
equivalent to their counterparts expressed in the
parent paper (Ia90).
For further details refer to Appendix \ref{a:limca}.

For heteroscedastic data, the combination
of Eqs.\,(\ref{eq:acw}) and (\ref{eq:abS}),
requiring $a_{\rm C}=a_{\rm B}$, after a lot
of algebra involving the roots
of a second-degree equation, yields:
\begin{lefteqnarray}
\label{eq:c2B2}
&& c^2=(a_{\rm B})^2\frac{a_{\rm X}\mp a_{\rm B}}{a_{\rm B}}
\left(\frac{a_{\rm B}\mp a_{\rm Y}^\prime}{a_{\rm Y}^\prime}\right)^{-1}~~;
\end{lefteqnarray}
where $\hat{a}_{\rm X}=(\widetilde{w_x})_{02}/
(\widetilde{w_x})_{11}$; $\hat{a}_{\rm Y}^\prime=
(\widetilde{w_x})_{11}/(\widetilde{w_x})_{20}$;
and the parasite solution
must be disregarded.   Accordingly,
Eqs.\,(\ref{eq:xyc2}) and
(\ref{eq:Wic2}) also hold.

The extension of the above results to
heteroscedastic data via
Eqs.\,(\ref{eq:abS})-(\ref{eq:bbS}) and
(\ref{eq:vacb2})-(\ref{eq:cbXaB}) reads:
\begin{lefteqnarray}
\label{eq:bbw}
&& \hat{b}_{\rm B}=\widetilde{Y}-\hat{a}_{\rm B}\widetilde{X}~~; \\
\label{eq:vaBw}
&& [(\hat{\sigma}_{\hat{a}_{\rm B}})_{\rm N}]^2=\frac{(\hat{a}_{\rm B})^2}
{n-2}\left[\frac{n-2}n\frac{R_{\rm B}}{\hat{a}_{\rm B}}\frac
{(\widetilde{w_x})_{00}}{(\widetilde{w_x})_{11}}+
\Theta(\hat{a}_{\rm B},\hat{a}_{\rm Y}^\prime,\hat{a}_{\rm X})\right]
\nonumber \\
&& \phantom{[(\hat{\sigma}_{\hat{a}_{\rm B}})_{\rm N}]^2}=
\frac{(\hat{a}_{\rm B})^2}{n-2}\left[
\frac{\hat{a}_{\rm X}-\hat{a}_{\rm B}}{\hat{a}_{\rm B}}+
\frac{\hat{a}_{\rm B}-\hat{a}_{\rm Y}^\prime}{\hat{a}_{\rm Y}^\prime}+
\Theta(\hat{a}_{\rm B},\hat{a}_{\rm Y}^\prime,\hat{a}_{\rm X})\right]~~; \\
\label{eq:vbBw}
&& [(\hat{\sigma}_{\hat{b}_{\rm B}})_{\rm N}]^2=\left[\frac1
{\hat{a}_{\rm B}}\frac{(\widetilde{w_x})_{11}}{(\widetilde{w_x})_{00}}+
(\widetilde{X})^2\right][(\hat{\sigma}_{\hat{a}_{\rm B}})_{\rm N}]^2-
\frac{\hat{a}_{\rm B}}{n-2}\frac{(\widetilde{w_x})_{11}}
{(\widetilde{w_x})_{00}}\Theta(\hat{a}_{\rm B},\hat{a}_{\rm Y}^\prime,
\hat{a}_{\rm X});\qquad \\
\label{eq:vaBx}
&& (\hat{\sigma}_{\hat{a}_{\rm B}})^2=\frac{(\hat{a}_{\rm B})^2}
{(\hat{a}_{\rm Y}+\hat{a}_{\rm X})^2}\left[\frac{a_u^2+(\hat{a}_{\rm X})^2}
{a_u^2+(\hat{a}_{\rm Y})^2}(\hat{\sigma}_{\hat{a}_{\rm Y}})^2+
\frac{a_u^2+(\hat{a}_{\rm Y})^2}{a_u^2+(\hat{a}_{\rm X})^2}
(\hat{\sigma}_{\hat{a}_{\rm X}})^2+2\hat{\sigma}_{\hat{a}_{\rm Y}
\hat{a}_{\rm X}}\right]; \\
\label{eq:vbBx}
&& (\hat{\sigma}_{\hat{b}_{\rm B}})^2=\frac{\hat{a}_{\rm B}}n\left[
\frac{\hat{a}_{\rm X}-\hat{a}_{\rm B}}{\hat{a}_{\rm B}}+
\frac{\hat{a}_{\rm B}-\hat{a}_{\rm Y}^\prime}{\hat{a}_{\rm Y}^\prime}
\right]\frac{(\widetilde{w_x})_{11}}{(\widetilde{w_x})_{00}}
+(\widetilde{X})^2(\hat{\sigma}_{\hat{a}_{\rm B}^\prime})^2 \nonumber \\
&& \phantom{(\hat{\sigma}_{\hat{b}_{\rm B}})^2=}
-\frac2n\widetilde{X}
(\hat{\sigma}_{\hat{b}_{\rm Y}\hat{a}_{\rm B}}+
\hat{\sigma}_{\hat{b}_{\rm X}\hat{a}_{\rm B}})~~; \\
\label{eq:WraYaX}
%
%
\label{eq:WbaBY}
&& \hat{\sigma}_{\hat{b}_{\rm Y}\hat{a}_{\rm B}}=\frac{\hat{a}_{\rm B}\sqrt
{a_u^2+(\hat{a}_{\rm X})^2}}{(\hat{a}_{\rm Y}+\hat{a}_{\rm X})\sqrt
{a_u^2+(\hat{a}_{\rm Y})^2}}
\frac{(\widetilde{w_x})_{12}+\hat{a}_{\rm Y}\hat{a}_{\rm B}(\widetilde{w_x})_
{30}-(\hat{a}_{\rm Y}+\hat{a}_{\rm B})(\widetilde{w_x})_{21}}
{(\widetilde{w_x})_{20}};\qquad \\
\label{eq:WbaBX}
&& \hat{\sigma}_{\hat{b}_{\rm X}\hat{a}_{\rm B}}=\frac{\hat{a}_{\rm B}\sqrt
{a_u^2+(\hat{a}_{\rm Y})^2}}{(\hat{a}_{\rm Y}+\hat{a}_{\rm X})\sqrt
{a_u^2+(\hat{a}_{\rm X})^2}}
\frac{(\widetilde{w_x})_{03}+\hat{a}_{\rm X}\hat{a}_{\rm B}
(\widetilde{w_x})_{21}-(\hat{a}_{\rm X}+\hat{a}_{\rm B})(\widetilde{w_x})_
{12}}{(\widetilde{w_x})_{11}};
\end{lefteqnarray}
and, in addition:
\begin{lefteqnarray}
\label{eq:vaBxp}
&& (\hat{\sigma}_{\hat{a}_{\rm B}^\prime})^2=\frac{(\hat{a}_{\rm B})^2}
{(\hat{a}_{\rm Y}+\hat{a}_{\rm X})^2}\left[\frac{a_u^2+(\hat{a}_{\rm X})^2}
{a_u^2+(\hat{a}_{\rm Y})^2}(\hat{\sigma}_{\hat{a}_{\rm Y}^\prime})^2+
\frac{a_u^2+(\hat{a}_{\rm Y})^2}{a_u^2+(\hat{a}_{\rm X})^2}
(\hat{\sigma}_{\hat{a}_{\rm X}})^2+2\hat{\sigma}_{\hat{a}_{\rm Y}
\hat{a}_{\rm X}}\right];\qquad
\end{lefteqnarray}
where $\hat{a}_{\rm Y}=(\widetilde{w_y})_{11}/
(\widetilde{w_y})_{20}$; $\hat{a}_{\rm Y}^\prime=
(\widetilde{w_x})_{11}/(\widetilde{w_x})_{20}$;
$\hat{a}_{\rm X}=(\widetilde{w_x})_{02}/
(\widetilde{w_x})_{11}$;
$R$ is defined in Appendix \ref{a:esr2},
$\hat{\sigma}_{\hat{a}_{\rm Y}\hat{a}_{\rm X}}$,
$(\hat{\sigma}_{\hat{a}_{\rm Y}^\prime})^2$,
are expressed by Eqs.\,(\ref{eq:WaYaX}),
(\ref{eq:vaYxp}), respectively, and
$\Theta$ is formulated in terms of
$n(\widetilde{w_x})_{pq}/(\widetilde{w_x})_{00}$
instead of $S_{pq}$.

In absence of a rigorous proof,
Eqs.\,(\ref{eq:vaBw})-(\ref{eq:WbaBX}) must be
considered as approximate results.

\subsection{Extension to structural models}
\label{exts}

\noindent\noindent

A nontrivial question is to what extent the
above results, valid for extreme structural models,
can be extended to generic structural models.   In
general, assumptions related to generic structural
models are different from their counterparts
related to extreme structural models (e.g., Buonaccorsi,
2006; 2010, Chap.\,6, \S 6.4.5) but, on the
other hand, they could coincide for a special
subclass.

In any case, whatever different
assumptions and models can be made with regard
to generic and extreme structural models, results
from the former are expected to tend to their
counterparts from the latter when the instrumental
scatter is negligible with respect to the
intrinsic scatter.
It is worth noticing that most
work on linear regression by astronomers
involves the situation where both intrinsic
scatter and heteroscedastic data are present
(e.g., AB96; Tremaine et al., 2002; Kelly,
2007, 2011).

A special subclass of structural models with
normal residuals can be defined where, for
a selected regression estimator, the regression line slope
and intercept variance estimators are independent of the
amount of instrumental and intrinsic scatter, including
the limit of null intrinsic scatter (functional models)
and null instrumental scatter (extreme structural models).
More specifically, the dependence occurs only via the
total (instrumental + intrinsic) scatter.
In this view, the whole subclass of structural models
under consideration could be related to functional
modelling (Carroll et al., 2006, Chap.\,2, \S 2.1).
For further details refer to the parent paper (C11).

\section{An example of astronomical application}
\label{apfm}

\noindent\noindent

\subsection{Astronomical introduction}
\label{asti}

\noindent\noindent

Heavy elements are synthesised within stars and (partially or
totally) returned to the interstellar medium via supernovae.
In an ideal situation where the initial stellar mass function
(including binary and multiple systems) is universal and the
gas returned after star death is instantaneously and uniformly
mixed with the interstellar medium, the abundance ratio of
primary elements produced mainly by large-mass ($m\appgeq8m_\odot$,
where $m_\odot$ is the solar mass) stars maintains unchanged,
which implies a linear relation.   This is why large-mass stars
have a short lifetime with respect to the age of the universe,
and related ejecta (due to type II supernovae) may be considered
as instantaneously returned to the interstellar medium.

A linear relation also holds if low-mass ($m\appleq8m_\odot$)
stars are considered, where the stellar lifetime can no longer
be neglected with respect to the age of the universe.   Close
binary systems including a white dwarf with masses, $m_{\rm WD}+
m_{\rm C}>m_{\rm Ch}$, are (type Ia) supernovae progenitors,
where $m_{\rm WD}$ is the white dwarf mass, $m_{\rm C}$ is the
companion mass, and $m_{\rm Ch}\approx\,1.44m_\odot$ is the
Chandrasekhar upper mass limit for stable white dwarfs.   An
additional restriction, for a linear relation between two generic
primary elements in the interstellar medium, is a constant number
ratio of type II to type Ia supernovae at any epoch.   For further
details refer to Appendix \ref{a:lrpe}.

Restricting to iron and oxygen, the generic linear relation,
Eq.\,(\ref{eq:relg}), reads:
\begin{equation}
\label{eq:OFe}
{\rm [O/H]}^\ast=a{\rm[Fe/H]}^\ast+b~~;
\end{equation}
where ${\rm [O/H]}$, ${\rm[Fe/H]}$, are logarithmic number
abundances normalized to the solar value (e.g., Caimmi 2007)
and the asterisks denote the ideal situation.   More specifically,
oxygen and iron abundance determinations performed on ideal stars
by use of ideal instruments yield coordinates of points lying on
the straight line defined by Eq.\,(\ref{eq:OFe}).

The intrinsic dispersion outside or along the ideal regression
line may be owing to several processes, such as fluctuations in
the stellar initial mass function (including binary and multiple
systems) and inhomogeneous mixing of stellar ejecta with the
interstellar medium, at different rates for different elements.
Accordingly, ideal points, ${\sf P}_i^\ast\equiv({\rm[Fe/H]}_i^\ast,
{\rm [O/H]}_i^\ast)$, are shifted towards actual points,
${\sf P}_{{\rm S}i}\equiv({\rm[Fe/H]}_{{\rm S}i},{\rm [O/H]}_{{\rm S}i})$.

More specifically, coeval ideal stars are represented by a single point on
the ideal regression line, while related actual stars correspond
to points which, in general, are shifted to a different extent
outside or along the ideal regression line.   Conversely, stars
with different age could be represented by a same actual point,
${\sf P}_{{\rm S}i}$.   The occurrence of instrumental scatter,
related to iron and oxygen abundance determination on a star
sample, makes actual points, ${\sf P}_{{\rm S}i}$, be shifted
towards observed points, ${\sf P}_i\equiv({\rm[Fe/H]}_i,{\rm [O/H]}_i)$.

With regard to the ideal regression line, there is a one-to-one
correspondence between the coordinates, ${\rm[Fe/H]}$ and ${\rm [O/H]}$,
while the contrary holds for actual points and observed points.
In the limit of extreme structural models, where instrumental
scatter is negligible with respect to intrinsic scatter, observed
points are very close to actual points (if otherwise, any linear
dependence would be hidden).   The latter, to a first extent, may
be determined along the following steps.
\begin{description}
\item[(1)]
Estimate a plausible regression line.
\item[(2)]
Calculate the mean distance of observed
points from the estimated regression line,
parallel to each coordinate axis.
\item[(3)]
Subdivide each coordinate axis into bins
of width equal to the related mean distance
calculated in (2).
\item[(4)]
Evaluate the intrinsic scatter within each
bin, using the method described in an earlier
attempt (AB96).
\item[(5)]
Minimize the loss function and determine the
regression line slope and intercept estimators.
\item[(6)]
Verify the absolute difference between previous
and current regression line slope and intercept
estimators is less than a previously assigned
tolerance value.   If otherwise, return to (2)
taking into consideration the current estimated
regression line.
\end{description}
In general, the total scatter, $\sigma_{{\rm[W/H]}_i}^2=
\sigma_{{\rm[W/H]}_{{\rm F}i}}^2+\sigma_{{\rm[W/H]}_{{\rm S}i}}^2$,
should be used for evaluating the weights, $w_{x_i}$, $w_{y_i}$,
$1\le i\le n$, appearing in the sum of the squared residuals,
expressed by Eq.\,(\ref{eq:R2ga}), which implies the knowledge
of the instrumental covariance matrix (e.g., AB96; Kelly 2011).

\subsection{Statistical results}
\label{star}

\noindent\noindent

An astronomical application performed in an earlier
attempt (C11) with regard to functional models, shall
be repeated here for extreme structural
models.   Accordingly, related samples will be left unchanged but
with the additional assumptions: (i) the intrinsic scatter is dominant 
with respect to the instrumental scatter, and (ii) uncertainties
mentioned in the parent papers and reported below are related to
the intrinsic scatter.

More specifically, the
following samples related to the [O/H]-[Fe/H] relation
shall be considered: RB09 (Rich and Boesgaard, 2009),
$n=49$, heteroscedastic data; Fa09 (Fabbian et al.,
2009), $n=44$, homoscedastic data with three
different [O/H] determinations, namely LTE
(standard local thermodynamical equilibrium for
one-dimensional hydrostatic model atmospheres), SH0
(three-dimensional
hydrostatic model atmospheres in absence of LTE with
no account taken of the inelastic collisions via neutral
H atoms, $S_{\rm H}=0)$, SH1
(three-dimensional hydrostatic model atmospheres in absence
of LTE with due account taken of the inelastic collisions
via neutral H atoms, $S_{\rm H}=1)$; Sa09 (Schmidt et
al., 2009), $n=63$, heteroscedastic data.
For further details refer to the parent paper (Caimmi, 2010).
In any case, [Fe/H] and [O/H] are determined independently
for each sample star.

The [O/H]-[Fe/H] empirical relations are
interpolated using the regression models,
G, Y, X, O, R, B, for heteroscedastic data
(FB09 and Sa09 samples) and Y, X, O, R,
B, for homoscedastic data (Fa09 sample,
cases LTE, SH0, SH1) and heteroscedastic
data where intrinsic scatters are
taken equal to the typical uncertainties
mentioned in the parent papers (FB09, Sa09),
$\sigma_{\rm [Fe/H]}=0.15$, $\sigma_{\rm
[O/H]}=0.15$, for both FB09 and Sa09
samples.   Model G relates to a general
case where the slope and intercept
estimators are determined via
Eqs.\,(\ref{eq:pccu}) and (\ref{eq:bar}),
respectively.   For further details refer
to the parent papers (Y66; Y69; C11).
Slope and intercept estimators
together with related dispersion estimators
are listed in Tables \ref{t:soga},
\ref{t:sogb}, and \ref{t:soa}, \ref{t:sob},
for heteroscedastic and
homoscedastic data, respectively.

Owing to high difficulties
intrinsic to the determination of slope
and intercept dispersion estimators
for G models, related calculations
were not performed, leaving
only approximate expressions (Y66)
and asymptotic formulae [Appendix
\ref{a:Theta}, Eq.\,(\ref{eq:saFB4})
related to G models].
\begin{table}
\caption{Regression line slope estimators,
$\hat{a}$, and related dispersion estimators,
$\hat{\sigma}_{\hat{a}}$, for heteroscedastic
models, G, Y, X, O, R, B, applied
to the [O/H]-[Fe/H] empirical relation
deduced from the following samples (from up 
to down): RB09, Sa09.   Dispersion column
captions: ENNR - extreme structural models
with non normal residuals (Ia90); ENRR -
extreme structural models with normal residuals
(FB92); FNRR - functional models with normal
residuals (Y66; Y69; C11); YANR - approximate
formula for normal residuals (Y66; Y69; C11);
AFNR - asymptotic formula for normal residuals
[Appendix \ref{a:Theta}, Eq.\,(\ref{eq:saFB4})
related to the appropriate model].
For G models, exact expressions of slope estimators
were not evaluated in the present attempt.
For Y models and normal residuals, different slope
dispersion estimators yield coinciding values,
as expected.}
\label{t:soga}
\begin{center}
\begin{tabular}{cccccccc} \hline
\multicolumn{1}{c}{$m$} &
\multicolumn{1}{c}{$\hat{a}$} &
\multicolumn{5}{c}{$\hat{\sigma}_{\hat{a}}$} &
\multicolumn{1}{c}{sample} \\
  &        &  ENNR  &  ENRR  &  FNRR  &  YANR  &  AFNR  &       \\
\hline
  &        &        &        &        &        &        &       \\
G & 0.7279 &        &        &        & 0.0294 & 0.0288 & RB09  \\
Y & 0.6714 & 0.0302 & 0.0314 & 0.0314 & 0.0314 & 0.0314 &       \\
X & 0.7305 & 0.0271 & 0.0290 & 0.0290 & 0.0279 & 0.0290 &       \\
O & 0.6964 & 0.0277 & 0.0276 & 0.0278 & 0.0271 & 0.0274 &       \\
R & 0.7050 & 0.0254 & 0.0280 & 0.0282 & 0.0272 & 0.0277 &       \\
B & 0.7005 & 0.0254 & 0.0278 & 0.0280 & 0.0271 & 0.0275 &       \\ 
  &        &        &        &        &        &        &       \\
G & 0.6383 &        &        &        & 0.0435 & 0.0582 & Sa09 \\
Y & 0.6167 & 0.0810 & 0.0398 & 0.0398 & 0.0398 & 0.0398 &       \\
X & 0.8652 & 0.0772 & 0.0833 & 0.0829 & 0.0664 & 0.0829 &       \\
O & 0.6355 & 0.0753 & 0.0609 & 0.0637 & 0.0541 & 0.0580 &       \\
R & 0.6927 & 0.0677 & 0.0664 & 0.0700 & 0.0560 & 0.0626 &       \\
B & 0.7336 & 0.0662 & 0.0704 & 0.0738 & 0.0579 & 0.0666 &       \\
\hline
\end{tabular}                     
\end{center}                      
\end{table}                       
\begin{table}
\caption{Regression line intercept estimators,
$\hat{b}$, and related dispersion estimators,
$\hat{\sigma}_{\hat{b}}$, for heteroscedastic
models, G, Y, X, O, R, B, applied
to the [O/H]-[Fe/H] empirical relation
deduced from the following samples (from up 
to down): RB09, Sa09.   Dispersion column
captions: ENNR - extreme structural models
with non normal residuals (Ia90); ENRR -
extreme structural models with normal residuals
(FB92); FNRR - functional models with normal
residuals (Y66; Y69; C11); YANR - approximate
formula for normal residuals (Y66; Y69; C11);
AFNR - asymptotic formula for normal residuals
[via Appendix \ref{a:Theta}, Eq.\,(\ref{eq:saFB4})
related to the appropriate model].
For G models, exact expressions of intercept estimators
were not evaluated in the present attempt.
For Y models and normal residuals, different intercept
dispersion estimators yield coinciding values,
as expected.}
\label{t:sogb}
\begin{center}
\begin{tabular}{cccccccc} \hline
\multicolumn{1}{c}{$m$} &
\multicolumn{1}{c}{$\hat{b}$} &
\multicolumn{5}{c}{$\hat{\sigma}_{\hat{b}}$} &
\multicolumn{1}{c}{sample} \\
  &        &  ENNR  &  ENRR  &  FNRR  &  YANR  &  AFNR  &         \\
\hline
  &           &        &        &        &        &        &      \\
G & $+$0.0043 &        &        &        & 0.0672 & 0.0660 & RB09 \\
Y & $-$0.1121 & 0.0608 & 0.0675 & 0.0675 & 0.0675 & 0.0675 &      \\
X & $+$0.0316 & 0.0636 & 0.0736 & 0.0735 & 0.0712 & 0.0735 &      \\
O & $-$0.0512 & 0.0523 & 0.0702 & 0.0707 & 0.0689 & 0.0697 &      \\
R & $-$0.0305 & 0.0521 & 0.0710 & 0.0725 & 0.0693 & 0.0704 &      \\
B & $-$0.0413 & 0.0522 & 0.0706 & 0.0711 & 0.0691 & 0.0700 &      \\
  &           &        &        &        &        &        &       \\
G & $+$0.0619 &        &        &        & 0.0251 & 0.0336 & Sa09 \\
Y & $+$0.0439 & 0.0105 & 0.0198 & 0.0198 & 0.0198 & 0.0198 &       \\
X & $+$0.3080 & 0.0712 & 0.0676 & 0.0673 & 0.0575 & 0.0673 &       \\
O & $+$0.1461 & 0.0396 & 0.0509 & 0.0525 & 0.0469 & 0.0491 &       \\
R & $+$0.1864 & 0.0436 & 0.0547 & 0.0549 & 0.0485 & 0.0524 &       \\
B & $+$0.2153 & 0.0455 & 0.0575 & 0.0603 & 0.0501 & 0.0552 &       \\
\hline
\end{tabular}                     
\end{center}                      
\end{table}                       
\begin{table}
\caption{Regression line slope estimators,
$\hat{a}$, and related dispersion estimators,
$\hat{\sigma}_{\hat{a}}$, for homoscedastic
models, Y, X, O, R, B, applied
to the [O/H]-[Fe/H] empirical relation
deduced from the following samples (from up 
to down): RB09, Sa09, Fa09, cases LTE, SH0, 
SH1.   Dispersion column
captions: ENNR - extreme structural models
with non normal residuals (Ia90); ENRR -
extreme structural models with normal residuals
(FB92); FNRR - functional models with normal
residuals (Y66; Y69; C11); YANR - approximate
formula for normal residuals (Y66; Y69; C11);
AFNR - asymptotic formula for normal residuals
[Appendix \ref{a:Theta}, Eq.\,(\ref{eq:saFB4})
related to the appropriate model].
For Y models and normal residuals, different slope
dispersion estimators yield coinciding values,
as expected.}
\label{t:soa}
\begin{center}
\begin{tabular}{cccccccc} \hline
\multicolumn{1}{c}{$m$} &
\multicolumn{1}{c}{$\hat{a}$} &
\multicolumn{5}{c}{$\hat{\sigma}_{\hat{a}}$} &
\multicolumn{1}{c}{sample} \\
  &        &  ENNR  &  ENRR  &  FNRR  &  YANR  &  AFNR  &       \\
\hline
  &        &        &        &        &        &        &       \\
Y & 0.6917 & 0.0268 & 0.0317 & 0.0317 & 0.0317 & 0.0317 & RB09  \\
X & 0.7600 & 0.0326 & 0.0349 & 0.0348 & 0.0332 & 0.0348 &       \\
O & 0.7143 & 0.0282 & 0.0327 & 0.0331 & 0.0319 & 0.0324 &       \\
R & 0.7251 & 0.0278 & 0.0332 & 0.0336 & 0.0321 & 0.0328 &       \\
B & 0.7253 & 0.0278 & 0.0333 & 0.0336 & 0.0321 & 0.0329 &       \\ 
  &        &        &        &        &        &        &       \\
Y & 0.5868 & 0.0596 & 0.0461 & 0.0461 & 0.0461 & 0.0461 & Sa09 \\
X & 0.8077 & 0.0563 & 0.0637 & 0.0635 & 0.0541 & 0.0635 &       \\
O & 0.6476 & 0.0620 & 0.0509 & 0.0526 & 0.0468 & 0.0491 &       \\
R & 0.6885 & 0.0523 & 0.0541 & 0.0562 & 0.0479 & 0.0519 &       \\
B & 0.6916 & 0.0513 & 0.0544 & 0.0565 & 0.0480 & 0.0521 &       \\
  &        &        &        &        &        &        &       \\
Y & 0.8961 & 0.0333 & 0.0303 & 0.0303 & 0.0303 & 0.0303 & Fa09 \\ 
X & 0.9381 & 0.0294 & 0.0318 & 0.0317 & 0.0310 & 0.0317 & (LTE) \\ 
O & 0.9150 & 0.0319 & 0.0310 & 0.0311 & 0.0305 & 0.0308 &       \\ 
R & 0.9168 & 0.0310 & 0.0310 & 0.0312 & 0.0305 & 0.0308 &       \\ 
B & 0.9169 & 0.0310 & 0.0310 & 0.0312 & 0.0305 & 0.0308 &       \\
  &        &        &        &        &        &        &       \\
Y & 1.2261 & 0.0459 & 0.0432 & 0.0432 & 0.0432 & 0.0432 & Fa09 \\
X & 1.2884 & 0.0434 & 0.0454 & 0.0454 & 0.0443 & 0.0454 & (SH0) \\
O & 1.2640 & 0.0449 & 0.0445 & 0.0448 & 0.0436 & 0.0443 &       \\
R & 1.2569 & 0.0441 & 0.0443 & 0.0445 & 0.0435 & 0.0440 &       \\
B & 1.2568 & 0.0441 & 0.0443 & 0.0445 & 0.0435 & 0.0440 &       \\
  &        &        &        &        &        &        &       \\
Y & 1.0492 & 0.0358 & 0.0341 & 0.0341 & 0.0341 & 0.0341 & Fa09 \\
X & 1.0946 & 0.0315 & 0.0356 & 0.0356 & 0.0348 & 0.0356 & (SH1) \\
O & 1.0732 & 0.0337 & 0.0349 & 0.0350 & 0.0343 & 0.0347 &       \\
R & 1.0716 & 0.0332 & 0.0348 & 0.0350 & 0.0343 & 0.0346 &       \\
B & 1.0716 & 0.0332 & 0.0348 & 0.0350 & 0.0343 & 0.0346 &       \\
\hline
\end{tabular}                     
\end{center}                      
\end{table}                       
\begin{table}
\caption{Regression line intercept estimators,
$\hat{b}$, and related dispersion estimators,
$\hat{\sigma}_{\hat{b}}$, for homoscedastic
models, Y, X, O, R, B, applied
to the [O/H]-[Fe/H] empirical relation
deduced from the following samples (from up 
to down): RB09, Sa09, Fa09, cases LTE, SH0, 
SH1.   Dispersion column
captions: ENNR - extreme structural models
with non normal residuals (Ia90); ENRR -
extreme structural models with normal residuals
(FB92); FNRR - functional models with normal
residuals (Y66; Y69; C11); YANR - approximate
formula for normal residuals (Y66; Y69; C11);
AFNR - asymptotic formula for normal residuals
[via Appendix \ref{a:Theta}, Eq.\,(\ref{eq:saFB4})
related to the appropriate model].
For Y models and normal residuals, different intercept
dispersion estimators yield coinciding values,
as expected.}
\label{t:sob}
\begin{center}
\begin{tabular}{cccccccc} \hline
\multicolumn{1}{c}{$m$} &
\multicolumn{1}{c}{$\hat{b}$} &
\multicolumn{5}{c}{$\hat{\sigma}_{\hat{b}}$} &
\multicolumn{1}{c}{sample} \\
  &        &  ENNR  &  ENRR  &  FNRR  &  YANR  &  AFNR  &         \\
\hline
  &        &        &        &        &        &        &         \\
Y & $-$0.0766 & 0.0598 & 0.0737 & 0.0737 & 0.0737 & 0.0737 & RB09 \\
X & $+$0.0742 & 0.0811 & 0.0807 & 0.0806 & 0.0773 & 0.0806 &      \\
O & $-$0.0268 & 0.0658 & 0.0759 & 0.0766 & 0.0741 & 0.0752 &      \\
R & $-$0.0030 & 0.0664 & 0.0770 & 0.0778 & 0.0746 & 0.0762 &      \\
B & $-$0.0025 & 0.0665 & 0.0770 & 0.0778 & 0.0746 & 0.0762 &      \\
  &        &        &        &        &           &        &       \\
Y & $+$0.0908 & 0.0211 & 0.0338 & 0.0338 & 0.0338 & 0.0338 & Sa09 \\
X & $+$0.2011 & 0.0423 & 0.0431 & 0.0430 & 0.0397 & 0.0430 &       \\
O & $+$0.1212 & 0.0268 & 0.0357 & 0.0363 & 0.0343 & 0.0351 &       \\
R & $+$0.1416 & 0.0279 & 0.0373 & 0.0381 & 0.0352 & 0.0365 &       \\
B & $+$0.1431 & 0.0282 & 0.0375 & 0.0382 & 0.0352 & 0.0367 &       \\
  &        &        &        &        &           &        &       \\
Y & $+$0.5476 & 0.0761 & 0.0663 & 0.0663 & 0.0663 & 0.0663 & Fa09 \\ 
X & $+$0.6366 & 0.0665 & 0.0693 & 0.0693 & 0.0678 & 0.0693 & (LTE) \\ 
O & $+$0.5877 & 0.0725 & 0.0676 & 0.0680 & 0.0666 & 0.0673 &       \\ 
R & $+$0.5916 & 0.0706 & 0.0678 & 0.0681 & 0.0667 & 0.0674 &       \\ 
B & $+$0.5916 & 0.0706 & 0.0678 & 0.0681 & 0.0667 & 0.0674 &       \\
  &        &        &        &        &           &        &       \\
Y & $+$0.8717 & 0.1017 & 0.0945 & 0.0945 & 0.0945 & 0.0945 & Fa09 \\
X & $+$1.0037 & 0.1003 & 0.0992 & 0.0991 & 0.0968 & 0.0991 & (SH0) \\
O & $+$0.9519 & 0.1019 & 0.0973 & 0.0978 & 0.0953 & 0.0967 &       \\
R & $+$0.9369 & 0.0998 & 0.0967 & 0.0973 & 0.0950 & 0.0961 &       \\
B & $+$0.9367 & 0.0998 & 0.0967 & 0.0973 & 0.0950 & 0.0961 &       \\
  &        &        &        &        &           &        &       \\
Y & $+$0.6518 & 0.0808 & 0.0745 & 0.0745 & 0.0745 & 0.0745 & Fa09 \\
X & $+$0.7479 & 0.0730 & 0.0777 & 0.0777 & 0.0761 & 0.0777 & (SH1) \\
O & $+$0.7027 & 0.0772 & 0.0762 & 0.0765 & 0.0750 & 0.0758 &       \\
R & $+$0.6993 & 0.0760 & 0.0761 & 0.0764 & 0.0750 & 0.0757 &       \\
B & $+$0.6993 & 0.0760 & 0.0761 & 0.0764 & 0.0749 & 0.0757 &       \\
\hline
\end{tabular}                     
\end{center}                      
\end{table}                       
For the remaining models,
the regression line slope and intercept
estimators and related dispersion estimators
are calculated using
Eqs.\,(\ref{eq:aYu})-(\ref{eq:SbaY})
and (\ref{eq:aYw})-(\ref{eq:WbaY}),
case Y, homoscedastic and heteroscedastic
data, respectively;
Eqs.\,(\ref{eq:aXu})-(\ref{eq:SbaX}) and
(\ref{eq:aXw})-(\ref{eq:WbaX}),
case X, homoscedastic and heteroscedastic
data, respectively;
Eqs.\,(\ref{eq:acS})-(\ref{eq:cbXaC}) and
(\ref{eq:acw})-(\ref{eq:WbaCX}), $c^2=1$,
case O, homoscedastic and heteroscedastic
data, respectively;
Eqs.\,(\ref{eq:arS})-(\ref{eq:cbXaR}) and
(\ref{eq:arw})-(\ref{eq:WbaRX}),
case R, homoscedastic and heteroscedastic
data, respectively;
Eqs.\,(\ref{eq:abS})-(\ref{eq:cbXaB}) and
(\ref{eq:bbw})-(\ref{eq:WbaBX}),
case B, homoscedastic and heteroscedastic
data, respectively.

\begin{figure*}[t]
\begin{center}      
\includegraphics[scale=0.8]{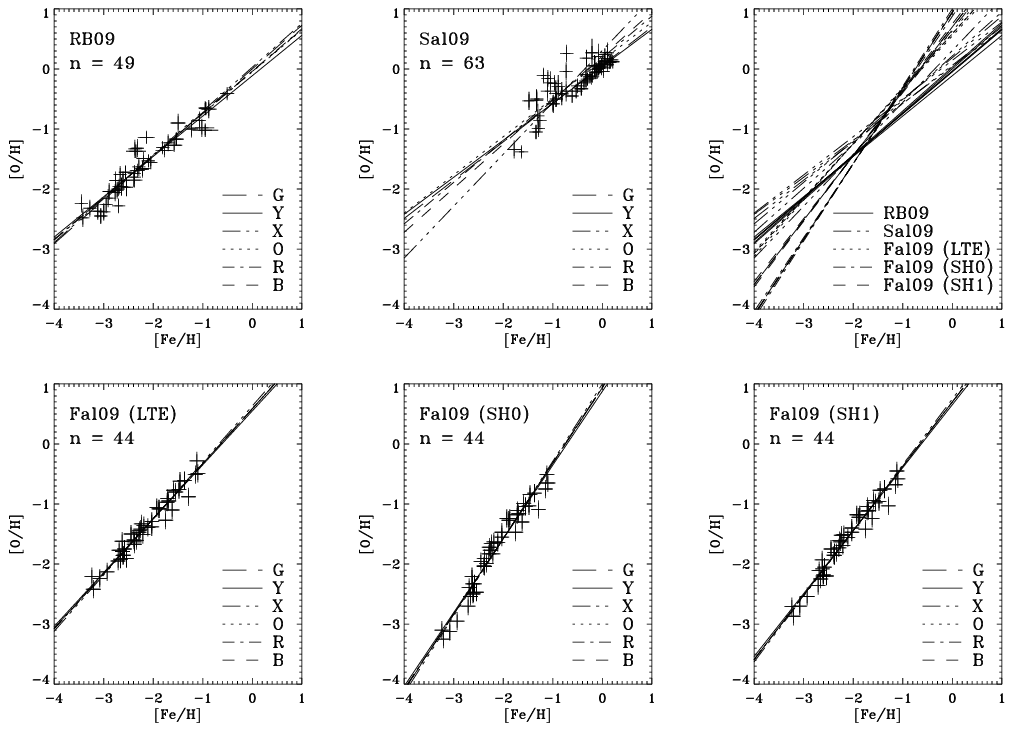}                      
\caption{Regression lines related to
[O/H]-[Fe/H] empirical relations
deduced from two samples with heteroscedastic
data, RB09 and Sa09, and three samples with
homoscedastic data (using the computer code
for heteroscedastic data), Fa09, cases LTE, SH0,
and SH1, indicated on each panel together
with related population and model captions.
The regression lines
related to six different methods are shown for
each sample on the top right panel.   For further
details refer to the text.}
\label{f:sog}
\end{center}       
\end{figure*}

\begin{figure*}[t]
\begin{center}      
\includegraphics[scale=0.8]{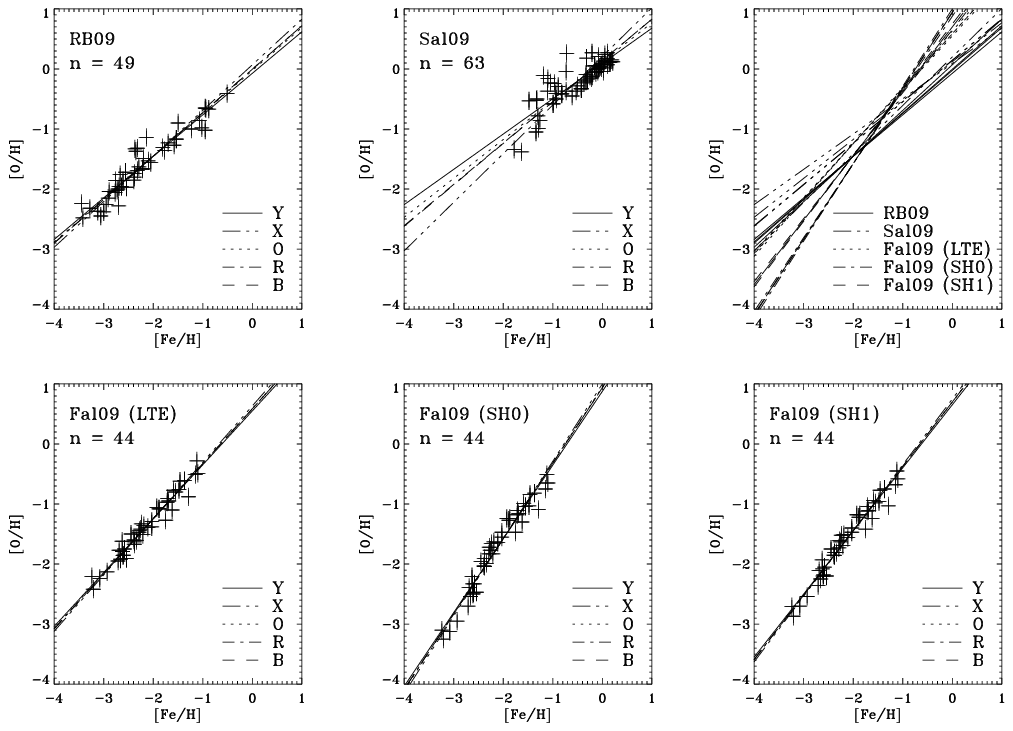}                      
\caption{Regression lines related to
[O/H]-[Fe/H] empirical relations
deduced from two samples with heteroscedastic
data (with instrumental scatters taken equal
to related typical values), RB09 and Sa09,
and three samples with
homoscedastic data, Fa09, cases LTE, SH0,
and SH1, indicated on each panel together
with related population and model captions.
The regression lines
related to five different methods are shown for
each sample on the top right panel.   For further
details refer to the text.}
\label{f:so}
\end{center}       
\end{figure*}

The regression lines determined by
use of the above mentioned methods
are plotted in Figs.\,\ref{f:sog}
and \ref{f:so} for heteroscedastic
and homoscedastic data, respectively,
where sample denomination and
population are indicated on each panel
together with model captions.
Homoscedastic data
are conceived as a special case of
heteroscedastic data in Fig.\,\ref
{f:sog} to test the computer code, which
is different for heteroscedastic and
homoscedastic data.   It can be seen
that lower panels of Figs.\,\ref{f:sog}
and \ref{f:so} coincide, and the regression
lines related to models G and O in
lower panels of Figs.\,\ref{f:sog} 
also coincide, as expected.
The whole set of regression lines for
all methods and all samples is shown
in the upper right panel of Figs.\,\ref
{f:sog} and \ref{f:so}.

Regression line slope and intercept
estimators have the same expression
for both structural and funcional
models.   Accordingly, Figs.\,\ref{f:sog}
and \ref{f:so} maintain unchanged
with respect to their counteparts
shown in an earlier attempt (C11)
where, on the other hand, B models
were not included.

An inspection of Tables \ref{t:soga}-\ref{t:sob}
and Figs.\,\ref{f:sog}-\ref{f:so} discloses
the following.
\begin{description}
\item[(1)]
Either of the inequalities (Ia90):
\begin{leftsubeqnarray}
\slabel{eq:disa}
&& \hat{a}_{\rm Y}<\hat{a}_{\rm O}<\hat{a}_{\rm R}<\hat{a}_{\rm B}<
\hat{a}_{\rm X}~~;\qquad \hat{a}_{\rm B}<a_u~~;\qquad S_{11}>0~~;\qquad \\
\slabel{eq:disb}
&& \hat{a}_{\rm Y}<\hat{a}_{\rm B}<\hat{a}_{\rm R}<\hat{a}_{\rm O}<
\hat{a}_{\rm X}~~;\qquad \hat{a}_{\rm B}>a_u~~;\qquad S_{11}>0~~;\qquad
\label{seq:dis}
\end{leftsubeqnarray}
where $a_u$ is the unit slope,
is satisfied for  homoscedastic data
but the contrary holds for
heteroscedastic data.   In
particular, $\hat{a}_{\rm B}<
\hat{a}_{\rm R}<a_u$ for RB09
sample, see Table \ref{t:soga}.
In addition,
$\hat{a}_{\rm Y}<\hat{a}_{\rm G}<
\hat{a}_{\rm X}$
for heteroscedastic data, but a
counterexample is provided in an
earlier attempt (Y66).
\item[(2)]
Slope and intercept estimators
from O, R and B models
are in agreement within $\mp\sigma$.
The extension of the above result to
slope and intercept estimators
from Y and X models holds
for samples with lower dispersion (Fa09).
An increasing dispersion yields marginal
(RB09) or no (Sa09) agreement within
$\mp\sigma$, for 
both heteroscedastic and homoscedastic data.
\item[(3)]
For normal residuals, slope and intercept
dispersion estimators related to functional
and structural models yield slightly different
results, as expected from the fact that related
asymptotic formulae coincide
[Appendix \ref{a:Theta}, Eq.\,(\ref{eq:saFB4})
related to the appropriate model].
Asymptotic formulae used in the current attempt
make a better fit with respect to earlier
approximations (Y66; Y69; C11).
\item[(4)]
Systematic variations due to different sample data
are dominant with respect to the intrinsic
scatter.
\end{description}
In conclusion, regression lines deduced
from different sample data represent
correct (from the standpoint of regression
models considered in the current attempt)
[O/H]-[Fe/H] relations, but
no definitive choice can be made until
systematic errors due to different methods
and/or spectral lines in determining oxygen
abundance, are alleviated.

\section{Discussion}\label{disc}

\noindent\noindent

For an assigned sample, structural
models belonging to a special subclass
are indistinguishable from extreme
structural models, as outlined in
an earlier attempt (C11).
Accordingly, the results of the current
paper also apply to structural
models of the kind considered.   The
expression of regression line slope and
intercept estimators and related variance
estimators in terms of weighted
deviation traces, for heteroscedastic
and homoscedastic data, makes a second step
towards a unified formalism of 
bivariate least squares linear regression.

Exact expressions of regression line slope
and intercept estimators and related variance
estimators have been rewritten in a more
compact form with respect to an earlier
attempt (FB92) in the limit of oblique
regression i.e. $(\sigma_{yy})_i/
(\sigma_{xx})_i=c^2$, $1\le i\le n$.
It is noteworthy
that a constant variance ratio, $c^2$, for
all data points, does not
necessarily imply equal variances,
$(\sigma_{xx})_i=\sigma_{xx}=$ const,
$(\sigma_{yy})_i=\sigma_{yy}=$ const,
$1\le i\le n$.   While regression line
slope and intercept estimators attain
a coinciding expression in different
attempts (Y66; Y69; Ia90; FB92), the
results of the current paper show that
the contrary holds for related variance
estimators.   The same holds for both
reduced major-axis and bisector regression.

Approximate expressions provided in
earlier attempts for normal residuals
(Y66; Y69) make (at
least in computed cases) a lower limit
to their exact counterparts, as shown
in Tables \ref{t:soga}-\ref{t:sob},
YANR vs. ENRR, FNRR.   The same holds,
to a better extent, for the asymptotic
expressions determined in the current
paper, as shown
in Tables \ref{t:soga}-\ref{t:sob},
AFNR vs. ENRR, FNRR.   Related
fractional discrepancies for
low-dispersion data (RB09, Fa09)
do not exceed a few percent, which
grows up to about 10\% in presence
of large-dispersion data (Sa09).

It is well known that the regression line
slope and intercept estimators 
are biased towards zero for Y models (e.g.,
F87, Chap.\,1, \S 1.1.1; Carroll et al., 2006,
Chap.\,3, \S 3.2; Kelly, 2007, 2011; Buonaccorsi,
2010, Chap.\,4, \S 4.4).
Biases can be explicitly expressed in the
special case of homoscedastic models with
normal residuals.
More specifically, the condition $1-\rho_{20}
\ll1$ ensures bias effects are negligible,
where $\rho_{20}$ is the reliability ratio:
\begin{equation}
\label{eq:rho20}
\rho_{20}=\frac{S_{20}}{S_{20}+(n-1)\sigma_{xx}}~~;
\end{equation}
which implies $0\le\rho_{20}\le1$.   For
further details refer to specific monographies
(e.g., F87, Chap.\,1, \S 1.1.1;
Carroll et al., 2006, Chap.\,3, \S 3.2.1;
Buonaccorsi, 2010, Chap.\,4, \S 4.4).

Similarly, it can be seen that regression
line slope and intercept variance estimators
are biased towards infinity for X models.
In the special case of homoscedastic models
with normal residuals, the condition
$1-\rho_{02}\ll1$ ensures bias effects are
negligible, where $\rho_{02}$ is the
reliability ratio:
\begin{equation}
\label{eq:rho02}
\rho_{02}=\frac{S_{02}}{S_{02}+(n-1)\sigma_{yy}}~~;
\end{equation}
which implies $0\le\rho_{02}\le1$ (e.g., C11).

Accordingly, slopes are understimated in
Y models and overstimated in X models by
a factor, $\rho_{20}$ and $1/\rho_{02}$,
respectively.   For C models (oblique
regression), O models (orthogonal regression),
R models (reduced major-axis regression),
B models (bisector regression), the regression
line slope estimators lie between their
counterparts related to Y and X models,
according to Eqs.\,(\ref{eq:rho20}) and
(\ref{eq:rho02}), which
implies bias corrections (e.g., Carroll et al.,
2006, Chap.\,3, \S 3.4.2).   Though there is
skepticism about an indiscriminate use of
oblique regression estimators,
still it is accepted the method is viable
provided both instrumental and intrinsic
covariance matrix are known (e.g., Carroll et al.,
2006, Chap.\,3, \S 3.4.2; Buonaccorsi,
2010, Chap.\,4, \S 4.5).

With regard to heteroscedastic data,
an inspection of Tables \ref{t:soga}-\ref{t:sob}
shows that for lower data dispersion
(RB09 sample) the values of regression
line slope and intercept estimators,
deduced for weighted (Tables \ref{t:soga}-\ref{t:sogb})
and unweighted (Tables \ref{t:soa}-\ref{t:sob}) data,
are systematically smaller in the former
case with respect to the latter, but are
still in agreement within $\mp\sigma$.
For larger dispersion data (Sa09 sample)
no systematic trend of the kind considered
appears, but the values of regression line
slope and intercept estimators are still
in agreement within $\mp\sigma$ for O, R,
and B models.   It may be a general property
of the regression models considered in the
current attempt or, more realistically,
intrinsic to the samples selected for the
application performed in subsection \ref{star}.

The reliability ratios, Eqs.\,(\ref{eq:rho20})
and (\ref{eq:rho02}),
have been calculated for all sample data
and the inequalities, $\rho_{20}>0.92$,
$\rho_{02}>0.91$, hold in any case except
$\rho_{02}>0.86$ for the Sa09 sample,
which implies poorly biased regression line
slope and intercept estimators for the samples
considered using Y and X models and, a fortiori,
using C, O, R, and B models.

Numerical simulations can determine the performance
of the regression coefficients in presence of small
samples and large scatter, and evaluate whether the
approximations made in deriving variances are accurate.
According to the results of a classical paper (Ia90),
the uncertainties to the slope predicted by O models
are, on average, larger than those predicted by Y, R,
or B models.   For this reason, skepticism is expressed
towards O models and, in any case, caution is urged in
interpreting slopes when small samples and large scatter
are involved (Ia90).

On the other hand, O models are special cases of C
models, which could also include R and B models, and
the predicted slopes lie between their counterparts
related to the limiting cases of Y and X models.
Extended numerical simulations should be used for
searching a relation between the family of C models,
$c^2=c_{\rm min}^2$, with the lowest uncertainty to
the slope, and values of population variances and
covariance, namely $c_{\rm min}^2=f(\sigma_{XX},
\sigma_{YY},\sigma_{XY})$.   In this view, it should
be recommended use of C models where $c^2=c_{\rm min}^2$
for assigned sample variances and covariance, which
estimate their counterparts related to the parent population.

Concerning samples listed in Tables \ref{t:soga}-\ref{t:sob}
and represented in Figs.\,\ref{f:sog}-\ref{f:so}, 
the slope uncertainty predicted by O models is slightly
larger than the slope predicted by R and B models for
non normal residuals (ENNR), while the reverse occurs
for normal residuals (ENNR).   In addition, the slope
uncertainty predicted by G models (the general case),
when estimated, is close to the slope uncertainty
predicted by O, R, and B models.

\section{Conclusion}
\label{conc}

\noindent\noindent

From the standpoint of a unified analytic formalism of
bivariate least squares linear regression, extreme
structural models have been conceived as a limiting
case where the instrumental scatter is
negligible (ideally null) with respect to
the intrinsic scatter.

Within the framework of a variant of the classical
additive error model (e.g., Carroll et al., 2006, Chap.\,1,
\S 1.2, Chap.\,3, \S 3.2.1; Buonaccorsi, 2010,
Chap.\,4, \S 4.3; Kelly 2011), the classical results
presented in earlier papers (Ia90; FB92)
have been rewritten in a more compact form
using a new formalism in terms
of weighted deviation traces
which, for homoscedastic data, reduce to usual
quantities, leaving aside an
unessential (but dimensional) multiplicative
factor.

Regression line slope and intercept
estimators, and related variance estimators,
have been expressed in the special case of
uncorrelated errors in $X$ and in $Y$ for
the following models: (Y) errors
in $X$ negligible (ideally null) with
respect to errors in $Y$; (X) errors
in $Y$ negligible (ideally null) with
respect to errors in $X$; (C) oblique
regression; (O) orthogonal regression;
(R) reduced major-axis regression; (B)
bisector regression.   Related variance
estimators have been expressed for both
non normal and normal residuals and
compared to their counterparts determined
for functional models (C11).

Under the assumption that regression line
slope and intercept variance estimators
for homoscedastic and heteroscedastic data
are connected to a similar extent in
functional and structural models, the
above mentioned results have been extended
from homoscedastic to heteroscedastic data.
In absence of a rigorous proof, related
expressions have been considered as
approximate results.

An example of astronomical application
has been considered, concerning the
[O/H]-[Fe/H] empirical relations deduced
from five samples related to different
populations and/or different methods of
oxygen abundance determination.   For
low-dispersion samples and assigned
methods, different regression models
have been found to yield  results which
are in agreement within the errors $(\mp\sigma)$
for both heteroscedastic and homoscedastic
data, while the contrary has been shown
to hold for large-dispersion samples.
In any case, samples related to different
methods have been found to produce
discrepant results, due to the presence
of (still undetected) systematic errors,
which implies no definitive statement can
be made at present.

Asymptotic expressions have been found to
approximate regression line slope and
intercept variance estimators, for normal
residuals, to a better extent with respect
to earlier attempts (Y66; Y69).   Related
fractional discrepancies have been shown
to be not exceeding a few percent for
low-dispersion data, which has grown up
to about 10\% in presence of large-dispersion
data.

An extension of the formalism to generic structural
models has been left to a forthcoming paper.

\section*{Acknowledgements}
Thanks are due to G.J. Babu, E.D. Feigelson,
M.A. Bershady, I. Lavagnini, S.J. Schmidt for fruitful
e-mail correspondance on their 
quoted papers (FB92; AB96; Lavagnini
and Magno, 2007; Sa09;
respectively).
The author is indebted to G.J. Babu and E.D. Feigelson
for having kindly provided the erratum of their quoted paper
(Feigelson and Babu, 1992) before publication
(Feigelson and Babu, 2011).

\appendix
\section*{Appendix}

\section{Euclidean and statistical squared residual sum}
\label{a:esr2}

\noindent\noindent

For homoscedastic data, the sum of squared (dimensional)
Euclidean distances
between observed points, ${\sf P}_i(X_i,Y_i)$, and
adjusted points on the estimated regression line,
$\hat{\sf P}_i(x_i,y_i)$, $y_i=\hat{a}x_i+\hat{b}$,
is expressed as (e.g., F87, Chap.\,1, \S 1.3.3;
FB92; C11):
\begin{lefteqnarray}
\label{eq:Rn}
&& (n-2)R=\sum_{i=1}^n\left[(Y_i-\overline{Y})-\hat{a}(X_i-\overline{X})
\right]^2=S_{02}+(\hat{a})^2S_{20}-2\hat{a}S_{11}~~;\qquad
\end{lefteqnarray}
where $R$ is denoted as $s_{vv}$ in the
earlier quotation (F87).

The sum of squared (dimensionless) statistical distances
(e.g., F87, Chap.\,1, \S 1.3.3) between
the above mentioned points, ${\sf P}_i(X_i,Y_i)$
and $\hat{\sf P}_i(x_i,y_i)$, reads (C11):
\begin{lefteqnarray}
\label{eq:FS}
&& T_{\overline{R}}=W\left[S_{02}+
(\hat{a})^2S_{20}-2\hat{a}S_{11}\right]~~;
\end{lefteqnarray}
which, for heteroscedastic data, takes
the general expression (C11):
\begin{lefteqnarray}
\label{eq:sR23}
&& T_{\widetilde{R}}=\widetilde{W}_{02}+
(\hat{a})^2\widetilde{W}_{20}-2\hat{a}\widetilde{W}_{11}~~;
\end{lefteqnarray}
accordingly, the extension of Eq.\,(\ref{eq:Rn})
to heteroscedastic data reads:
\begin{lefteqnarray}
\label{eq:RW}
&& (n-2)R=\frac{n[\widetilde{W}_{02}+(\hat{a})^2\widetilde{W}_{20}-
2\hat{a}\widetilde{W}_{11}]}{\widetilde{W}_{00}}~~;
\end{lefteqnarray}
which, in the limit of homoscedastic data,
$W_i=W=\overline{W}$, $1\le i\le n$,
$\widetilde{W}_{00}=n\overline{W}=nW$,
$\widetilde{W}_{pq}=WS_{pq}$, via
Eqs.\,(\ref{eq:wgc}), (\ref{eq:w00}), reduces to
Eq.\,(\ref{eq:Rn}), as expected.

\section{Equivalence between earlier and current
formulation}
\label{a:Theta}

\noindent\noindent

Let oblique regression models be
taken into consideration under the following
restrictive assumptions: (1) homoscedastic data;
(2) uncorrelated errors in $Y$ and in $X$; (3) normal
residuals.
Accordingly, the regression
line slope variance estimator is expressed by
Eq.\,(\ref{eq:vacc2}) where the function,
$\Theta(\hat{a}_{\rm C},\hat{a}_{\rm Y},\hat{a}_{\rm X})$,
may be different for different methods and/or models,
as shown in
Table \ref{t:Theta}.   Aiming to a formal demonstration,
some preliminary relations are needed.

In terms of dimensionless ratios, using Eqs.\,(\ref{eq:aYu})
and (\ref{eq:aXu}), Eq.\,(\ref{eq:Rn})
translates into:
\begin{lefteqnarray}
\label{eq:Ra}
&& \frac{(n-2)R}{\hat{a}S_{11}}=\frac{\hat{a}_{\rm X}}{\hat{a}}+\frac{\hat{a}}
{\hat{a}_{\rm Y}}-2=\frac{\hat{a}_{\rm X}-\hat{a}}{\hat{a}}+
\frac{\hat{a}-\hat{a}_{\rm Y}}{\hat{a}_{\rm Y}}~~;
\end{lefteqnarray}
%
where the following identities:
\begin{lefteqnarray}
\label{eq:id1}
&& \frac{\hat{a}_{\rm X}-\hat{a}}{\hat{a}}+\frac{\hat{a}-
\hat{a}_{\rm Y}}{\hat{a}_{\rm Y}}=
\frac{\hat{a}_{\rm X}-\hat{a}_{\rm Y}}{\hat{a}_{\rm Y}}-
\frac{\hat{a}_{\rm X}-\hat{a}}{\hat{a}}
\frac{\hat{a}-\hat{a}_{\rm Y}}{\hat{a}_{\rm Y}}~~;  \\
\label{eq:id2}
&& \frac{\hat{a}_{\rm X}-\hat{a}}{\hat{a}} 
\frac{\hat{a}-\hat{a}_{\rm Y}}{\hat{a}_{\rm Y}}=
\frac{\hat{a}_{\rm X}-\hat{a}_{\rm Y}}{\hat{a}_{\rm Y}}-
\frac{\hat{a}_{\rm X}-\hat{a}}{\hat{a}}-
\frac{\hat{a}-\hat{a}_{\rm Y}}{\hat{a}_{\rm Y}}=
\frac{\hat{a}_{\rm X}-\hat{a}}{\hat{a}_{\rm Y}}-
\frac{\hat{a}_{\rm X}-\hat{a}}{\hat{a}};\qquad
\end{lefteqnarray}
may easily be verified.

In the case under discussion of oblique
regression models, $\hat{a}=
\hat{a}_{\rm C}$, the following inequalities
hold (Ia90):
\begin{lefteqnarray}
\label{eq:in1}
&& \hat{a}_{\rm X}\ge\hat{a}_{\rm C}\ge\hat{a}_{\rm Y}~~;\qquad S_{11}>0~~;
\\
\label{eq:in2}
&& \hat{a}_{\rm X}\le\hat{a}_{\rm C}\le\hat{a}_{\rm Y}~~;\qquad S_{11}<0~~;
\end{lefteqnarray}
which makes the left-hand side of
Eq.\,(\ref{eq:Ra}) always positive
provided $S_{11}\ne0$.

Using the method of partial differentiation,
the regression line slope variance estimator
in the case under discussion is (C11):
\begin{lefteqnarray}
\label{eq:saC}
&& (\hat{\sigma}_{\hat{a}_{\rm C}})^2=\frac{(\hat{a}_{\rm C})^2}{n-2}\left[2
\frac{S_{02}S_{20}-(S_{11})^2}{(S_{11})^2}+2-\frac{S_{02}+(\hat{a}_{\rm C})^2
S_{20}}{\hat{a}_{\rm C}S_{11}}\right]~~;
\end{lefteqnarray}
and the substitution of Eqs.\,(\ref{eq:aYu})
and (\ref{eq:aXu}) into (\ref{eq:saC}),
using (\ref{eq:id1}) and (\ref{eq:id2})
yields after some algebra:
\begin{lefteqnarray}
\label{eq:saC2}
&& (\hat{\sigma}_{\hat{a}_{\rm C}})^2=\frac{(\hat{a}_{\rm C})^2}{n-2}\left[
\frac{\hat{a}_{\rm X}-\hat{a}_{\rm C}}{\hat{a}_{\rm C}}+
\frac{\hat{a}_{\rm C}-\hat{a}_{\rm Y}}{\hat{a}_{\rm Y}}+2
\frac{\hat{a}_{\rm X}-\hat{a}_{\rm C}}{\hat{a}_{\rm C}}
\frac{\hat{a}_{\rm C}-\hat{a}_{\rm Y}}{\hat{a}_{\rm Y}}\right]~~;
\end{lefteqnarray}
from which the following is inferred by
comparison with Eq.\,(\ref{eq:vacc2}):
\begin{lefteqnarray}
\label{eq:TC}
&& \Theta(\hat{a}_{\rm C},\hat{a}_{\rm Y},\hat{a}_{\rm X})=2
\frac{\hat{a}_{\rm X}-\hat{a}_{\rm C}}{\hat{a}_{\rm C}}
\frac{\hat{a}_{\rm C}-\hat{a}_{\rm Y}}{\hat{a}_{\rm Y}}~~;
\end{lefteqnarray}
as listed in Table \ref{t:Theta}.

Using the method of moments estimators,
the elements sample covariance matrix
are:
\begin{lefteqnarray}
\label{eq:mS}
&&     m_{XX}=       \frac{S_{20}}{n-1}~~;
\qquad m_{YY}=       \frac{S_{02}}{n-1}~~;
\qquad m_{XY}=m_{YX}=\frac{S_{11}}{n-1}~~;
\qquad
\end{lefteqnarray}
which, in terms of the variance
estimators, $(\hat{\sigma}_{xx})_{\rm S}$
(intrinsic $x$ error distribution),
$(\hat{\sigma}_{xx})_{\rm F}$
(instrumantal $x$ error distribution),
$(\hat{\sigma}_{yy})_{\rm F}$
(instrumental $y$ error distribution)
via $c_{\rm F}^2=({\sigma}_{yy})_{\rm F}/
({\sigma}_{xx})_{\rm F}$, and regression
line slope estimator, $\hat{a}_{\rm C}$,
are expressed as:
\begin{leftsubeqnarray}
\slabel{eq:mXYa}
&& m_{XX}=(\hat{\sigma}_{xx})_{\rm S}+(\hat{\sigma}_{xx})_{\rm F}~~; \\
\slabel{eq:mXYb}
&& m_{YY}=(\hat{a}_{\rm C})^2(\hat{\sigma}_{xx})_{\rm S}+
          (c_{\rm F})^2(\hat{\sigma}_{xx})_{\rm F}~~; \\
\slabel{eq:mXYc}
&& m_{XY}=m_{YX}=\hat{a}_{\rm C}(\hat{\sigma}_{xx})_{\rm S}~~;
\label{seq:mXY}
\end{leftsubeqnarray}
for further details and specification of
the model refer to the parent paper (F87,
Chap.\,1, \S 1.3.2).

The substitution of Eqs.\,(\ref{eq:mS})
and (\ref{seq:mXY}) into (\ref{eq:Rn})
yields:
\begin{lefteqnarray}
\label{eq:Rac}
&& (n-2)R_{\rm C}=(n-1)[(\hat{a}_{\rm C})^2+(c_{\rm F})^2](\hat{\sigma}_{xx})_
{\rm F}~~;
\end{lefteqnarray}
where the variance ratio, $(c_{\rm F})^2$,
may explicitly be expressed using
Eqs.\,(\ref{eq:mS}) and (\ref{seq:mXY}).
The result is:
\begin{lefteqnarray}
\label{eq:cF2}
&& (c_{\rm F})^2=\frac{\hat{a}_{\rm C}(S_{02}-\hat{a}_{\rm C}S_{11})}
{\hat{a}_{\rm C}S_{20}-S_{11}}~~;
\end{lefteqnarray}
which, using Eq.\,(\ref{eq:Rn}) and
performing some algebra, takes the
equivalent form:
\begin{lefteqnarray}
\label{eq:acF2}
&& (\hat{a}_{\rm C})^2+(c_{\rm F})^2=\frac{\hat{a}_{\rm C}(n-2)R_{\rm C}}
{\hat{a}_{\rm C}S_{20}-S_{11}}~~;
\end{lefteqnarray}
finally, the substitution of
Eq.\,(\ref{eq:acF2}) into (\ref{eq:Rac})
yields:
\begin{lefteqnarray}
\label{eq:sxxF}
&& (\hat{\sigma}_{xx})_{\rm F}=\frac{\hat{a}_{\rm C}S_{20}-S_{11}}
{(n-1)\hat{a}_{\rm C}}~~;
\end{lefteqnarray}
where the dependence on the variance
ratio, $(c_{\rm F})^2$, has been
eliminated.

In the limit of large samples ($n\gg1$,
ideally $n\to+\infty$) where, in addition,
$S_{11}\ne0$, the regression line slope
variance estimator is (F87, Chap.\,1, \S
1.3.2):
\begin{lefteqnarray}
\label{eq:saF1}
&& (\hat{\sigma}_{\hat{a}_{\rm C}})^2=\frac1{n-1}\frac1{[(\hat{\sigma}_{xx})_
{\rm S}]^2}\left\{[(\hat{\sigma}_{xx})_{\rm S}+(\hat{\sigma}_{xx})_{\rm F}]
R_{\rm C}-(\hat{a}_{\rm C})^2[(\hat{\sigma}_{xx})_{\rm F}]^2\right\}~~;\qquad
\end{lefteqnarray}
and the substitution of 
Eqs.\,(\ref{eq:Ra}), (\ref{eq:mS}),
(\ref{seq:mXY}), (\ref{eq:sxxF}),
into (\ref{eq:saF1}) yields after
some algebra:
\begin{lefteqnarray}
\label{eq:saF}
&& (\hat{\sigma}_{\hat{a}_{\rm C}})^2=\frac{(\hat{a}_{\rm C})^2}{n-2}
\nonumber \\
&&
\times\left[
\frac{\hat{a}_{\rm X}-\hat{a}_{\rm C}}{\hat{a}_{\rm C}}+
\frac{\hat{a}_{\rm C}-\hat{a}_{\rm Y}}{\hat{a}_{\rm Y}}+
\frac{\hat{a}_{\rm X}-\hat{a}_{\rm C}}{\hat{a}_{\rm C}}
\frac{\hat{a}_{\rm C}-\hat{a}_{\rm Y}}{\hat{a}_{\rm Y}}+
\frac1{n-1}\left(
\frac{\hat{a}_{\rm C}-\hat{a}_{\rm Y}}{\hat{a}_{\rm Y}}
\right)^2\right]~~;\qquad
\end{lefteqnarray}
from which the following is inferred by
comparison with Eq.\,(\ref{eq:vacc2}):
\begin{lefteqnarray}
\label{eq:TF}
&& \Theta(\hat{a}_{\rm C},\hat{a}_{\rm Y},\hat{a}_{\rm X})=
\frac{\hat{a}_{\rm X}-\hat{a}_{\rm C}}{\hat{a}_{\rm C}}
\frac{\hat{a}_{\rm C}-\hat{a}_{\rm Y}}{\hat{a}_{\rm Y}}+
\frac1{n-1}\left(
\frac{\hat{a}_{\rm C}-\hat{a}_{\rm Y}}{\hat{a}_{\rm Y}}
\right)^2~~;
\end{lefteqnarray}
as listed in Table \ref{t:Theta}.

On the other hand, the regression
line slope variance estimator
reported in an earlier attempt
[FB92, Eq.\,(4) therein] reads:
\begin{lefteqnarray}
\label{eq:saFB}
&& (\hat{\sigma}_{\hat{a}_{\rm C}})^2=\frac{(\hat{a}_{\rm C})^2}{n-2}
\left[\frac{(n-2)R_{\rm C}}{\hat{a}_{\rm C}S_{11}}+\frac{\hat{a}_{\rm C}
(S_{02}-\hat{a}_{\rm C}S_{11})}{(c_{\rm S})^2}\frac{(n-2)R_{\rm C}}
{\hat{a}_{\rm C}(S_{11})^2}\right. \nonumber \\
&& \phantom{(\hat{\sigma}_{\hat{a}_{\rm C}})^2=\frac{(\hat{a}_{\rm C})^2}
{n-2}\left[\right.}\left.
-\frac{n-2}{n-1}\frac{(\hat{a}_{\rm C})^2}{(S_{11})^2}\left(\frac
{S_{02}-\hat{a}_{\rm C}S_{11}}{(c_{\rm S})^2}\right)^2\right]~~;\qquad
\end{lefteqnarray}
where $c_{\rm S}=c$ and the counterpart
of Eq.\,(\ref{eq:cF2}) holds (C11):
\begin{lefteqnarray}
\label{eq:cS2}
&& c^2=c_{\rm S}^2=\frac{\hat{a}_{\rm C}(S_{02}-\hat{a}_{\rm C}S_{11})}
{\hat{a}_{\rm C}S_{20}-S_{11}}~~;
\end{lefteqnarray}
and the substitution of Eqs.\,(\ref{eq:Ra}),
(\ref{eq:cS2}), into (\ref{eq:saFB}), after
some algebra yields Eq.\,(\ref{eq:saF}).
Then the regression line slope variance
estimator, expressed by Eq.\,(\ref{eq:saFB}),
coincides with its counterpart deduced by use
of the method of moment estimators, expressed
by Eq.\,(\ref{eq:saF}).

Using the method of least squares estimation,
under the assumption that the entire instrumental covariance
matrix is known, the regression line slope
variance estimator reads (F87, Chap.\,1, \S
1.3.3):
\begin{lefteqnarray}
\label{eq:saL}
&& (\hat{\sigma}_{\hat{a}_{\rm C}})^2=\frac1{n-1}\frac1{(\hat{m}_{XX})^2}
\left\{\hat{m}_{XX}\hat{\sigma}_{vv}+({\sigma}_{xx})_{\rm F}\hat{\sigma}_{vv}-
(\hat{\sigma}_{xv})^2\right\}~~; \\
\label{eq:svv}
&& \hat{\sigma}_{vv}=({\sigma}_{yy})_{\rm F}+(\hat{a}_{\rm C})^2
({\sigma}_{xx})_{\rm F}-2\hat{a}_{\rm C}({\sigma}_{xy})_{\rm F}~~; \\
\label{eq:sxv}
&& \hat{\sigma}_{xv}=(\sigma_{xy})_{\rm F}-\hat{a}_{\rm C}
({\sigma}_{xx})_{\rm F}~~;
\end{lefteqnarray}
where $\hat{m}_{XX}$ is the maximum
likelihood estimator for $({\sigma}_{xx})_{\rm S}$,
$\hat{m}_{XX}=(\hat{\sigma}_{xx})_{\rm S}$.

In the special case under consideration,
$({\sigma}_{yy})_{\rm F}=(c_{\rm F})^2
({\sigma}_{xx})_{\rm F}$, $({\sigma}_
{xy})_{\rm F}=0$, Eqs.\,(\ref{eq:saL}),
(\ref{eq:svv}), (\ref{eq:sxv}),  reduce to:
\begin{lefteqnarray}
\label{eq:saL2}
&& (\hat{\sigma}_{\hat{a}_{\rm C}})^2=\frac1{n-1}\frac1{(\hat{m}_{XX})^2}
\left\{\left[\hat{m}_{XX}+({\sigma}_{xx})_{\rm F}\right]\hat{\sigma}_{vv}-
(\hat{a}_{\rm C})^2[({\sigma}_{xx})_{\rm F}]^2\right\}~~; \\
\label{eq:svv2}
&& \hat{\sigma}_{vv}=[(\hat{a}_{\rm C})^2+(c_{\rm F})^2]({\sigma}_{xx})_
{\rm F}~~; \\
\label{eq:sxv2}
&& \hat{\sigma}_{xv}=-\hat{a}_{\rm C}({\sigma}_{xx})_{\rm F}~~;
\end{lefteqnarray}
if, in addition, least squares estimators
are proportional to corresponding
moments estimators, the following relations
hold:
\begin{leftsubeqnarray}
\slabel{eq:CXa}
&& [(\hat{\sigma}_{xx})_{\rm U}]_{\rm lsc}=
C_{X_{\rm U}}[(\hat{\sigma}_{xx})_{\rm U}]_{\rm mme}~~;\qquad{\rm U}={\rm F,S}
~~; \\
\slabel{eq:CXb}
&& (\hat{\sigma}_{vv})_{\rm lsc}=C_v(\hat{\sigma}_{vv})_{\rm mme}~~;
\label{seq:CX}
\end{leftsubeqnarray}
where $C_{X_{\rm U}}$, $C_v$, are constants and the indices, lsc, mme,
mean least squares estimators and methods of moments estimators, respectively.

The substitution of Eq.\,(\ref{seq:CX})
into (\ref{eq:saL2}) yields:
\begin{lefteqnarray}
\label{eq:saL3}
&& (\hat{\sigma}_{\hat{a}_{\rm C}})^2=\frac1{n-1}\frac1{(C_{X_{\rm S}})^2
[(\hat{\sigma}_{xx})_{\rm S}]^2}
\left\{\left[C_{X_{\rm S}}(\hat{\sigma}_{xx})_{\rm S}+C_{X_{\rm F}}
(\hat{\sigma}_{xx})_{\rm F}\right]C_v[(\hat{a}_{\rm C})^2+(c_{\rm F})^2]
\right. \nonumber \\
&& \phantom{(\hat{\sigma}_{\hat{a}_{\rm C}})^2=\frac1{n-1}\frac1
{(C_{X_{\rm S}})^2}\qquad\qquad}
\times\left.
(\hat{\sigma}_{xx})_{\rm F}-(\hat{a}_{\rm C})^2[C_{X_{\rm F}}
(\hat{\sigma}_{xx})_{\rm F}]^2\right\};
\end{lefteqnarray}
where the index, mme, has been omitted
for simplifying the notation.

The substitution of Eq.\,(\ref{eq:Rac})
into (\ref{eq:saL3}) produces:
\begin{lefteqnarray}
\label{eq:saL4}
&& (\hat{\sigma}_{\hat{a}_{\rm C}})^2=\frac1{n-1}\frac1
{[(\hat{\sigma}_{xx})_{\rm S}]^2} \nonumber \\
&& \times\left\{\frac{C_v}{C_{X_{\rm S}}}\left[(\hat{\sigma}_{xx})_{\rm S}+
\frac{C_{X_{\rm F}}}{C_{X_{\rm S}}}(\hat{\sigma}_{xx})_{\rm F}\right]\frac
{n-2}{n-1}R_{\rm C}-(\hat{a}_{\rm C})^2\frac{(C_{X_{\rm F}})^2}
{(C_{X_{\rm S}})^2}[(\hat{\sigma}_{xx})_{\rm F}]^2\right\};\qquad
\end{lefteqnarray}
where the estimators, $(\hat{\sigma}_{xx})_{\rm F}$
and $(\hat{\sigma}_{xx})_{\rm S}$, are expressed by
Eqs.\,(\ref{eq:mS}), (\ref{seq:mXY}), (\ref{eq:cF2}),
(\ref{eq:sxxF}).   Accordingly, the explicit expression
of Eq.\,(\ref{eq:saL4}) after some algebra reads:
\begin{lefteqnarray}
\label{eq:saL5}
&& (\hat{\sigma}_{\hat{a}_{\rm C}})^2=\frac{(\hat{a}_{\rm C})^2}{(S_{11})^2}
\left\{\frac{C_v}{C_{X_{\rm S}}}\left[\frac{S_{11}}{\hat{a}_{\rm C}}+
\frac{C_{X_{\rm F}}}{C_{X_{\rm S}}}\frac{S_{02}-\hat{a}_{\rm C}S_{11}}
{(c_{\rm F})^2}\right]\frac{n-2}{n-1}R_{\rm C}\right. \nonumber \\
&& \phantom{(\hat{\sigma}_{\hat{a}_{\rm C}})^2=\frac{(\hat{a}_{\rm C})^2}
{(S_{11})^2}\left\{\right.}\left.
-\frac{(C_{X_{\rm F}})^2}{(C_{X_{\rm S}})^2}\frac
{(\hat{a}_{\rm C})^2}{n-1}\left[\frac{S_{02}-\hat{a}_{\rm C}S_{11}}
{(c_{\rm F})^2}\right]^2\right\};\qquad
\end{lefteqnarray}
where the restrictive assumptions:
\begin{lefteqnarray}
\label{eq:Csp}
&& \frac{C_{X_{\rm F}}}{C_{X_{\rm S}}}=1~~;\qquad\frac{C_v}{C_{X_{\rm S}}}=
\frac{n-1}{n-2}~~; 
\end{lefteqnarray}
make Eq.\,(\ref{eq:saL5}) reduce to:
\begin{lefteqnarray}
\label{eq:saL6}
&& (\hat{\sigma}_{\hat{a}_{\rm C}})^2=\frac{(\hat{a}_{\rm C})^2}{(S_{11})^2}
\left\{\left[\frac{S_{11}}{\hat{a}_{\rm C}}+\frac{S_{02}-\hat{a}_{\rm C}
S_{11}}{(c_{\rm F})^2}\right]R_{\rm C}
-\frac{(\hat{a}_{\rm C})^2}{n-1}\left[\frac{S_{02}-\hat{a}_{\rm C}S_{11}}
{(c_{\rm F})^2}\right]^2\right\};\qquad
\end{lefteqnarray}
which formally coincides with the result of an
earlier attempt where $c_{\rm S}=c$ appears
instead of $c_{\rm F}$ [FB92, Eq.\,(4) therein].

Finally, the substitution of Eqs.\,(\ref{eq:Ra})
and (\ref{eq:cF2}) into (\ref{eq:saL6}) yields
after some algebra:
\begin{lefteqnarray}
\label{eq:saFB1}
&& (\hat{\sigma}_{\hat{a}_{\rm C}})^2=\frac{(\hat{a}_{\rm C})^2}{n-2}
\nonumber \\
&&
\times\left[
\frac{\hat{a}_{\rm X}-\hat{a}_{\rm C}}{\hat{a}_{\rm C}}+
\frac{\hat{a}_{\rm C}-\hat{a}_{\rm Y}}{\hat{a}_{\rm Y}}+
\frac{\hat{a}_{\rm X}-\hat{a}_{\rm C}}{\hat{a}_{\rm C}}
\frac{\hat{a}_{\rm C}-\hat{a}_{\rm Y}}{\hat{a}_{\rm Y}}+
\frac1{n-1}\left(
\frac{\hat{a}_{\rm C}-\hat{a}_{\rm Y}}{\hat{a}_{\rm Y}}
\right)^2\right]~~;\qquad
\end{lefteqnarray}
from which the following is inferred by
comparison with Eq.\,(\ref{eq:vacc2}):
\begin{lefteqnarray}
\label{eq:TFB1}
&& \Theta(\hat{a}_{\rm C},\hat{a}_{\rm Y},\hat{a}_{\rm X})=
\frac{\hat{a}_{\rm X}-\hat{a}_{\rm C}}{\hat{a}_{\rm C}}
\frac{\hat{a}_{\rm C}-\hat{a}_{\rm Y}}{\hat{a}_{\rm Y}}+
\frac1{n-1}\left(
\frac{\hat{a}_{\rm C}-\hat{a}_{\rm Y}}{\hat{a}_{\rm Y}}
\right)^2~~;
\end{lefteqnarray}
as listed in Table \ref{t:Theta}.

The revised version of the regression line
variance estimator reported in an earlier
attempt [FB92, Eq.\,(4) therein, erratum
2011] reads:
\begin{lefteqnarray}
\label{eq:saFB2}
&& (\hat{\sigma}_{\hat{a}_{\rm C}})^2=\frac{(\hat{a}_{\rm C})^2}{n-2}
\left\{\frac{(n-2)R_{\rm C}}{\hat{a}_{\rm C}S_{11}}\right. \nonumber \\
&& \phantom{(\hat{\sigma}_{\hat{a}_{\rm C}})^2=}
\left.+\frac1{n-1}\frac
{(\hat{a}_{\rm C})^2}{(\hat{a}_{\rm C})^2+c^2}\left[1-\frac{n-2}{n-1}\frac
{(\hat{a}_{\rm C})^2}{(\hat{a}_{\rm C})^2+c^2}\right]\left[\frac{(n-2)R_
{\rm C}}{\hat{a}_{\rm C}S_{11}}\right]^2\right\}~~;\qquad
\end{lefteqnarray}
and the substitution of Eqs.\,(\ref{eq:Ra}),
(\ref{eq:cS2}), into (\ref{eq:saFB2}), after
a lot of algebra yields:
\begin{lefteqnarray}
\label{eq:saFB3}
&& (\hat{\sigma}_{\hat{a}_{\rm C}})^2=\frac{(\hat{a}_{\rm C})^2}{n-2}
\left\{\frac{\hat{a}_{\rm X}-\hat{a}_{\rm C}}{\hat{a}_{\rm C}}+
\frac{\hat{a}_{\rm C}-\hat{a}_{\rm Y}}{\hat{a}_{\rm Y}}\right. \nonumber \\
&& \phantom{(\hat{\sigma}_{\hat{a}_{\rm C}})^2=}
\left.+\frac1{n-1}\left[
\frac{\hat{a}_{\rm X}-\hat{a}_{\rm C}}{\hat{a}_{\rm C}}
\frac{\hat{a}_{\rm C}-\hat{a}_{\rm Y}}{\hat{a}_{\rm Y}}+\frac1{n-1}\left(
\frac{\hat{a}_{\rm C}-\hat{a}_{\rm Y}}{\hat{a}_{\rm Y}}
\right)^2\right]\right\}~~;\qquad
\end{lefteqnarray}
from which the following is inferred by
comparison with Eq.\,(\ref{eq:vacc2}):
\begin{lefteqnarray}
\label{eq:TFB2}
&& \Theta(\hat{a}_{\rm C},\hat{a}_{\rm Y},\hat{a}_{\rm X})=\frac1{n-1}\left[
\frac{\hat{a}_{\rm X}-\hat{a}_{\rm C}}{\hat{a}_{\rm C}}
\frac{\hat{a}_{\rm C}-\hat{a}_{\rm Y}}{\hat{a}_{\rm Y}}+
\frac1{n-1}\left(
\frac{\hat{a}_{\rm C}-\hat{a}_{\rm Y}}{\hat{a}_{\rm Y}}
\right)^2\right]~~;\qquad
\end{lefteqnarray}
as listed in Table \ref{t:Theta}.

The asymptotic expression $(n\to+\infty)$ of
Eq.\,(\ref{eq:saFB3}) is obtained neglecting
the terms of higher order with respect to
$1/n$.  The result is:
\begin{lefteqnarray}
\label{eq:saFB4}
&& (\hat{\sigma}_{\hat{a}_{\rm C}})^2=\frac{(\hat{a}_{\rm C})^2}{n-2}
\left\{\frac{\hat{a}_{\rm X}-\hat{a}_{\rm C}}{\hat{a}_{\rm C}}+
\frac{\hat{a}_{\rm C}-\hat{a}_{\rm Y}}{\hat{a}_{\rm Y}}\right\}~~;
\end{lefteqnarray}
which implies $\Theta(\hat{a}_{\rm C},\hat{a}_{\rm Y},\hat{a}_{\rm X})=0$,
as listed in Table \ref{t:Theta}.    The
asymptotic formula, Eq.\,(\ref{eq:saFB4}),
coincides with an approximation reported
in earlier attempts (Y66; Y69) for Y
models and makes a better approximation
for X, C, O, R, and B models.

\section{Data-independent residuals}
\label{a:daire}

\noindent\noindent

Let $u_{\rm A}$, $u_{\rm B}$, be
independent random variables,
$f_{\rm A}(u_{\rm A})\diff u_{\rm A}$,
$f_{\rm B}(u_{\rm B})\diff u_{\rm B}$,
related distributions, $u_{\rm A}^\ast$,
$u_{\rm B}^\ast$, related expectation
values, and $\hat{u}_{\rm A}$,
$\hat{u}_{\rm B}$, related estimators.
The random variable, $u=u_{\rm A}
u_{\rm B}$, obeys the distribution,
$f(u)\diff u=\int_{\rm U}f_{\rm A}(u_{\rm A})
f_{\rm B}(u_{\rm B})\diff u_{\rm A}
\diff u_{\rm B}$, where U is the domain
for which the product, $u_{\rm A}u_{\rm B}$,
equals a fixed $u$.   According to a
theorem of statistics, the expectation
value is $u^\ast=(u_{\rm A}u_{\rm B})^
\ast=u_{\rm A}^\ast u_{\rm B}^\ast$
and the related estimator is $\hat{u}=
\widehat{u_{\rm A}u_{\rm B}}\approx
\hat{u}_{\rm A}\hat{u}_{\rm B}$.

The special case of the arithmetic
mean reads $\overline{u}=\overline
{u_{\rm A}u_{\rm B}}\approx\overline
{u_{\rm A}}~\overline{u_{\rm B}}$ or:
\begin{lefteqnarray}
\label{eq:maAB}
&& \frac1n\sum_{i=1}^n(u_{\rm A})_i(u_{\rm B})_i\approx\frac1n\sum_{i=1}^n
(u_{\rm A})_i~\frac1n\sum_{i=1}^n(u_{\rm B})_i~~;
\end{lefteqnarray}
with regard to $u_{\rm A}$ and $u_{\rm B}$
samples with population equal to $n$.

With these general results in mind, let
Eqs.\,(\ref{eq:vaYv}), (\ref{eq:vaXv}),
(\ref{eq:caYaX}), be rewritten into the
explicit form
[Ia90, Eqs.\,(A4)-(A6) therein]%
\footnote{With regard to the above
quoted Eqs.\,(A4)-(A6), it is worth
noticing $a_{\rm Y}$,
$a_{\rm X}$, are denoted as $\beta_1$,
$\beta_2$, respectively, and $\beta_1$
has to be replaced by $(\beta_1)^{-1}$
in Eq.\,(A6) to get the right dimensions
and to be consistent with the expression
of the covariance term (Ia90, note to
Table 1 therein).}:
\begin{lefteqnarray}
\label{eq:saYe}
&& (\hat{\sigma}_{\hat{a}_{\rm Y}})^2=\frac1{(S_{20})^2}\sum_{i=1}^n\left\{
(X_i-\overline{X})^2[(Y_i-\overline{Y})-\hat{a}_{\rm Y}(X_i-\overline{X})]^2
\right\}~~; \\
\label{eq:saXe}
&& (\hat{\sigma}_{\hat{a}_{\rm X}})^2=\frac1{(S_{11})^2}\sum_{i=1}^n\left\{
(Y_i-\overline{Y})^2[(Y_i-\overline{Y})-\hat{a}_{\rm X}(X_i-\overline{X})]^2
\right\}~~; \\
\label{eq:saYXe}
&& \hat{\sigma}_{\hat{a}_{\rm Y}\hat{a}_{\rm X}}=\frac1{S_{20}S_{11}}
\sum_{i=1}^n\left\{(X_i-\overline{X})(Y_i-\overline{Y})[(Y_i-\overline{Y})-
\hat{a}_{\rm Y}(X_i-\overline{X})]\right. \nonumber \\
&& \phantom{\hat{\sigma}_{\hat{a}_{\rm Y}\hat{a}_{\rm X}}=\frac1{S_{20}S_{11}}
\left\{\sum_{i=1}^n\right.}\times\left.
[(Y_i-\overline{Y})-\hat{a}_{\rm X}(X_i-\overline{X})]\right\}~~;
\end{lefteqnarray}
where (dimensional) residuals related to
Y and X models are enclosed in square
brackets via Eqs.\,(\ref{eq:bYu}) and
(\ref{eq:bXu}), respectively.

If residuals are independent of
coordinates of observed points,
${\sf P}_i\equiv(X_i,Y_i)$,
$1\le i\le n$, then the particularization
of Eq.\,(\ref{eq:maAB}) to
$u_{\rm A}=(X_i-\overline{X})^2$,
$(Y_i-\overline{Y})^2$,
$(X_i-\overline{X})(Y_i-\overline{Y})$;
$u_{\rm B}=[(Y_i-\overline{Y})-\hat{a}_{\rm Y}(X_i-\overline{X})]^2$,
$[(Y_i-\overline{Y})-\hat{a}_{\rm X}(X_i-\overline{X})]^2$,
$[(Y_i-\overline{Y})-\hat{a}_{\rm Y}(X_i-\overline{X})]
[(Y_i-\overline{Y})-\hat{a}_{\rm X}(X_i-\overline{X})]$;
respectively, makes Eqs.\,(\ref{eq:saYe})-(\ref{eq:saYXe})
reduce to:
\begin{lefteqnarray}
\label{eq:saYt}
&& (\hat{\sigma}_{\hat{a}_{\rm Y}})^2=\frac1n\frac1{(S_{20})^2}\sum_{i=1}^n
(X_i-\overline{X})^2\sum_{i=1}^n[(Y_i-\overline{Y})-\hat{a}_{\rm Y}
(X_i-\overline{X})]^2~~; \\
\label{eq:saXt}
&& (\hat{\sigma}_{\hat{a}_{\rm X}})^2=\frac1n\frac1{(S_{11})^2}\sum_{i=1}^n
(Y_i-\overline{Y})^2\sum_{i=1}^n[(Y_i-\overline{Y})-\hat{a}_{\rm X}
(X_i-\overline{X})]^2~~; \\
\label{eq:saYXt}
&& \hat{\sigma}_{\hat{a}_{\rm Y}\hat{a}_{\rm X}}=\frac1n\frac1{S_{20}S_{11}}
\sum_{i=1}^n(X_i-\overline{X})(Y_i-\overline{Y})\sum_{i=1}^n
[(Y_i-\overline{Y})-\hat{a}_{\rm Y}(X_i-\overline{X})] \nonumber \\
&& \phantom{\hat{\sigma}_{\hat{a}_{\rm Y}\hat{a}_{\rm X}}=\frac1{S_{20}S_{11}}
\sum_{i=1}^n}\times
[(Y_i-\overline{Y})-\hat{a}_{\rm X}(X_i-\overline{X})]~~;
\end{lefteqnarray}
as outlined in an earlier attempt (Ia90).

Using Eqs.\,(\ref{eq:Spq}), (\ref{eq:aYu}),
(\ref{eq:aXu}), while performing some algebra,
Eqs.\,(\ref{eq:saYt})-(\ref{eq:saYXt}) may
be cast into the form:
\begin{lefteqnarray}
\label{eq:saYc}
&& (\hat{\sigma}_{\hat{a}_{\rm Y}})^2=\frac{(\hat{a}_{\rm Y})^2}n\frac
{\hat{a}_{\rm X}-\hat{a}_{\rm Y}}{\hat{a}_{\rm Y}}~~; \\
\label{eq:saXc}
&& (\hat{\sigma}_{\hat{a}_{\rm X}})^2=\frac{(\hat{a}_{\rm X})^2}n\frac
{\hat{a}_{\rm X}-\hat{a}_{\rm Y}}{\hat{a}_{\rm Y}}~~; \\
\label{eq:saYXc}
&& \hat{\sigma}_{\hat{a}_{\rm Y}\hat{a}_{\rm X}}=\frac{(\hat{a}_{\rm Y})^2}n
\frac{\hat{a}_{\rm X}-\hat{a}_{\rm Y}}{\hat{a}_{\rm Y}}~~;
\end{lefteqnarray}
which provide correct asymptotic
$(n\to+\infty)$ formulae but
understimate the true regression
coefficient uncertainty in samples
with low $(n\appleq50)$ or weakly
correlated population (FB92).

An inspection of Table \ref{t:Theta} shows
Eq.\,(\ref{eq:vaYu}) and the asymptotic $(n\to+\infty)$
expression of Eq.\,(\ref{eq:vaXu}) match Eqs.\,(\ref{eq:saYc})
and (\ref{eq:saXc}), respectively,
provided $n$ therein is replaced by
$(n-2)$.   Accordingly, Eqs.\,(\ref{eq:saYc})-(\ref{eq:saYXc})
translate into:
\begin{lefteqnarray}
\label{eq:saYa}
&& (\hat{\sigma}_{\hat{a}_{\rm Y}})^2=\frac{(\hat{a}_{\rm Y})^2}{n-2}\frac
{\hat{a}_{\rm X}-\hat{a}_{\rm Y}}{\hat{a}_{\rm Y}}~~; \\
\label{eq:saXa}
&& (\hat{\sigma}_{\hat{a}_{\rm X}})^2=\frac{(\hat{a}_{\rm X})^2}{n-2}\frac
{\hat{a}_{\rm X}-\hat{a}_{\rm Y}}{\hat{a}_{\rm Y}}~~; \\
\label{eq:saYXa}
&& \hat{\sigma}_{\hat{a}_{\rm Y}\hat{a}_{\rm X}}=\frac{(\hat{a}_{\rm Y})^2}
{n-2}\frac{\hat{a}_{\rm X}-\hat{a}_{\rm Y}}{\hat{a}_{\rm Y}}~~;
\end{lefteqnarray}
which are expected to yield improved values
for samples with low or weakly correlated
population.

With regard to oblique
regression models, the substitution of
Eqs.\,(\ref{eq:saYa})-(\ref{eq:saYXa})
into (\ref{eq:vaCv2}) yields after
some algebra:
\begin{lefteqnarray}
\label{eq:saCe}
&& (\hat{\sigma}_{\hat{a}_{\rm C}})^2=\frac{(\hat{a}_{\rm C})^2}{n-2}
\nonumber \\
&& \phantom{(\hat{\sigma}_{\hat{a}_{\rm C}})^2=}\times
\left\{A_{\rm XY}+\frac{2(\hat{a}_{\rm Y})^2(\hat{a}_{\rm C})^2
(A_{\rm XY})^2A_{\rm XC}A_{\rm CY}}{4(\hat{a}_{\rm Y})^2(\hat{a}_{\rm C})^2
A_{\rm XC}A_{\rm CY}+[\hat{a}_{\rm Y}\hat{a}_{\rm X}A_{\rm CY}-
(\hat{a}_{\rm C})^2A_{\rm XC}]^2}\right\};\qquad
\end{lefteqnarray}
where the identity:
\begin{lefteqnarray}
\label{eq:AXYC}
&& A_{\rm XY}=A_{\rm XU}+A_{\rm UY}+A_{\rm XU}A_{\rm UY}~~;
\end{lefteqnarray}
may easily be verified, being U = C in the case under discussion.
Accordingly, Eq.\,(\ref{eq:saCe}) may be cast under the form:
\begin{lefteqnarray}
\label{eq:saCf}
&& (\hat{\sigma}_{\hat{a}_{\rm C}})^2=\frac{(\hat{a}_{\rm C})^2}{n-2}
\left[A_{\rm XC}+A_{\rm CY}+\Theta(\hat{a}_{\rm C},\hat{a}_{\rm Y},
\hat{a}_{\rm X})\right]~~; \\
\label{eq:ThC}
&& \Theta(\hat{a}_{\rm C},\hat{a}_{\rm Y},\hat{a}_{\rm X})=A_{\rm XC}
A_{\rm CY} \nonumber \\
&& \phantom{\Theta(\hat{a}_{\rm C},\hat{a}_{\rm Y},\hat{a}_{\rm X})}\times
\left\{1+\frac{2(\hat{a}_{\rm Y})^2(\hat{a}_{\rm C})^2
(A_{\rm XY})^2}{4(\hat{a}_{\rm Y})^2(\hat{a}_{\rm C})^2
A_{\rm XC}A_{\rm CY}+[\hat{a}_{\rm Y}\hat{a}_{\rm X}A_{\rm CY}-
(\hat{a}_{\rm C})^2A_{\rm XC}]^2}\right\}\qquad
\end{lefteqnarray}
where Eqs.\,(\ref{eq:saCf}) and (\ref{eq:vacc2}) coincide
in the limit, $\Theta\to0$. 

With regard to reduced major axis regression models,
the substitution of Eqs.\,(\ref{eq:saYa})-(\ref{eq:saYXa})
into (\ref{eq:vaRv}) yields after some algebra:
\begin{lefteqnarray}
\label{eq:saRa}
&& (\hat{\sigma}_{\hat{a}_{\rm R}})^2=\frac{(\hat{a}_{\rm R})^2}{n-2}
\frac{A_{\rm XY}}2\left(1+\frac{\hat{a}_{\rm Y}}{\hat{a}_{\rm X}}\right)~~;
\end{lefteqnarray}
where the identity:
\begin{lefteqnarray}
\label{eq:aYaX}
&& \frac{\hat{a}_{\rm Y}}{\hat{a}_{\rm X}}=\frac1{A_{\rm XY}+1}~~;
\end{lefteqnarray}
may easily be verified.   Accordingly,
Eq.\,(\ref{eq:saRa}) via (\ref{eq:AXYC}) may be cast into
the form:
\begin{lefteqnarray}
\label{eq:saRf}
&& (\hat{\sigma}_{\hat{a}_{\rm R}})^2=\frac{(\hat{a}_{\rm R})^2}{n-2}
\left[A_{\rm XR}+A_{\rm RY}+\Theta(\hat{a}_{\rm R},\hat{a}_{\rm Y},
\hat{a}_{\rm X})\right]~~; \\
\label{eq:ThR}
&& \Theta(\hat{a}_{\rm R},\hat{a}_{\rm Y},\hat{a}_{\rm X})=A_{\rm XR}
A_{\rm RY}-\frac12\frac{(A_{\rm XY})^2}{A_{\rm XY}+1}~~;
\end{lefteqnarray}
where Eqs.\,(\ref{eq:saRf}) and (\ref{eq:vacr2})
coincide in the limit, $\Theta\to0$.

With regard to bisector regression models,
the substitution of Eqs.\,(\ref{eq:saYa})-(\ref{eq:saYXa})
into (\ref{eq:vaBv}) yields after some algebra:
\begin{lefteqnarray}
\label{eq:saBa}
&& (\hat{\sigma}_{\hat{a}_{\rm B}})^2=\frac{(\hat{a}_{\rm B})^2}{n-2}
\frac{A_{\rm XY}(\hat{a}_{\rm Y})^2}{(\hat{a}_{\rm Y}+\hat{a}_{\rm X})^2}
\left[\frac{a_u^2+(\hat{a}_{\rm X})^2}{a_u^2+(\hat{a}_{\rm Y})^2}+
\frac{a_u^2+(\hat{a}_{\rm Y})^2}{a_u^2+(\hat{a}_{\rm X})^2}
\frac{(\hat{a}_{\rm X})^2}{(\hat{a}_{\rm Y})^2}+2\right]~~;\qquad
\end{lefteqnarray}
which, using Eqs.\,(\ref{eq:AXYC}) and (\ref{eq:aYaX}), may be
cast under the form:
\begin{lefteqnarray}
\label{eq:saBf}
&& (\hat{\sigma}_{\hat{a}_{\rm B}})^2=\frac{(\hat{a}_{\rm B})^2}{n-2}
\left[A_{\rm XB}+A_{\rm BY}+\Theta(\hat{a}_{\rm B},\hat{a}_{\rm Y},
\hat{a}_{\rm X})\right]~~; \\
\label{eq:ThB}
&& \Theta(\hat{a}_{\rm B},\hat{a}_{\rm Y},\hat{a}_{\rm X})=\frac{A_{\rm XY}}
{(A_{\rm XY}+2)^2}
\left[\frac{a_u^2+(\hat{a}_{\rm X})^2}{a_u^2+(\hat{a}_{\rm Y})^2}+
\frac{a_u^2+(\hat{a}_{\rm Y})^2}{a_u^2+(\hat{a}_{\rm X})^2}
\frac{(\hat{a}_{\rm X})^2}{(\hat{a}_{\rm Y})^2}+2\right]\qquad~~ \nonumber \\
&& \phantom{(\Theta(\hat{a}_{\rm B},\hat{a}_{\rm Y},\hat{a}_{\rm X})=}
-A_{\rm XY}+A_{\rm XB}A_{\rm BY}~~;
\end{lefteqnarray}
where Eqs.\,(\ref{eq:saBf}) and (\ref{eq:vacb2})
coincide in the limit, $\Theta\to0$.

\section{Special cases of oblique regression}
\label{a:limca}

\noindent\noindent

With regard to homoscedastic data,
special cases of oblique regression
may be considered starting from the expression of
regression line slope and intercept estimators,
Eqs.\,(\ref{eq:acS}) and (\ref{eq:bcS}), and related
variance estimators, Eqs.\,(\ref{eq:vacc2}) and
(\ref{eq:vbcc2}) for normal residuals or (\ref{eq:vaCv})
and (\ref{eq:vbCv}) for non normal residuals.   As
outlined in the parent paper (FB92), the special cases,
$c\to+\infty$, $c\to0$, $c\to1$,
correspond to errors in $X$ negligible with respect
to errors in $Y$, errors in $Y$ negligible with
respect to errors in $X$, and orthogonal
regression, respectively.   In addition, the limiting case,
$c\to c_{\rm max}=\sqrt{a_{\rm X}a_{\rm Y}}$,
corresponds to reduced major-axis regression
(e.g., Ia90; C11).   An exhaustive discussion
related to regression line slope and intercept estimators,
can be found in an earlier attempt (C11).   Finally, the
limiting case, $c\to c_{\rm bis}$, where $c_{\rm bis}$
is expressed by Eq.\,(\ref{eq:c2B}),
corresponds to bisector regression.   The result is:
\begin{leftsubeqnarray}
\slabel{eq:aCYXOa}
&& \lim_{c\to+\infty}\hat{a}_{\rm C}=\hat{a}_{\rm Y}~;\quad
   \lim_{c\to0}\hat{a}_{\rm C}=\hat{a}_{\rm X}~;\quad
   \lim_{c\to1}\hat{a}_{\rm C}=\hat{a}_{\rm O}~;\quad
   \lim_{c\to c_{\rm max}}\hat{a}_{\rm C}=
   \hat{a}_{\rm R}~;\qquad \\
\slabel{eq:aCYXOb}
&& \lim_{c\to c_{\rm bis}}\hat{a}_{\rm C}=\hat{a}_{\rm B}~;\quad
\label{eq:aCYXO}
\end{leftsubeqnarray}
where related models are denoted by the indices,
Y, X, O, R, B, respectively.

Concerning regression line slope variance
estimators for normal residuals,
the following relations can be inferred
from Eq.\,(\ref{eq:vacc2}):
\begin{lefteqnarray}
\label{eq:aCYN}
&& \lim_{c\to+\infty}[(\hat{\sigma}_{\hat{a}_{\rm C}})_{\rm N}]^2=
\frac{(\hat{a}_{\rm Y})^2}{n-2}\left[\frac{\hat{a}_{\rm X}-\hat{a}_{\rm Y}}
{\hat{a}_{\rm Y}}+\Theta(\hat{a}_{\rm Y},\hat{a}_{\rm Y},\hat{a}_{\rm X})
\right]~~; \\
\label{eq:aCXN}
&& \lim_{c\to0}[(\hat{\sigma}_{\hat{a}_{\rm C}})_{\rm N}]^2=
\frac{(\hat{a}_{\rm X})^2}{n-2}\left[\frac{\hat{a}_{\rm X}-\hat{a}_{\rm Y}}
{\hat{a}_{\rm Y}}+\Theta(\hat{a}_{\rm X},\hat{a}_{\rm Y},\hat{a}_{\rm X})
\right]~~; \\
\label{eq:aCON}
&& \lim_{c\to1}[(\hat{\sigma}_{\hat{a}_{\rm C}})_{\rm N}]^2=
\frac{(\hat{a}_{\rm O})^2}{n-2}\left[\frac{\hat{a}_{\rm X}-\hat{a}_{\rm O}}
{\hat{a}_{\rm O}}+\frac{\hat{a}_{\rm O}-\hat{a}_{\rm Y}}{\hat{a}_{\rm Y}}+
\Theta(\hat{a}_{\rm O},\hat{a}_{\rm Y},\hat{a}_{\rm X})\right]~~; \\
\label{eq:aRON}
&& \lim_{c\to c_{\rm max}}[(\hat{\sigma}_{\hat{a}_
{\rm C}})_{\rm N}]^2=
\frac{(\hat{a}_{\rm R})^2}{n-2}\left[\frac{\hat{a}_{\rm X}-\hat{a}_{\rm R}}
{\hat{a}_{\rm R}}+\frac{\hat{a}_{\rm R}-\hat{a}_{\rm Y}}{\hat{a}_{\rm Y}}+
\Theta(\hat{a}_{\rm R},\hat{a}_{\rm Y},\hat{a}_{\rm X})\right]~~;\qquad \\
\label{eq:aBON}
&& \lim_{c\to c_{\rm bis}}[(\hat{\sigma}_{\hat{a}_{\rm C}})_{\rm N}]^2=
\frac{(\hat{a}_{\rm B})^2}{n-2}\left[\frac{\hat{a}_{\rm X}-\hat{a}_{\rm B}}
{\hat{a}_{\rm B}}+\frac{\hat{a}_{\rm B}-\hat{a}_{\rm Y}}{\hat{a}_{\rm Y}}+
\Theta(\hat{a}_{\rm B},\hat{a}_{\rm Y},\hat{a}_{\rm X})\right]~~;\qquad
\end{lefteqnarray}
where the function, $\Theta$, is listed
in Table \ref{t:Theta} for different
methods and/or models.

A comparison between Eqs.\,(\ref{eq:vaYu}),
(\ref{eq:vaXu}), and (\ref{eq:aCYN}),
(\ref{eq:aCXN}), respectively, yields:
\begin{lefteqnarray}
\label{eq:saCYN}
&& \lim_{c\to+\infty}[(\hat{\sigma}_{\hat{a}_{\rm C}})_{\rm N}]^2=
[(\hat{\sigma}_{\hat{a}_{\rm Y}})_{\rm N}]^2~~; \\
\label{eq:saCXN}
&& \lim_{c\to0}[(\hat{\sigma}_{\hat{a}_{\rm C}})_{\rm N}]^2=
[(\hat{\sigma}_{\hat{a}_{\rm X}})_{\rm N}]^2~~;
\end{lefteqnarray}
and, on the other hand:
\begin{lefteqnarray}
\label{eq:saCON}
&& \lim_{c\to1}[(\hat{\sigma}_{\hat{a}_{\rm C}})_{\rm N}]^2=
[(\hat{\sigma}_{\hat{a}_{\rm O}})_{\rm N}]^2~~; \\
\label{eq:saCRN}
&& \lim_{c\to c_{\rm max}}[(\hat{\sigma}_{\hat{a}_
{\rm C}})_{\rm N}]^2=[(\hat{\sigma}_{\hat{a}_{\rm R}})_{\rm N}]^2~~; \\
\label{eq:saCBN}
&& \lim_{c\to c_{\rm bis}}[(\hat{\sigma}_{\hat{a}_
{\rm C}})_{\rm N}]^2=[(\hat{\sigma}_{\hat{a}_{\rm B}})_{\rm N}]^2~~;
\end{lefteqnarray}
by definition of orthogonal regression
(e.g., Carroll et al., 2006, Chap.\,3,
\S 4.4.2), reduced major-axis regression
(e.g., Ia90; C11), and bisector regression
(e.g., Ia90).

Concerning regression line intercept variance
estimators for normal residuals, the following
relations can be inferred from Eq.\,(\ref{eq:vbcc2}):
\begin{lefteqnarray}
\label{eq:bCYN}
&& \lim_{c\to+\infty}[(\hat{\sigma}_{\hat{b}_{\rm C}})_{\rm N}]^2=
\frac{\hat{a}_{\rm Y}}{n-2}\frac{\hat{a}_{\rm X}-\hat{a}_{\rm Y}}
{\hat{a}_{\rm Y}}\frac{S_{11}}{S_{00}}+(\overline{X})^2
[(\hat{\sigma}_{\hat{a}_{\rm Y}})_{\rm N}]^2~~; \\
\label{eq:bCXN}
&& \lim_{c\to0}[(\hat{\sigma}_{\hat{b}_{\rm C}})_{\rm N}]^2=
\frac{\hat{a}_{\rm X}}{n-2}\frac{\hat{a}_{\rm X}-\hat{a}_{\rm Y}}
{\hat{a}_{\rm Y}}\frac{S_{11}}{S_{00}}+(\overline{X})^2
[(\hat{\sigma}_{\hat{a}_{\rm X}})_{\rm N}]^2~~; \\
\label{eq:bCON}
&& \lim_{c\to1}[(\hat{\sigma}_{\hat{b}_{\rm C}})_{\rm N}]^2=
\frac{\hat{a}_{\rm O}}{n-2}\left[\frac{\hat{a}_{\rm X}-\hat{a}_{\rm O}}
{\hat{a}_{\rm O}}+\frac{\hat{a}_{\rm O}-\hat{a}_{\rm Y}}{\hat{a}_{\rm Y}}
\right]\frac{S_{11}}{S_{00}}+(\overline{X})^2
[(\hat{\sigma}_{\hat{a}_{\rm O}})_{\rm N}]^2~~;\qquad \\
\label{eq:bCRN}
&& \lim_{c\to c_{\rm max}}[(\hat{\sigma}_{\hat{b}_
{\rm C}})_{\rm N}]^2=
\frac{\hat{a}_{\rm R}}{n-2}\left[\frac{\hat{a}_{\rm X}-\hat{a}_{\rm R}}
{\hat{a}_{\rm R}}+\frac{\hat{a}_{\rm R}-\hat{a}_{\rm Y}}{\hat{a}_{\rm Y}}
\right]\frac{S_{11}}{S_{00}}
+(\overline{X})^2[(\hat{\sigma}_{\hat{a}_{\rm R}})_{\rm N}]^2~~;\quad \\
\label{eq:bCBN}
&& \lim_{c\to c_{\rm bis}}[(\hat{\sigma}_{\hat{b}_{\rm C}})_{\rm N}]^2=
\frac{\hat{a}_{\rm B}}{n-2}\left[\frac{\hat{a}_{\rm X}-\hat{a}_{\rm B}}
{\hat{a}_{\rm B}}+\frac{\hat{a}_{\rm B}-\hat{a}_{\rm Y}}{\hat{a}_{\rm Y}}
\right]\frac{S_{11}}{S_{00}}+(\overline{X})^2
[(\hat{\sigma}_{\hat{a}_{\rm B}})_{\rm N}]^2~~;\qquad
\end{lefteqnarray}
due to Eqs.\,(\ref{eq:aCYXO}) and (\ref{eq:saCYN})-(\ref{eq:saCBN}).

A comparison between  Eqs.\,(\ref{eq:vbYu}), (\ref{eq:vbXu}),
and (\ref{eq:bCYN}), (\ref{eq:bCXN}), respectively, yields:
\begin{lefteqnarray}
\label{eq:sbCYN}
&& \lim_{c\to+\infty}[(\hat{\sigma}_{\hat{b}_{\rm C}})_{\rm N}]^2=
[(\hat{\sigma}_{\hat{b}_{\rm Y}})_{\rm N}]^2~~; \\
\label{eq:sbCXN}
&& \lim_{c\to0}[(\hat{\sigma}_{\hat{b}_{\rm C}})_{\rm N}]^2=
[(\hat{\sigma}_{\hat{b}_{\rm X}})_{\rm N}]^2~~;
\end{lefteqnarray}
and, on the other hand:
\begin{lefteqnarray}
\label{eq:sbCON}
&& \lim_{c\to1}[(\hat{\sigma}_{\hat{b}_{\rm C}})_{\rm N}]^2=
[(\hat{\sigma}_{\hat{b}_{\rm O}})_{\rm N}]^2~~; \\
\label{eq:sbCRN}
&& \lim_{c\to c_{\rm max}}[(\hat{\sigma}_{\hat{b}_
{\rm C}})_{\rm N}]^2=[(\hat{\sigma}_{\hat{b}_{\rm R}})_{\rm N}]^2~~; \\
\label{eq:sbCBN}
&& \lim_{c\to c_{\rm bis}}[(\hat{\sigma}_{\hat{b}_
{\rm C}})_{\rm N}]^2=[(\hat{\sigma}_{\hat{b}_{\rm B}})_{\rm N}]^2~~;
\end{lefteqnarray}
by definition of orthogonal regression
(e.g., Carroll et al., 2006, Chap.\,3,
\S 4.4.2), reduced major-axis regression
(e.g., Ia90; C11), and bisector regression
(e.g., Ia90).


Concerning regression line slope variance estimators for non normal residuals,
the following relations can be inferred from Eq.\,(\ref{eq:vaCv}):
\begin{lefteqnarray}
\label{eq:saCY}
&& \lim_{c\to+\infty}(\hat{\sigma}_{\hat{a}_{\rm C}})^2=
(\hat{\sigma}_{\hat{a}_{\rm Y}})^2~~; \\
\label{eq:saCX}
&& \lim_{c\to0}(\hat{\sigma}_{\hat{a}_{\rm C}})^2=
(\hat{\sigma}_{\hat{a}_{\rm X}})^2~~; \\
\label{eq:aCO}
&& \lim_{c\to1}(\hat{\sigma}_{\hat{a}_{\rm C}})^2=(\hat{a}_{\rm O})^2\frac
{a_u^4(\hat{\sigma}_{\hat{a}_{\rm Y}})^2+(\hat{a}_{\rm Y})^4
(\hat{\sigma}_{\hat{a}_{\rm X}})^2+2(\hat{a}_{\rm Y})^2a_u^2
\hat{\sigma}_{\hat{a}_{\rm Y}\hat{a}_{\rm X}}}{(\hat{a}_{\rm Y})^2
[4(\hat{a}_{\rm Y})^2a_u^2+(\hat{a}_{\rm Y}\hat{a}_{\rm X}-a_u^2)^2]}~~;
\end{lefteqnarray}
where $a_u=1$ is the (dimensional) unit
slope, according to their counterparts expressed
in the parent paper (Ia90) provided
$\vert\hat{a}_{\rm Y}\vert$ is replaced
by $\hat{a}_{\rm Y}$ therein.

On the other hand, the following relation holds:
\begin{lefteqnarray}
\label{eq:saCO}
&& \lim_{c\to1}(\hat{\sigma}_{\hat{a}_{\rm C}})^2=(\hat{\sigma}_{\hat{a}_
{\rm O}})^2~~;
\end{lefteqnarray}
by definition of orthogonal regression
(e.g., Carroll et al., 2006, Chap.\,3,
\S 4.4.2).

The intercept variance estimators for
special cases of oblique
regression, are expressed by a single
formula characterized by different
dimensionless coefficients, $\gamma_{1k}$,
$\gamma_{2k}$, where $k=1,2,4,$ for Y,
X, O, models, respectively (Ia90).
The extended expressions for oblique
regression, where $k=6$ for
C models, read:
\begin{lefteqnarray}
\label{eq:ga16}
&& \gamma_{16}=\frac{\hat{a}_{\rm C}c^2}
{\hat{a}_{\rm Y}[4(\hat{a}_{\rm Y})^2c^2+(\hat{a}_{\rm Y}\hat{a}_{\rm X}-
c^2)^2]^{1/2}}~~; \\
\label{eq:ga26}
&& \gamma_{26}=\frac{\hat{a}_{\rm C}
\hat{a}_{\rm Y}}{[4(\hat{a}_{\rm Y})^2c^2+(\hat{a}_{\rm Y}\hat{a}_{\rm X}-
c^2)^2]^{1/2}}~~;
\end{lefteqnarray}
which, for the above mentioned special
cases, reduce to:
\begin{lefteqnarray}
\label{eq:ga11}
&& \gamma_{11}=\lim_{c\to+\infty}\gamma_{16}=1~~; \\
\label{eq:ga21}
&& \gamma_{21}=\lim_{c\to+\infty}\gamma_{26}=0~~; \\
\label{eq:ga12}
&& \gamma_{12}=\lim_{c\to0}\gamma_{16}=0~~; \\
\label{eq:ga22}
&& \gamma_{22}=\lim_{c\to0}\gamma_{26}=1~~;
\end{lefteqnarray}
according to their counterparts expressed
in the parent paper (Ia90) and, in addition:
\begin{lefteqnarray}
\label{eq:ga14}
&& \gamma_{14}=\lim_{c\to1}\gamma_{16}=\frac{\hat{a}_{\rm O}a_u^2}
{\hat{a}_{\rm Y}[4(\hat{a}_{\rm Y})^2a_u^2+(\hat{a}_{\rm Y}\hat{a}_{\rm X}-
a_u^2)^2]^{1/2}}~~; \\
\label{eq:ga24}
&& \gamma_{24}=\lim_{c\to1}\gamma_{26}=\frac{\hat{a}_{\rm O}
\hat{a}_{\rm Y}}{[4(\hat{a}_{\rm Y})^2a_u^2+(\hat{a}_{\rm Y}\hat{a}_{\rm X}-
a_u^2)^2]^{1/2}}~~;
\end{lefteqnarray}
where $a_u=1$ is the (dimensional) unit
slope, according to their counterparts expressed
in the parent paper (Ia90) provided
$\vert\hat{a}_{\rm Y}\vert$ is replaced
by $\hat{a}_{\rm Y}$ therein.


The validity of Eqs.\,(\ref{eq:ga11})-(\ref{eq:ga24})
implies the validity of the following relations:
\begin{lefteqnarray}
\label{eq:sbCY}
&& \lim_{c\to+\infty}(\hat{\sigma}_{\hat{b}_{\rm C}})^2=
(\hat{\sigma}_{\hat{b}_{\rm Y}})^2~~; \\
\label{eq:sbCX}
&& \lim_{c\to0}(\hat{\sigma}_{\hat{b}_{\rm C}})^2=
(\hat{\sigma}_{\hat{b}_{\rm X}})^2~~; \\
\label{eq:sbCO}
&& \lim_{c\to1}(\hat{\sigma}_{\hat{b}_{\rm C}})^2=
(\hat{\sigma}_{\hat{b}_{\rm O}})^2~~;
\end{lefteqnarray}
in the general case of non normal residuals.

The above results cannot be extended to R
and B models i.e. $k=5,3,$ respectively,
due to use of the $\delta$-method for
determining variance estimators (Ia90),
which implies $\lim_{u_{\rm C}\to u_{\rm U}}
(\hat{\sigma}_
{\hat{u}_{\rm C}})^2\ne(\hat{\sigma}_
{\hat{u}_{\rm U}})^2$; $u=a,b$; U=R,\,B.

With regard to heteroscedastic data, the
above results can be extended starting
from the expression of regression line
slope and intercept estimators,
Eqs.\,(\ref{eq:vaCw}) and (\ref{eq:vbCw})
for normal residuals, which yields counterparts
of Eqs.\,(\ref{eq:bCYN})-(\ref{eq:bCRN})
where $n(\widetilde{w_x})_{pq}/
(\widetilde{w_x})_{00}$ appears
in place of $S_{pq}$ and $\hat{a}_{\rm Y}^
\prime=(\widetilde{w_x})_{11}/
(\widetilde{w_x})_{20}$ in place of
$\hat{a}_{\rm Y}=(\widetilde{w_y})_{11}/
(\widetilde{w_y})_{20}$.    A similar
procedure can be used for non normal residuals,
starting from Eqs.\,(\ref{eq:vaCx})
and (\ref{eq:vbCx}).

\section{C11 erratum}
\label{a:C11e}

\noindent\noindent

Due to the occurrence of printing errors,
Eqs.\,(147) and (152) in an earlier attempt
(C11) were lacking of a dimensionless factor
and must be corrected as follows:
\begin{lefteqnarray*}
&& (\hat{\sigma}_{\hat{a}_{\rm R}})^2=\frac2{n-2}
\frac{(\widetilde{w_x})_{02}}{(\widetilde{w_x})_{20}}\left\{\frac1
{(\lambda_{w_x})^2}-\sgn[(\widetilde{w_x})_{11}]\frac1{\lambda_{w_x}}\right\}
~~;\hspace{33mm}(147) \\
&& (\hat{\sigma}_{\hat{a}_{\rm R}})^2=\frac2{n-2}\frac{S_{02}}{S_{20}}\left[
\frac1{(\lambda_S)^2}-\sgn(S_{11})\frac1{\lambda_S}\right]~~;\hspace{48mm}
(152)
\end{lefteqnarray*}
which are equivalent to their alternative
expressions, Eqs.\,(149) and (154) therein,
respectively.

Sample FB09 listed in Table 2 therein has
to be read as RB09.

\section{A linear relation between primary elements from stellar
nucleosynthesis}
\label{a:lrpe}

\noindent\noindent

The composition of the interstellar medium, from which a star generation was
born, remains locked in stellar atmospheres.   Attention shall be restricted
to primary elements i.e. those synthesized in stellar cores starting from
hydrogen and helium regardless of the initial composition.   Conversely,
secondary elements can be synthesized only in presence of heavier (with
respect to hydrogen and helium) nuclides, which are called metals in
astrophysics.

The ideal situation, where a linear relation holds between primary elements in
stellar atmospheres, is defined by the following assumptions.
\begin{description}
\item[(i)]
The initial stellar mass function is universal, which implies the star
distribution in mass (including binary and multiple systems), normalized
to unity, maintains
unchanged regardless of the formation place and the formation epoch.
\item[(ii)]
Gas returned after star death is instantaneously and uniformly mixed with the
interstellar medium.
\item[(iii)]
The yield of primary elements synthesized within a star depends only on the
mass regardless of the initial composition.
\item[(iv)]
Supernovae may occur as either type II $(m>8{\rm m}_\odot)$ or type Ia
$(m\le8{\rm m}_\odot)$.
\end{description}
Accordingly, the composition of the interstellar medium is due to the
accretion of newly synthesized material via supernovae.   With regard
to a sufficiently short time step, $\Delta t$, let $n_{\rm I}$ and
$n_{\rm II}$ be the number of type Ia and type II supernovae, respectively;
in addition,
let $\delta m_{\rm W\,I}$ and $\delta m_{\rm W\,II}$ be the mean mass in the
primary element, W, newly synthesized and returned to the interstellar medium
via type Ia and type II supernovae, respectively.

Within a time range equal to the first $\ell$ steps,
$t_\ell-t_0=\ell\Delta t$, the interstellar
medium has been enriched by a mass in the primary element, W, as:
\begin{lefteqnarray}
\label{eq:mW}
&& m_{\rm W}=\sum_{k=1}^\ell[(n_{\rm II})_k\delta m_{\rm W\,II}+(n_{\rm I})_k
\delta m_{\rm W\,I}]= \nonumber \\
&& \phantom{m_{\rm W}}=
\sum_{k=1}^\ell(n_{\rm II})_k\delta m_{\rm W\,II}\left[1+\frac{(n_{\rm I})_k}
{(n_{\rm II})_k}\frac{\delta m_{\rm W\,I}}{\delta m_{\rm W\,II}}\right]~~;
\end{lefteqnarray}
where $\delta m_{\rm W\,I}$ and $\delta m_{\rm W\,II}$ may be considered, to
a good extent, as time independent.

The further assumption of time independent number ratios:
\begin{equation}
\label{eq:nIII}
\frac{(n_{\rm I})_k}{(n_{\rm II})_k}=\frac{n_{\rm I}}{n_{\rm II}}~~;
\end{equation}
makes Eq.\,(\ref{eq:mW}) reduce to:
\begin{equation}
\label{eq:mW2}
m_{\rm W}=\sum_{k=1}^\ell(n_{\rm II})_k\delta m_{\rm W\,II}\left[1+\frac
{n_{\rm I}}{n_{\rm II}}\frac{\delta m_{\rm W\,I}}{\delta m_{\rm W\,II}}
\right]~~;
\end{equation}
which implies an abundance ratio of two generic primary elements, A and B,
in stellar atmospheres i.e. interstellar medium, as:
\begin{lefteqnarray}
\label{eq:AB}
&& \frac{\exp_{10}{\rm[A/H]}}{\exp_{10}{\rm[B/H]}}\approx\frac{\phi_{\rm A}}
{\phi_{\rm B}}=\frac{Z_{\rm A}/(Z_{\rm A})_\odot}{Z_{\rm B}/(Z_{\rm B})_\odot}
=\frac{m_{\rm A}}{m_{\rm B}}\frac{(Z_{\rm B})_\odot}{(Z_{\rm A})_\odot}
\nonumber \\
&& \phantom{\frac{\exp_{10}{\rm[A/H]}}{\exp_{10}{\rm[B/H]}}\approx\frac
{\phi_{\rm A}}{\phi_{\rm B}}}=
\frac{\delta m_{\rm A\,II}}{\delta m_{\rm B\,II}}\,\displayfrac{1+\frac
{n_{\rm I}}{n_{\rm II}}\frac{\delta m_{\rm A\,I}}{\delta m_{\rm A\,II}}}{1+
\frac{n_{\rm I}}{n_{\rm II}}\frac{\delta m_{\rm B\,I}}{\delta m_{\rm B\,II}}}
\,\frac{(Z_{\rm B})_\odot}{(Z_{\rm A})_\odot}~~;
\end{lefteqnarray}
where [A/H], [B/H], are logarithmic number abundances normalized to the solar
value; $\phi_{\rm A}$, $\phi_{\rm B}$, are mass abundances normalized to the
solar value; $Z_{\rm A}$, $Z_{\rm B}$, are mass abundances; values are related
to the interstellar medium from which the star considered was born, with the
exception of $(Z_{\rm A})_\odot$, $(Z_{\rm B})_\odot$, denoting solar
abundances.   For further details refer to the parent paper (Caimmi 2007).

In terms of logarithmic number abundances, Eq.\,(\ref{eq:AB}) may be cast
under the form:
\begin{lefteqnarray}
\label{eq:reli}
&& {\rm[A/H]}={\rm[B/H]}+b~~; \\
\label{eq:bAB}
&& b=\log\left[
\frac{\delta m_{\rm A\,II}}{\delta m_{\rm B\,II}}\,\displayfrac{1+\frac
{n_{\rm I}}{n_{\rm II}}\frac{\delta m_{\rm A\,I}}{\delta m_{\rm A\,II}}}{1+
\frac{n_{\rm I}}{n_{\rm II}}\frac{\delta m_{\rm B\,I}}{\delta m_{\rm B\,II}}}
\,\frac{(Z_{\rm B})_\odot}{(Z_{\rm A})_\odot}\right]~~;
\end{lefteqnarray}
which is a linear relation with unit slope.   The general case:
\begin{equation}
\label{eq:relg}
{\rm[A/H]}=a{\rm[B/H]}+b~~;
\end{equation}
could arise under different assumptions.

\end{document}